\documentclass[twoside,12pt]{article}
\usepackage{epsfig}

\newcommand{\be}{\begin{equation}}
\newcommand{\ee}{\end{equation}}
\newcommand{\bea}{\begin{eqnarray}}
\newcommand{\eea}{\end{eqnarray}}

\usepackage{graphicx}
\usepackage[figuresright]{rotating}
\usepackage{amsmath}
\usepackage{cite}
\usepackage{amssymb}
\usepackage{amssymb}
\usepackage{amsfonts}

\newcommand{\veceps}{\mbox{\boldmath$\epsilon$}}
\newcommand{\vecsig}{\mbox{\boldmath$\sigma$}}
\newcommand{\bs}{\boldsymbol}
\graphicspath{{fig/}}
\begin{document}
\title{ \vspace{1cm} Polarizability of the Nucleon and Compton Scattering}
\author{Martin\ Schumacher\thanks{Supported by Deutsche Forschungsgemeinschaft
SPP(1034) and projects SCHU222 and 436RUS 113/510.
Email address: mschuma3@gwdg.de (Martin Schumacher)}
\\
Zweites Physikalisches Institut der Universit\"at G\"ottingen\\
Friedrich-Hund-Platz 1, D-37077 G\"ottingen}
\maketitle
\begin{abstract}
One of the central challenges of hadron physics in the regime of
strong (non-perturbative) QCD is to identify the relevant degrees
of freedom of the nucleon and to quantitatively explain experimental
data in terms of  these degrees of freedom. Among the processes
studied so far Compton scattering plays a prominent  role because of the
well understood properties of the electromagnetic interaction. 
Different approaches to describe Compton scattering have been discussed 
up to now.
It will be  shown that the  most appropriate ones  
are  provided by nonsubtracted dispersion theories  of the
fixed-$t$ and fixed-$\theta$ types, where the properties of these two
versions are complementary so that advantage can be taken from both of them.
In the frame of fixed-$t$ dispersion theory it was possible
to precisely reproduce experimental differential cross sections obtained
for the proton
in a wide angular range and for energies up to 1 GeV. At energies of the first
resonance region and below,  precise values for the
electromagnetic polarizabilities and  
spin-polarizabilities  have been determined for the proton and the neutron.
As a summary we give the following {\it recommended} experimental
values for the electromagnetic polarizabilities and backward
spin-polarizabilities of the nucleon: $\alpha_p=12.0\pm 0.6$,
$\beta_p=1.9\mp 0.6$, $\alpha_n= 12.5\pm 1.7$,
$\beta_n=2.7\mp  1.8$, in units of $10^{-4}{\rm fm}^3$ and
$\gamma^{(p)}_\pi=-38.7\pm 1.8$, 
$\gamma^{(n)}_\pi=58.6\pm 4.0$ in units of  $10^{-4}{\rm fm}^4$.
These data show that diamagnetism is a prominent property of nucleon structure.
It will be  shown that the largest part of diamagnetism, or equivalently
$(\alpha-\beta)$, is not related to the conventional isobar-meson
structure of the nucleon as showing up in meson photo-production.  Rather,
the underlying mechanism is a $t$-channel  $\sigma$-meson exchange,
with the constituent-quark-meson configuration remaining in its ground state.
The same is true for the backward spin-polarizability $\gamma_\pi$ where the
relevant meson is the $\pi^0$. Making the reasonable 
assumption that the quantities
$(\alpha-\beta)$ and $\gamma_\pi$ are related to the structure of the nucleon,
we come to the conclusion that the $\sigma$ and $\pi^0$ intermediate states 
are part of the structure of the nucleon. It is a challenge for further
research to integrate these degrees of freedom into a consistent description
of the structure of the nucleon.

\end{abstract}
\newpage
\section{Introduction}
The idea to apply the coherent elastic scattering of photons 
(Compton scattering) to  an  investigation of the internal structure
of the nucleon dates back to the early 1950's. From these early days
up to the present  the following  questions have been investigated:
\begin{itemize}
\item 
What are the appropriate degrees of freedom of the nucleon
if we probe it with (quasi-static) electromagnetic fields? 
\item
What are the appropriate theoretical tools to relate the experimental
observables  to the presumed degrees of freedom 
of the nucleon?
\item
How can one  measure differential cross sections for Compton scattering
very precisely
in view of the extremely small size of these quantities at energies
below $\pi$ photoproduction threshold  and in view
of the tremendously large background of photons due to photoproduction
of neutral pions with their subsequent decay into two photons at
energies above $\pi$ photoproduction threshold?
\end{itemize}

Though large progress has been made in these fields the work is not 
complete. The aim of this article, therefore, cannot
be to give final answers. Instead it may be considered as some 
intermediate status report written in a way that future researchers 
are given an easy access to the field.
This work is facilitated
by the fact that there are two other excellent recent reviews 
(i) the publication of Drechsel, Pasquini and Vanderhaeghen 
``Dispersion relations in real and virtual Compton scattering''
\cite{drechsel03} and the work of Wissmann  ``Compton Scattering:
Investigating the structure of the nucleon with real photons'' 
\cite{wissmann04}. The work of Drechsel et al. \cite{drechsel03}
emphasizes recent progress made in  dispersion
theories, the work  of Wissmann \cite{wissmann04} 
is concerned with experimental methods and recent experimental results
and their evaluation and interpretation in terms of the nonsubtracted
fixed-$t$ dispersion theory. Furthermore, the review articles of 
A.I. L'vov ``Theoretical
aspects of the polarizability of the nucleon'' \cite{lvov93} and
of V.A. Petrun'kin ``Electric and magnetic polarizability of hadrons''
\cite{petrunkin81} may be studied in parallel.

Using hydrogen and deuterium targets, Compton scattering by the proton 
and neutron has recently been studied at the tagged photon beam of the 
MAMI (Mainz) accelerator using  different experimental setups. 
Due to these investigations there are precise experimental
differential cross sections available for the proton in a large
angular range at energies below the $\pi$ photoproduction threshold
and for the first and second resonance region.
For the neutron, Compton differential cross sections
have been measured in the first resonance region using the method 
of quasi-free Compton scattering by the neutron bound in the deuteron.
When combined with an accurate theoretical description of the quasi-free
process these experiments are expected to provide differential cross
sections for Compton scattering by  the free neutron.  
The basic achievements of these experiments will be described in this
article.

The large improvement on  the data base achieved through these recent
experiments makes a reconsideration of dispersion theories advisable
which serve as the tools for data analysis and interpretation.
The appropriate versions of dispersion theories  are nonsubtracted
dispersion theories, either formulated for fixed scattering angle,
$\theta$, or for fixed momentum transfer squared, $t$. These two
approaches  have their technical advantages and disadvantages
and to some extent are complementary. They have in common that the
physics of subtractions inherent in the scattering amplitudes becomes an
essential part of the theory. It will become apparent in the following
that this is neither a shortcoming nor a disadvantage of the
dispersive approach.  On the contrary, the physics of subtractions
is an essential part of the physics of polarizability and Compton scattering 
and leads to new insights into the relevant degrees of freedom of the nucleon.

\subsection{The early history and basic ideas}

\begin{figure}[th]
\begin{center}
\includegraphics[scale=0.45]{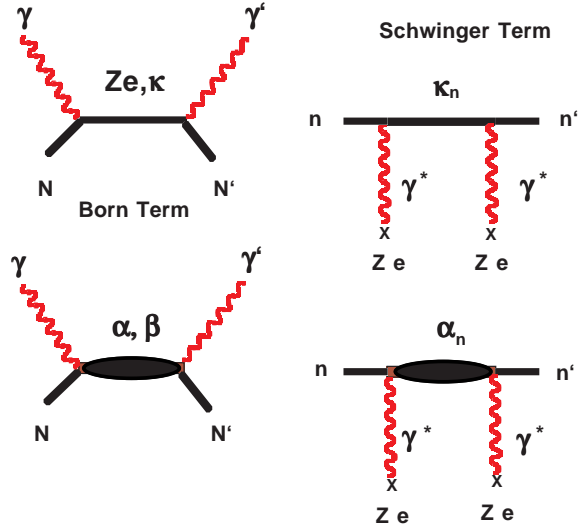}
\end{center}
\caption{Left panel: Born term  (upper part) and polarizability diagram 
(lower part) for
  Compton scattering by the nucleon. $Ze$, $\kappa$: electric charge and
  anomalous magnetic moment of the nucleon. $\alpha$, $\beta$:
  electric and magnetic polarizability.
Right panel: Schwinger term describing double Schwinger scattering due to
electric fields acting on the magnetic moment (upper part)
and 
polarizability diagram for electromagnetic scattering of a neutron in a Coulomb
field (lower part). $\kappa_n$: anomalous magnetic moment of the neutron. $Ze$:
electric charge of the scattering nucleus. $\gamma^*$: virtual photon.  
The crossed diagrams are not shown (see also \cite{lvov93}).}
\label{fig:ScatteringDiagrams}
\end{figure}

The possibility that the polarizability of the ``meson cloud'' of a proton
or neutron plays a role when the nucleon scatters a photon elastically
or  is scattered itself in a nuclear Coulomb field  has been
noted  by  several  authors \cite{sachs50,aleksandrov56} already in
the 1950's. 
A phenomenological description of the effect of nucleon
polarization on the processes
\begin{eqnarray}
&&\gamma + p   \to \gamma' + p',\\
&&\gamma^* + n \to \gamma^*{}' + n'
\end{eqnarray}
which are illustrated by the diagrams of  Figure \ref{fig:ScatteringDiagrams}
is  obtained by an 
extension of the  low-energy theorem of Low, Gell-Mann and
Goldberger \cite{low54,gell-mann54} for 
scattering amplitudes to quadratic terms in the photon energy $\omega$.
It was shown
\cite{klein55,baldin60,petrunkin61,petrunkin64,petrunkin68,shekhter68}
that allowance for the $\omega^2$-dependent terms in the amplitude for  Compton
scattering by the proton requires the introduction of 
two new constants in addition to the  charge $e$, mass $M$ and
anomalous magnetic moment $\kappa$ (for spin 1/2) of the target particle.
By analogy with the nonrelativistic theory of photon scattering by a
complex system (atom or nucleus) they have been named the 
electric and magnetic polarizabilities, $\alpha$ and 
$\beta$, respectively. 
\begin{table}[h]
\caption{Summary of results obtained in the 1950's -- 1970's  for the 
electromagnetic polarizabilities of the proton \cite{baranov00}
given without constraint of the Baldin sum rule \cite{baldin60}. 
The data are obtained \cite{baranov00} through a re-evaluation of the
differential cross sections available for energies below pion
photoproduction threshold.  The unit is  $10^{-4}{\rm fm}^3$.}
\begin{center}
\begin{tabular}{|l|c|c|c|}
\hline
Experiment&$\alpha_p$&$\beta_p$&$\alpha_p + \beta_p$\\
\hline\hline
Oxley(58) &17.0$\pm$ 8.1  &-6.7$\pm$3.7&   10.2$\pm$9.2\\
Hyman(59) &13.9$\pm$ 5.6  &-4.7$\pm$7.2&    9.2$\pm$6.1\\
Goldansky(60)&10.1$\pm$ 7.8&9.0$\pm$5.0&19.1$\pm$10.2\\
Bernardini(61)&11.4$\pm$ 2.9&2.6$\pm$2.9&$-$\\
Frish(67)&14.2$\pm 4.0$ &5.6$\pm$ 4.2& 19.8$\pm$4.3\\
Baranov(74)&11.4$\pm$ 1.4&-4.7$\pm$2.5&6.7$\pm$3.3\\
\hline
average&12.8$\pm$1.1&-0.3$\pm$1.6&12.5$\pm$2.2\\
\hline
\end{tabular}
\label{results50-70}
\end{center}
\end{table}

The first experimental data on electromagnetic polarizabilities analyzed
from  differential cross sections for
Compton  scattering by the proton were obtained by Gol'danskii et al.
in 1960 \cite{goldansky60}. 
The later  experiment of Baranov et al. \cite{baranov74} arrived at a 
good precision for the electric polarizability.
These two  experiments  were re-evaluated recently \cite{baranov00} 
and also other
experiments of the 1950's -- 1990's were given consideration.
The results of this re-evaluation are listed in 
Table \ref{results50-70}.

It should be noted that the data listed in Table  \ref{results50-70}
have been obtained without the tagging technique by which
quasi-monochromatic
photons are obtained and without including the
constraint of the Baldin sum rule \cite{baldin60} which allows to 
calculate $\alpha+\beta$ from the total photo-absorption cross section.
In view of this,  the data have to be
considered as remarkably precise. Nevertheless the important question remains 
unanswered whether  the magnetic polarizability is positive
or negative, i.e. the nucleon mainly behaves as a paramagnetic or
diamagnetic system. This question has been answered by modern experiments
described in Section 5.

Experiments on Compton scattering by the neutron appeared to be
impossible in these early days. Therefore, the electromagnetic scattering
of neutrons in a Coulomb field of heavy nuclei exploited 
in narrow-beam neutron transmission experiments
was given preference. Due to the absence of a magnetic field component
at low neutron velocities, only the electric polarizability is
measured in these experiments. 
The early history of these studies is summarized in \cite{aleksandrov92}.
The experimental results obtained in the 1980's are summarized in Table
\ref{results-80th}. 
\begin{table}[ht]
\caption{Summary of results obtained in the 1980's  for the 
electric polarizability  of the neutron 
using the method of electromagnetic scattering of neutrons in the
Coulomb field of heavy nuclei. The unit is $10^{-4}{\rm fm}^3$.}
\begin{center}
\begin{tabular}{|l|c|c|c|}
\hline
Experiment&$\alpha_n$\\
\hline
Alexandrov(86) \cite{alexandrov86}& 15 $\pm$ 33\\
Koester(86) \cite{koester86}& 30 $\pm$ 40 \\
Schmiedmayer(88) \cite{schmiedmayer88}& 12 $\pm$ 10\\
Koester(88) \cite{koester88}& 8 $\pm$ 10\\
\hline
\end{tabular}
\end{center}
\label{results-80th}
\end{table}
The data in Table \ref{results-80th} show that the errors  obtained 
for the electric polarizability of the neutron are of the order of 100\%.
This means that the electric polarizability of the neutron remained uncertain
in these early experiments. Improvements on the technique of
electromagnetic scattering of neutrons in the Coulomb field of heavy
nuclei have been made
recently. This will be described in Section 5.3.1. 

\subsection{Polarizabilities defined for real and quasi-real photons}

We include this subsection for the sake of completeness  and because
there are uncertainties about the definition of polarizabilities 
in the literature  up to recent 
times\footnote{E.g., the Review of Particle Physics of 2002 \cite{hagiwara02}
considers a  difference between the Compton
polarizabilities $\bar{\alpha}_p$ and $\bar{\beta}_p$ on the one
hand and the polarizabilities measured via static fields $\alpha_n$ 
and $\beta_n$ on the other, whereas 
the Review of Particle Physics of 2004 \cite{eidelman04}  consistently uses 
$\alpha$ and $\beta$.}.

The electric ($\alpha$) and magnetic ($\beta$) polarizabilities are 
defined through the second-order effective Hamiltonian $H^{(2)}$. 
Standard text books prefer the Gauss unit system \cite{landau90} whereas 
modern particle physics prefers the Heaviside system. Furthermore,
the Particle Data Group (PDG) \cite{eidelman04} promotes the SI system. 
A definition which is independent of the unit system reads
\begin{equation}
H^{(2)}= - 4 \pi \alpha \rho_{\rm el} - 4 \pi \beta \rho_{\rm mag}
\label{alphabetadefinition}
\end{equation}  
where $\rho_{\rm el}$ is the energy density due to the electric field, 
$\rho_{\rm mag}$ the energy density due to the magnetic field and
$H^{(2)}$ the energy change. The factor $4\pi$ indicates that the
electromagnetic polarizabilities have originally been defined in the
Gauss unit system.
Induced electric and magnetic dipole moments are given by the variational
derivatives 
\begin{equation}
{\bf d} = -\frac{\delta H^{(2)}}{\delta {\bf E}}, \quad
{\bf m} = -\frac{\delta H^{(2)}}{\delta {\bf H}},
\label{alphabetadefinition-1}
\end{equation}
for the Gauss and the Heaviside system and by 
\begin{equation}
{\bf d} = -\frac{\delta H^{(2)}}{\delta {\bf E}}, \quad
{\bf m} = -\frac{\delta H^{(2)}}{\delta {\bf B}},
\label{alphabetadefinition-2}
\end{equation}
for the SI system.

Table \ref{systems}
contains the relevant relations valid for static electric and magnetic
fields, given in the three unit systems.
\begin{table}[h]
\caption{The electromagnetic polarizabilities
$\alpha$, $\beta$  defined
in the Gauss unit system ($\hbar=c=1$) and their relations to electromagnetic
field strengths expressed in different unit
systems.  {\bf E}, {\bf H} are the electric and magnetic
field strength in
the respective unit system. In the SI system we also have  ${\bf
D}=\epsilon_0 {\bf E}$, ${\bf B}=\mu_0 {\bf H}$. {\bf d} and {\bf m}
are induced electric and magnetic dipole moments, respectively. }
\begin{center}
\begin{tabular}{|l|l|l|l|l|}
\hline
System&$H^{(2)}$&{\bf d}&{\bf m}& $\alpha_{e}=1/137.04$\\
\hline
Gauss&$-\frac12\, \alpha\,{\bf E}^2- \frac12 \, \beta\,{\bf H}^2$&
$\alpha{\bf E}$&$\beta{\bf H}$& $e^2$\\
Heaviside&$-\frac12 4\pi\, \alpha\, {\bf E}^2 -\frac12
4\pi\,\beta\, {\bf H}^2$&$4\pi \alpha{\bf E}$&$4\pi\beta{\bf
  H}$&$e^2/4\pi$\\
SI&$-\frac12 4\pi\, \alpha\, {\bf E}\cdot{\bf D} -\frac12
4\pi\,\beta\, {\bf H}\cdot {\bf B} $&
$4\pi \alpha {\bf D}$&$4\pi\beta {\bf H}$ &$e^2/4\pi\epsilon_0\hbar c$\\
\hline
\end{tabular}
\end{center}
\label{systems}
\end{table}
The corresponding 
${\cal O}(\omega^2)$ Compton differential cross section  for a
particle with charge $Ze$ but
without spin  becomes
\begin{eqnarray} 
&&\frac{d\sigma}{d\Omega}={( {\omega'}/{\omega}) }^2|T^{(2)}|^2,
 \nonumber\\
&&T^{(2)}={\boldsymbol \epsilon}\cdot {\boldsymbol \epsilon}'
\left(\frac{-Z^2 e^2}{M}+ \omega\omega'  \bar{\alpha}\right)
+({\boldsymbol \epsilon}\times {\bf \hat{k}})\cdot 
({\boldsymbol \epsilon}'\times {\bf \hat{k}}')\,\, \omega\omega'  \bar{\beta}+
{\cal O}(\omega^4)
\label{CompAmp}
\end{eqnarray}
if we use the Gauss system and if we denote -- for a temporary
distinction -- the electromagnetic polarizabilities by $\bar{\alpha}$
and $\bar{\beta}$. In the Heaviside system $\bar{\alpha}$ and
$\bar{\beta}$ carry a factor $4 \pi$ and correspondingly $|T^{(2)}|^2$
a factor $(1/4 \pi)^2$. As far as necessary we will make a statement about the
unit system. The general expression for the amplitude of
Compton scattering by a spin-$\frac12$ particle up to the
$\omega^2$ terms has been derived by Petrun'kin \cite{petrunkin81}.

The "dynamic" (Compton) electromagnetic
polarizabilities, $\bar{\alpha}$ and $\bar{\beta}$, measured in Compton
scattering experiments, and the "static" electromagnetic polarizabilities,
$\alpha$ and $\beta$,  measured with static fields or with quasi-real 
($Q^2 \to 0$) virtual
photons are identical  quantities. This is easily seen by calculating
the electric and magnetic field strengths provided by a real photon.
Simplifying the quantum-electrodynamic notation \cite{landau90} 
we write the  
vector potential of a real photon in the form
\begin{equation}
{\bf A}({\bf r},t)= {\boldsymbol \epsilon}
N e^{-i\omega t}e^{i {\bf k}\cdot {\bf r}}
+ c.c.
\label{vecpot}
\end{equation}
where ${\boldsymbol \epsilon}$ denotes the polarization vector of the
photon, $N$ a factor normalizing the 
integrated energy density to the energy $\omega$ of one photon 
and $c.c.$ stands for the complex conjugate term.
Using 
\begin{equation}
{\bf E}({\bf r},t)=- \frac{\partial}{\partial t}{\bf A}({\bf r},t),
\quad\quad {\bf H}({\bf r},t)=\nabla\times {\bf A}({\bf r},t), 
\label{EH}
\end{equation}
we arrive at the conclusion 
 that the  
factors $(\omega{\boldsymbol \epsilon})\cdot(\omega'{\boldsymbol \epsilon}')$
and $(\omega\, {\bf \hat{k}}\times{\boldsymbol \epsilon})\cdot
(\omega'\, {\bf \hat{k}}'\times{\boldsymbol \epsilon}')$ entering 
into (\ref{CompAmp}) may be interpreted as products 
${\bf E}\cdot {\bf E}'$ and ${\bf H}\cdot {\bf H}'$ of electric and magnetic
fields, respectively.

Static electric fields {\bf E}  are provided
with sufficient strength  by heavy nuclei. Therefore, use may be
made of the differential cross section for  electromagnetic scattering 
of slow neutrons in the Coulomb field of heavy nuclei \cite{lvov93}
\begin{equation}
\frac{{\rm d}\sigma_{\rm pol}}{{\rm d}\Omega}=
\pi M p (Z e)^2{\rm Re} a\left\{ \alpha_n \sin \frac{\theta}{2}
- \frac{e^2 \kappa^2_n}{2 M^3} \left(1 - \sin\frac{\theta}{2}\right)
\right\} \label{neutron}
\end{equation}
applied in its solid-angle integrated form
\begin{equation}
\sigma_{\rm pol}=\frac83
\pi^2 M p (Z e)^2{\rm Re} a\left\{ \alpha_n 
- \frac{e^2 \kappa^2_n}{4 M^3}
\right\} \label{neutron-1}
\end{equation}
to determine the neutron electric polarizability $\alpha_n$.
In  (\ref{neutron}) and  (\ref{neutron-1})  $p$ is the neutron 
momentum and $- a$ the
amplitude for hadronic scattering by the nucleus. The second term in
the braces is due to the Schwinger term, {\it i.e.} the term
describing neutron scattering in the Coulomb field due to the magnetic
moment of the neutron only. As outlined above the two different
definitions of electromagnetic polarizabilities are equivalent. This
means that the polarizabilities measured by quasi-static
electromagnetic fields, {\it i.e.},  either by using real photons 
to measure amplitudes for Compton scattering up to
${\cal O}(\omega^2)$ or by using virtual
photons in the initial state at the limit $Q^2 \to 0$ lead to the
same result. As explained above,
the magnetic polarizability cannot be measured by
electromagnetic scattering of slow neutrons in a Coulomb field.

\subsection{Calculations of electromagnetic polarizabilities from
models of the nucleon}

The electromagnetic polarizabilities of the nucleon
have attracted a great number of researchers to calculate these 
quantities from nucleon  models. These are the MIT bag model
\cite{hecking81}, the nonrelativistic quark model
\cite{dattoli77}
the chiral quark model
\cite{weiner85},
the chiral soliton model
\cite{scocola89}
and the Skyrme  model
\cite{nyman84}. 
In addition, model calculations for the charged pion have been carried
out
\cite{bernard88}.

The majority of these calculations apply the concept of internal
coordinates of the nucleon which are well defined in a nonrelativistic
approach.
In this  nonrelativistic approach
second-order perturbation theory leads
to the expressions \cite{petrunkin64,ericson73,friar75}
\begin{eqnarray}
&&\alpha=2\sum_{n\neq 0}\frac{|\langle n^{(i)}|D_z|0\rangle|^2}{
E^{(i)}_n-E^{(i)}_0}+Z^2\frac{e^2\langle r^2_E\rangle}{
3M},    \label{eq0}\\
&&\beta=2\sum_{n\neq 0}\frac{|\langle n^{(i)}|M_z|0\rangle|^2}{
E^{(i)}_n-E^{(i)}_0} -e^2\sum_i\frac{q^2_i}{6m_i}\langle 0|\rho^2_i|0\rangle
-\frac{\langle 0|{\bf D}^2|0\rangle}{2M}.
\label{eq1}
\end{eqnarray}
These equations contain the retardation correction 
$\Delta\alpha=Z^2 e^2 \langle r^2_E \rangle /3 M$ of the electric
polarizability and the
diamagnetic susceptibility 
$\beta_{\rm dia}= -e^2\sum_i (q^2_i/6m_i)\langle 0|\rho^2_i|0\rangle
  -\langle 0|{\bf D}^2|0\rangle/2M$ in addition to the
leading terms coming from second-order perturbation theory in the 
long wave-length limit. In (\ref{eq0}) and (\ref{eq1}) $Z$ and $M$ are
the  charge number and total mass, respectively, of the hadron and
$r^2_E$ the square of the quadratic charge radius. The quantity 
${\bf D}$ is the electric dipole moment and  $D_z$ and $M_z$ the
z-components of the electric and magnetic dipole moments, respectively.
The quantities $q_i$, $m_i$ and $\rho_i$ are the charge fraction, the 
mass and the internal coordinate of the constituents inside the hadron.
Recently, it has been shown that these relations  contain large uncertainties,
especially in the $r^2_E$ dependent  retardation correction
because there are other relativistic terms of at least the same order
\cite{lee01}. 
This will be outlined  in more detail in  Section 1.4.

A simple argument may  be first discussed here. One of several
equivalent derivations of (\ref{eq0}) makes use of the separation of
the Compton amplitude into a Thomson amplitude calculated classically
for an extended
charged sphere and an internal amplitude which -- in the long wavelength
limit -- is given by the first term on r.h.s. of (\ref{eq0}).
The classical treatment of Thomson  scattering of photons by an extended
charged sphere leads to
\begin{equation}
T^{\rm cl}=T^0 F_E({\bf k})F_E({\bf k'})= 
T^0 \left(1-\frac13 \langle r^2_E 
\rangle \omega\omega'
+{\cal O}(\omega^4) \right) 
  \label{R1}
\end{equation}
where  $T^0$ is the scattering amplitude of a point-like particle and 
\begin{equation}
F_E({\bf k})=\frac{1}{e} \int d^3r \rho({\bf r})
  e^{i{\bf k}\cdot{\bf r}}  \label{R2}
\end{equation}
is the form-factor with respect to the wave vector ${\bf k}$ of the 
photon. Then the
second term in the parentheses of (\ref{R1}) may be considered as part
of the electric polarizability so that the expression on the r.h.s.
of (\ref{eq0}) is obtained. It can be shown  \cite{schumacher94,huett00}
that the 
amplitude $T^0 F^2_E({\bf k})$
is not in agreement with forward-direction dispersion theory,
i.e. it violates causality.
As a consequence we conclude that when  using the expression of
(\ref{eq0})  it is necessary to take into account
relativistic corrections, because they substantionally modify the prediction.
An equivalent conclusion has previously been discussed by L'vov
\cite{lvov93}. 
A possible extension of the expressions given in
(\ref{eq0}) 
which includes relativistic corrections 
will be discussed in Section 1.4.

\subsection{Effects of the Breit Hamiltonian and
of the relativistic center-of-mass variable}

The difficulties discussed in the foregoing find an explanation in the
fact that the retardation correction $\Delta\alpha$ is only one of
several relativistic corrections.

In order to take into account these further  
relativistic corrections for the most simple case of two charged point
particles, we write
the two-particle  Hamiltonian with Coulomb
interaction \cite{lee01} in the form
\begin{equation}
\tilde{H}[{\bf A}]=\tilde{H}_{\rm nr}[{\bf A}]+\tilde{H}_{\rm B}[{\bf A}]
+\delta \tilde{H}[{\bf A}]
\label{tildeH}
\end{equation}
where ${\bf A}$ is the vector potential with the Coulomb field as its
time component, $\tilde{H}_{\rm nr}[{\bf A}]$ the nonrelativistic
Hamiltonian, $\tilde{H}_{\rm B}[{\bf A}]$ the Breit Hamiltonian and
$\delta \tilde{H}[{\bf A}]$ a spin dependent correction to both, the 
nonrelativistic  Hamiltonian and the Breit Hamiltonian. For 
a two-particle system the nonrelativistic Hamiltonian has the form  
\begin{equation}
\tilde{H}_{\rm nr}[{\bf A}] =\frac{{\boldsymbol \pi_1}^2}{2m_1}
+ \frac{{\boldsymbol \pi_2}^2}{2m_2} - \frac{g}{|{\bf r_1} - {\bf
    r_2}|}
\label{milstein-1}
\end{equation}
where  ${\boldsymbol \pi}_i={\bf p}_i  - e_i{\bf A}({\bf r}_i)$
and $g$ the coupling constant which, for two particles with electric
charges $e_1$ and $e_2$, is given by $g=-e_1 e_2>0$. This means that
the unit system $e^2=1/137.04$, $\hbar=c=1$ is used. The Breit
Hamiltonian takes into account relativistic corrections in lowest
order. These are corrections to the kinetic energies of the particles
and corrections to the Coulomb field which has to be supplemented by
the space component of the vector potential. This latter correction
frequently is termed ``magnetic quanta exchange''. For two particles
the Breit Hamiltonian may be written in the form \cite{pilkuhn79}
\begin{equation}
\tilde{H}_{\rm B}[{\bf A}]=-\frac{({\boldsymbol \pi}^2_1)^2}{8 m^3_1}
-\frac{({\boldsymbol \pi}^2_2)^2}{8 m^3_2}
+\frac{g}{2 m_1 m_2}\left( \frac{\delta^{ij}}{r}+
\frac{{r}^i{r}^j}{r^3}\right)
{\pi}^i_1{\pi}^j_2.
\label{milstein-2}
\end{equation}
The spin dependent correction  $\delta \tilde{H}[{\bf A}]$ takes into account
the effects of the  magnetic moments of the two particles
coming into play because of the non-vanishing space component of the 
vector potential. 
As discussed in detail in \cite{lee01} also the internal
dipole moment 
\begin{equation}
D=e_1{\bf r}_1 + e_2 {\bf r}_2 - (e_1+e_2){\bf R}_{\rm cm}
\label{dipole}
\end{equation}
needs a relativistic correction because of the 
center-of-mass vector ${\bf R}_{\rm cm}$ which has to be  
 defined to satisfy the relations
\begin{equation}
[{\bf R}_{\rm cm},{\bf P}]=i, \quad i[H_{\rm tot},{\bf R}_{\rm cm}]=
\frac{\bf P}{H_{\rm tot}}.
\label{milstein-3}
\end{equation}
Here, $H_{\rm tot}$ is the total relativistic Hamiltonian of the system
and 
{$\bf P$} the total momentum. This leads to the center of mass operator
\begin{equation}
{\bf R}_{\rm cm}={\bf R}+\frac{(m_2 - m_1)}{2M^2}
\left( \{ {\bf r},H_{\rm nr} \}+ g \frac{\bf r}{r}\right)
\label{milstein-4}
\end{equation}
with ${\bf r}_1 ={\bf R} + \frac{m_2}{M}{\bf r},\,\, 
{\bf r}_2 ={\bf R} - \frac{m_1}{M}{\bf r},\,\,
M =m_1+m_2$ and the notation $\{a,b\}=ab + ba$.

For a two-particle system with charges $e_1$ and $e_2$ the following
results are obtained
\begin{equation}
\alpha_{0 {\rm nr}}=\frac{9}{2 \mu g^4}\left( \frac{e_1}{m_1}- 
\frac{e_2}{m_2}\right)^2,
\label{nr}
\end{equation}
\begin{equation}
\Delta \alpha= \frac{e_1 + e_2}{M g^2}\left( \frac{e_1}{m^2_1} + 
\frac{e_2}{m^2_2} \right),
\label{Delta}
\end{equation}
\begin{equation}
\alpha_{0 {\rm B}}= - \frac{1}{g^2}\left(\frac{e_1}{m_1}-\frac{e_2}{m_2}
\right)^2 \left(\frac{121}{6 \mu} - \frac{113}{4 M}\right)
-\frac{e_1 + e_2}{M g^2}\left(\frac{e_1}{m_1}-\frac{e_2}{m_2}\right)
\frac{m_1-m_2}{2m_1m_2},
\label{breit}
\end{equation}
where $\mu = m_1 m_2/(m_1+m_2)$.
The first term (\ref{nr}) corresponds to the nonrelativistic result of 
second-order perturbation theory, the second (\ref{Delta}) to the retardation
correction and the third (\ref{breit}) to the Breit correction. The correction
due to spins has not been given consideration here.

Applying these formulae to a hydrogen-like atom with charge number $Z$
and mass $M$
we arrive at the following results
\begin{eqnarray}
&&\alpha_{0 {\rm nr}}=\frac92 \frac{1}{Z^4}\, a^3_{\rm H},\nonumber\\
&&\Delta\alpha= - \frac{m_e}{M}\frac{Z-1}{Z^2} \alpha^2_e 
\,a^3_{\rm H},\nonumber\\
&&\alpha_{0 {\rm B}} = -\frac{1}{Z^2}\frac{121}{6}\alpha^2_e
\,a^3_{\rm H},
\label{alphaHydrogen}
\end{eqnarray} 
where $m_e$ is the electron mass and 
$a_{\rm H}$  the Bohr radius of the hydrogen atom. 
The negative $\Delta\alpha$ is a consequence of the negative electron charge.
These formulae show that the retardation correction $\Delta\alpha$
is much smaller than the Breit correction $\alpha_{0 {\rm B}}$
for any charge number Z.
For $Z=1$ $\alpha_{0 {\rm nr}}$
provides  the by far largest contribution to the electric
polarizability.  With increasing  $Z$ the Breit correction leads to a
large negative contribution. Due to the cancellation between
$\alpha_{0 {\rm nr}}$ and $\alpha_{0 {\rm B}}$ the calculated 
electric polarizability becomes small and  
even becomes equal to zero for $Z=65$ or $Z\alpha=0.47$. 
We interpret this result as a consequence of the fact that the Breit
correction is valid for $Z\alpha\ll 1$ only.

{\it Summary:}
Electric polarizabilities  evaluated  from experiments using static electric
fields were occasionally   assumed to be identical with
the ``true'' electric polarizability $\alpha_0$,
a quantity  given by the first term on r.h.s. of (\ref{eq0}). 
The error contained in this identification has already been 
discussed  and corrected by L'vov  \cite{lvov93}.
This quantity merely is an incomplete nonrelativistic
decription of the electric polarizability but not a physical
observable. Furthermore, also the inclusion of the second term on the 
r.h.s. of (\ref{eq0}) does not lead to a reasonable  
expression because of missing
relativistic corrections. Finally, the polarizabilities $\alpha$ and
$\beta$ measured -- or at least defined -- by static electromagnetic fields
are identical to the polarizabilties $\bar \alpha$ and $\bar \beta$ measured
by Compton scattering.

\subsection{Chiral perturbation theory}

Electromagnetic polarizabilities have been calculated in chiral 
perturbation theory ($\chi$PT) in a series of papers 
\cite{bernard92}. In a leading-order relativistic $\chi$PT calculation, 
it was obtained 
\begin{eqnarray}
&&\alpha_p = \frac{e^2g^2_{\pi N}}{192 \pi^3 M^3}\left\{ \frac{5\pi}{2\mu}
+ 18\ln \mu +\frac{33}{2}+ \cal{O}(\mu)\right\}=\,\,7.4, \nonumber\\
&&\beta_p = \frac{e^2g^2_{\pi N}}{192 \pi^3 M^3}\left\{ \frac{\pi}{4\mu}
+ 18\ln \mu +\frac{63}{2}+ \cal{O}(\mu)\right\}=-2.0, \nonumber\\
&&\alpha_n = \frac{e^2g^2_{\pi N}}{192 \pi^3 M^3}\left\{ \frac{5\pi}{2\mu}
+ 6\ln \mu -\frac{3}{2}+ \cal{O}(\mu)\right\}=10.1,\nonumber\\
&&\beta_n = \frac{e^2g^2_{\pi N}}{192 \pi^3 M^3}\left\{ \frac{\pi}{4\mu}
+ 6\ln \mu +\frac{5}{2}+ \cal{O}(\mu)\right\}=-1.2,
\label{meis-4}
\end{eqnarray}
where $\mu=m_\pi/M$ is the pion-nucleon mass ratio. In (\ref{meis-4})
the numerical values correspond to the full calculation whereas the first
three terms in the braces represent a semi-relativistic expansion up to and
including next-to-leading order terms. This finding was 
explained \cite{lvov93a} by using dispersion relations. The analog
to chiral perturbation theory as underlying the results given 
in (\ref{meis-4}) is to carry
out a calculation in terms of a dispersion theory where recoil
corrections are included and the meson photoproduction amplitudes 
are used in the Born approximation.  The results obtained in this way
from dispersion theory
are $\alpha_p=7.3$, $\beta_p=-1.8$, $\alpha_n=9.8$, $\beta_n=-0.9$ in
reasonable agreement with the numbers in (\ref{meis-4}). 
  
To leading order  the results given in (\ref{meis-4}) have been
summarized  (see e.g. \cite{drechsel03,thomas01,bernard95}) in the form
\begin{equation}
\alpha_p=\alpha_n=10\beta_p=10\beta_n=\frac{5}{96 \pi}
\left( \frac{g_A}{f_\pi}\right)^2 \frac{\alpha_e}{m_\pi}
\simeq 12.6\cdot 10^{-4}{\rm fm}^3
\label{leading}
\end{equation}
where use has been made of the Goldberger-Treiman relation
$g_{\pi N}f_{\pi}=g_AM$, with $g_A$ being  the axial coupling of the nucleon 
and $f_\pi$ the pion-decay constant.
The relation given in (\ref{leading}) was found to be exactly 
identical to the result of a calculation in
heavy baryon $\chi$PT (HB$\chi$PT) \cite{bernard92a}.
The calculation carried out in 
HB$\chi$PT  was extended beyond the one-loop approach  \cite{bernard93}
yielding $\alpha_p=10.5 \pm 2.0$, $\beta_p=3.5\pm 3.6$, 
$\alpha_n= 13.4\pm 1.5$ and $\beta_n= 7.8 \pm 3.6$. 
All terms up to order ${\cal O}(q^4)$ were included. At this order counter
terms enter which were determined by several different procedures.

Subsequent
attempts  to overcome the shortcoming of the missing $\Delta$
contribution in terms of a  ``small scale expansion'' (SSE) 
\cite{hemmert98} led to 
$\alpha_p= 16.4$ and $\beta_p=9.1$ where both numbers are much larger
than the experimental results. 
Attempts to reproduce the predictions of Baldin's sum rule \cite{baldin60}
for $\alpha+\beta$ have been published in \cite{pascalutsa04}, emphasizing
the need for relativistic chiral EFT calculations. Recent calculations
on differential cross sections for Compton scattering by the nucleon
\cite{hildebrandt03,pascalutsa03} have in common that free parameters 
are introduced,
corresponding to counter-terms of unnatural size \cite{hildebrandt03}. 
A counter-term of unnatural
size signals the fact that the theory fails because of physics beyond the
theory \cite{hildebrandt03}. It remains to be a challenge for 
the $\chi$PT community to find out
what this missing physics is\footnote{Among other aspects it is of importance
  to know that $\chi$PT includes the graph f) of Figure \ref{ComptonGraphs} 
but does not
explicitly include the graph g).}

\subsection{Diagrammatic description of Compton scattering and
  dispersion theories}

The foregoing approaches have in common that they start from a given
set of degrees of freedom of the nucleon and calculate the electromagnetic
polarizabilities in theoretical frameworks which in general contain
further approximations. 
A more satisfactory approach of course would be to start from a
phenomenological description of the structure of the
nucleon and to relate the information obtained to polarizability 
and Compton scattering without further approximations. The appropriate
tools for this latter approach are provided by dispersion theories.

\begin{figure}[h]
\centering\includegraphics[scale=0.8]{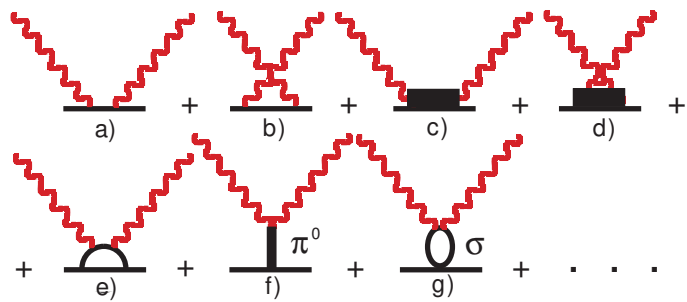}
\caption{a) and b): Born terms, c) and d): scattering through
isobar excitation, e) scattering through excitation of the pion
cloud,
f) scattering via exchanges of  pseudoscalar mesons in the $t$-channel, 
g) scattering via exchanges of
scalar mesons in the $t$-channel.}
\label{ComptonGraphs}
\end{figure}
\begin{figure}[h]
\centering\includegraphics[scale=0.95]{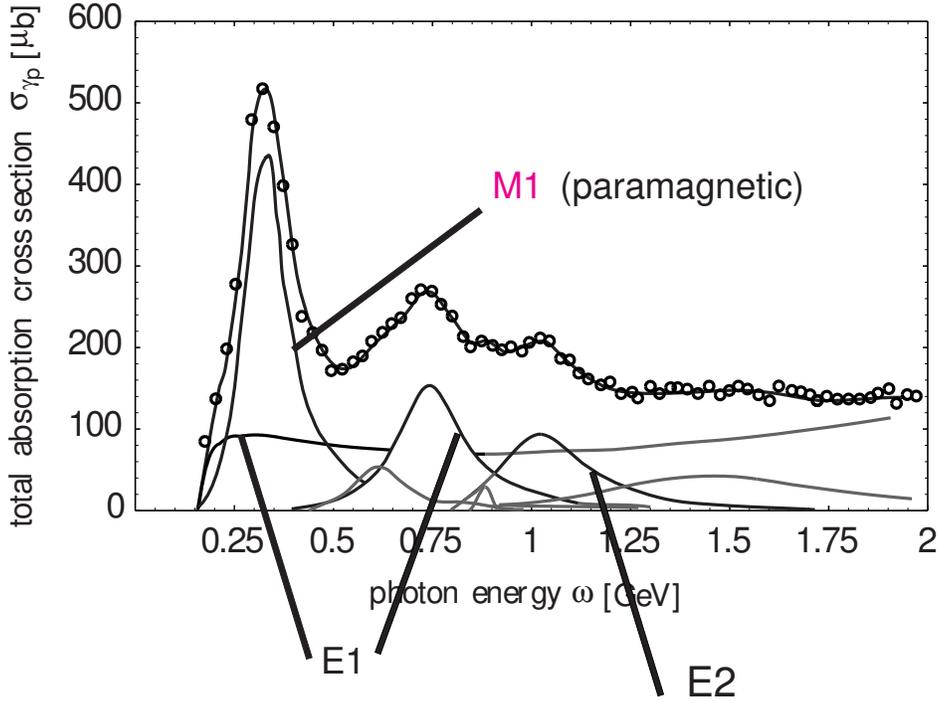}
\caption{Total photo-absorption cross section of the proton
  disentangled into multipole components.} 
\label{photoabsorption-1}
\end{figure}
\begin{figure}[h]
\centering\includegraphics[scale=1.1]{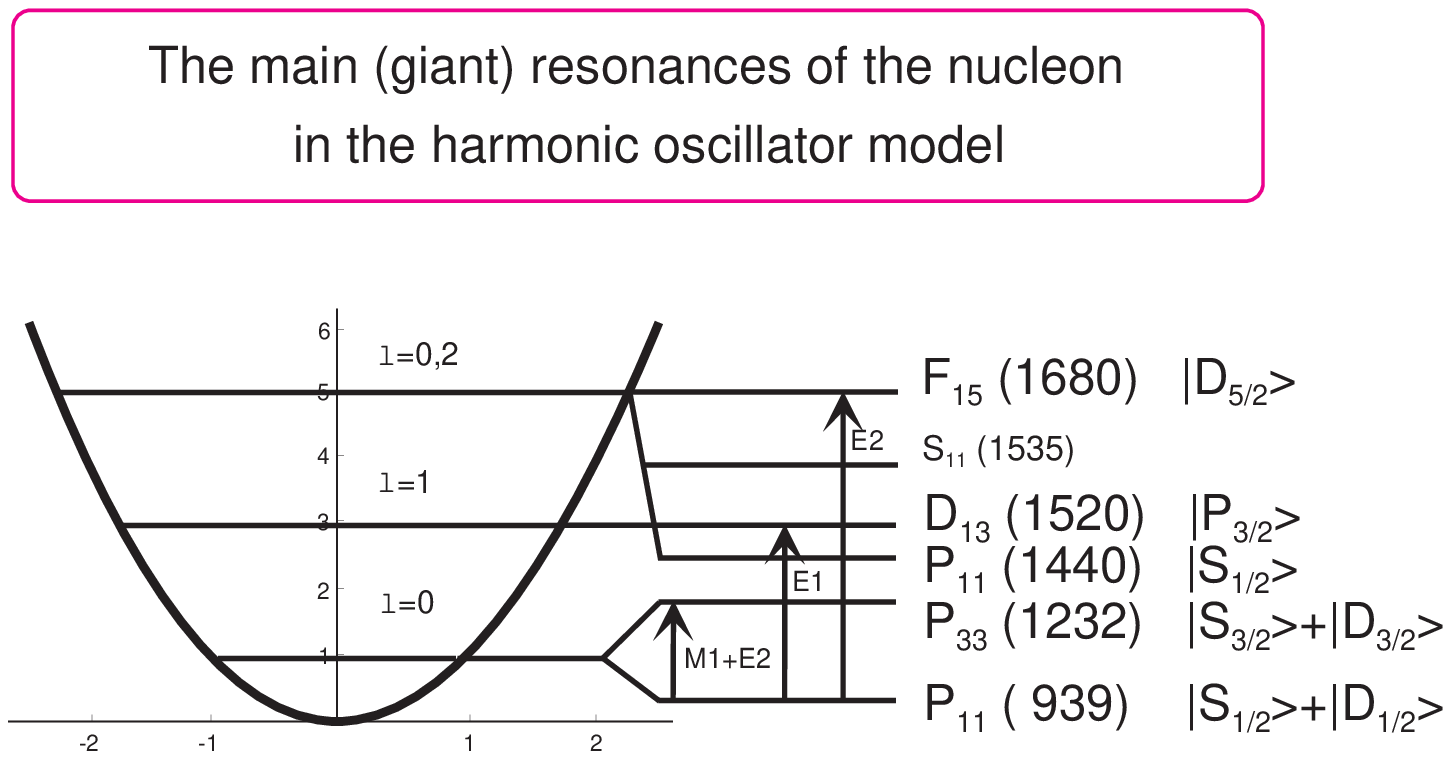}
\caption{
The main (giant) resonances of the nucleon in a diquark-quark
harmonic oscillator model. The lines inside the parabola denote the
harmonic oscillator states. The lines on the right side indicate the
centers of the resonant states. The observed widths of these states are
generated through emission and reabsorption of pions by the constituent 
quarks.} 
\label{photoabsorption-2}
\end{figure}

In a diagrammatic approach Compton scattering by the nucleon is
described by a series of graphs which give an overview over the
processes related to Compton scattering. An example is given in 
Figure \ref{ComptonGraphs} containing the   selection
of graphs given by Drechsel et al.  \cite{drechsel03}. 
These graphs denote  the Born term a) and its crossing
partner b), the scattering through the excitation of isobars c) and the
crossing  partner d), and the scattering through the meson
cloud e). This latter graph is one example of a larger 
number of graphs as discussed in chiral perturbation theory. The graphs
f) and g) depict scattering through exchanges of  pseudoscalar and scalar
mesons in the $t$-channel, respectively.

The Born terms a) and b) involve the scattering
through the electric charge and magnetic moment of the nucleon without
excitation of the internal structure. This process may be viewed 
as classical Thomson scattering, approximating a high-energy process where
the virtual creation and annihilation of a nucleon-antinucleon
pair in the field of the scattering nucleon is involved.

The physical meaning of graphs c) to e) is illustrated in Figures 
\ref{photoabsorption-1} and \ref{photoabsorption-2}. 
The M1-strength leading to a 
strong source of paramagnetism is provided
by the $P_{33}(1232)$ or $\Delta$ resonance. The $E1$-strength has a 
nonresonant component due to 
single-pion photoproduction, supplemented by a high-energy tail of
double- and many-pion photoproduction which presumably is also mainly
of $E1$ multipolarity. The nonresonant single-pion component may be
identified with the meson-cloud contribution to the electric
polarizability, illustrating the  common supposition that the electric
   polarizability is mainly generated  by the meson cloud.
A further smaller amount of $E1$-strength is provided by the 
$D_{13}(1520)$ resonance. The $E2$-strength of the
$F_{15}(1680)$-resonance makes contributions to the electric as well
as the magnetic polarizability.

Graph f) in Figure \ref{ComptonGraphs} represents the 
so-called pion pole term as introduced 
by Low in 1958 \cite{low58}. The idea is that not the isobar-meson structure 
of the nucleon is excited as discussed above. Instead,
a $\pi^0$ meson   is created in the intermediate state which then
couples to the incoming and outgoing photon on the one side and to the
nucleon on the other. Because of the
small widths of the $\pi^0$ meson resonance, this contribution 
to the Compton scattering process can be described by a pole located 
at positive $t=m^2_\pi$ in the Mandelstam plane ($t$-channel). 

The scalar analogue of the pseudoscalar $\pi^0$ pole term depicted in g)
has first been discussed by Hearn and Leader in 1962 \cite{hearn62}. These
authors pointed out  that in addition to the pion pole a continuum
of two-pion states should make a contribution to Compton scattering. 
Since these two-pion states do not form a particle with a narrow width,
the mathematical tools necessary for the description of this
contribution have to be somewhat different. Nevertheless,
it was discussed very early (see e.g. \cite{koeberle68})
that these two-pion states possibly may form some
kind of a broad resonance with an energy of about 600 MeV. 
Nowadays this resonance is identified with the 
pole-structure seen in 
$\pi$-$\pi$ phase-analyses of data obtained
in $\pi \, N\to  N\, \pi\pi$ scattering experiments.
It appears possible that the meaning of graph g) in Figure 2 goes
beyond the physics suggested by this graph.
As will be explained 
in more detail later, the  interpretation of the graph g) 
of Figure \ref{ComptonGraphs} in terms of correlated $\pi$ pairs 
is incomplete. Only part
of the strength seems to be 
exhausted by the $|\pi\pi\rangle$  component whereas an other part
correspond to a  ``core'' component of probably
$|q\bar{q}\rangle$ structure \cite{scadron04}. 
This important finding will be worked out and substantiated 
in more detail in our ongoing
work \cite{levchuk05}.

\section{Forward and backward Compton scattering and related phenomena}

The physics  of polarizability and Compton scattering can be understood  best
when  considering the extreme forward and extreme backward directions.
Indeed,  the relevant phenomena  show up very transparently  in these two cases
whereas for experimental purposes, of course, intermediate angles have
to be used. Our concept, therefore, is to first consider forward and
backward scattering and then to develop the tools to cover also
intermediate angles.

The differential cross section for Compton scattering
\begin{equation}
\gamma N \to \gamma' N'
\end{equation}
may be written in the form \cite{babusci98}
\begin{equation}
\frac{d \sigma}{d \Omega}= \Phi^2 |T_{fi}|^2
\end{equation}
with $\Phi=\frac{1}{8 \pi M}\frac{\omega'}{\omega}$ in the lab frame
and $\Phi=\frac{1}{8 \pi \sqrt{s}}$ in the c.m. frame 
($\sqrt{s}$ = total energy). For the
following discussion it is convenient to use the lab frame and to
consider special cases for the amplitude $T_{fi}$.
These special cases are
the extreme forward ($\theta=0$) and extreme backward ($\theta=\pi$)
direction  where the amplitudes for Compton scattering
may be written in the form
\cite{babusci98}  
\begin{eqnarray}
&&\frac{1}{8\pi M}[T_{fi}]_{\theta=0}
=f_0(\omega){\veceps}'{}^*\cdot{\veceps}+
g_0(\omega)\, \mbox{i}\, {\vecsig}\cdot({\veceps}'{}^*\times
{\veceps}), \label{T0}\\
&&\frac{1}{8\pi M}[T_{fi}]_{\theta=\pi}
=f_\pi(\omega){\veceps}'{}^*\cdot{\veceps}+
g_\pi(\omega)\, \mbox{i}\,{\vecsig}\cdot({{\veceps}}'{}^*
\times
{{\veceps}}). 
\label{Tpi}
\end{eqnarray}
For circularly polarized photons the polarization vectors $\veceps$
may be represented in terms of a linearly polarized basis 
$\hat{\bf e}_x$ and $\hat{\bf e}_y$.  The most frequently used  form 
of this representation is 
\begin{equation}
\veceps_{\lambda}= \frac{1}{\sqrt{2}}
(- \lambda \hat{\bf e}_x - i \hat{\bf e}_y),\quad\quad \lambda=\pm1, 
\label{epsilon}
\end{equation}
where $\lambda$ is the photon helicity \cite{walker69}.
Linearly polarized photons with electric vectors perpendicular or
parallel to the $x$-$z$ reaction plane have polarization vectors
$\veceps_\perp$ and $\veceps_\parallel$, respectively, where
\begin{eqnarray}
&&\veceps_\perp={\bf\hat{e}}_y=\frac{i}{\sqrt{2}}(
\veceps_+ + \veceps_- ),\nonumber\\
&&\veceps_\parallel ={\bf\hat{e}}_x=-\frac{1}{\sqrt{2}}(
\veceps_+ - \veceps_-). \label{epsparperp}
\end{eqnarray}
For linearly polarized photons the relations 
(\ref{T0}) and (\ref{Tpi}) remain valid if we interpret $\veceps$ 
and  $\veceps'$ as polarization vectors for linear polarization.

The process described in (\ref{T0}) is the
transmission of linearly polarized photons through a medium as
provided by a proton with spin vector $\vecsig$ parallel or
antiparallel to the direction of the incident photon
with rotation of the direction of linear
polarization \cite{gilman72}. The amplitudes   $f_0$ and $g_0$   correspond
to the  polarization components of the outgoing photon 
parallel and perpendicular to the direction of linear polarization
of the incoming photon. The
interpretation of (\ref{Tpi}) is the same as that of (\ref{T0})
except for the fact that the photon is reflected. In case of forward
scattering  (\ref{T0})   the origin of 
the rotation of the  direction of linear polarization may be
related to the alignment of the  internal magnetic dipole moments of
the  nucleon and in this sense the process may be considered as
a Faraday effect.  In case of backward
scattering  (\ref{Tpi}) a large portion of the relevant phenomenon
is of a completely different origin which will be studied in detail 
in the following.
It is important to realize that in contrast to 
frequent belief  
there is no flip of any spin in the two cases of Compton scattering.
The factor $\vecsig$ in the second terms of the two equations may be
interpreted as a spin dependence of scattering in the sense that the
direction of rotation of the electric vector changes sign (from $e.g.$
clock-wise to anti clock-wise) when the spin of the target nucleon
is reversed. 

The two scattering processes may also be related to the two states of
circular polarization, i.e. helicity amplitudes for forward and
backward Compton scattering. Here, we first restrict the discussion to
the well known case of forward scattering  \cite{gilman72}. 
The equally important case of backward scattering is more complicated
and will be considered later.
If the photon and nucleon spins are parallel
(photon helicity $\lambda_\gamma=+1$, nucleon helicity 
$\lambda_N= -\frac12$, and net helicity
in the photon direction 
$\lambda =\lambda_\gamma-\lambda_N= \frac32$) 
then the amplitude is
\begin{equation}
f^{3/2}_0(\omega)=f_0(\omega)-g_0(\omega), \label{f32}
\end{equation}
while for  the spins being  anti-parallel (photon helicity 
$\lambda_\gamma=+1$, nucleon
helicity $\lambda_N=+\frac12$, and net helicity along the photon direction
$\lambda=\lambda_\gamma-\lambda_N=+\frac12$) the amplitude is
\begin{equation}
f^{1/2}_0(\omega)=f_0(\omega)+g_0(\omega). \label{f12}
\end{equation}
The amplitudes $f^{3/2}_0$ and $f^{1/2}_0$ are related by the optical
theorem to the total cross sections $\sigma_{3/2}$ and $\sigma_{1/2}$
for the reaction 
\begin{displaymath}
\gamma+N \to N^* \to N + {\rm mesons} + \mbox{radiative decay of N$^*$}
\end{displaymath}
(see Section 4.1)
when the photon spin is parallel or anti-parallel to the nucleon spin:
\begin{eqnarray}
&&{\rm Im}f^{3/2}_0(\omega) =\frac{\omega}{4\pi}\sigma_{3/2}(\omega),\\
&&{\rm Im}f^{1/2}_0(\omega) =\frac{\omega}{4\pi}\sigma_{1/2}(\omega).
\label{optical-1}
\end{eqnarray}
Therefore,
\begin{eqnarray}
&&{\rm Im}\, f_0(\omega)
=\frac{\omega}{4\pi}\, \frac{\sigma_{1/2}(\omega)+\sigma_{3/2}(\omega)}{2}
=\frac{\omega}{4\pi}\sigma_{\rm tot}(\omega),\label{optical-3}\\
&&{\rm Im}\, g_0(\omega)=\, \frac{\omega}{4\pi}\frac{\sigma_{1/2}(\omega)
-\sigma_{3/2}(\omega)}{2}=\frac{\omega}{8\pi}\,\Delta\sigma(\omega),
\label{optical-4}
\end{eqnarray}
where $\sigma_{\rm tot}(\omega)=(\sigma_{1/2}+\sigma_{3/2})/2$
is the spin averaged total cross section and $\Delta\sigma=
\sigma_{1/2}-\sigma_{3/2}$.

\subsection{Definition of polarizabilities}

Following Babusci et al. \cite{babusci98} the equations (\ref{T0})
and (\ref{Tpi}) can be used to define the electromagnetic
polarizabilities and spin-polarizabilities as the lowest-order
coefficients in an $\omega$-dependent development of the
nucleon-structure dependent parts of the scattering amplitudes:
 \begin{eqnarray}
f_0(\omega) & = & - ({e^2}/{4 \pi M})q^2 + 
{\omega}^2 ({\alpha}
+{\beta}) + {\cal O}({\omega}^4) \label{f0}\\
g_0( \omega) &=&  \omega\left[ - ({e^2}/{8 \pi M^2})\,   
{\kappa}^2 
+ {\omega}^2
{\gamma_0}  + {\cal O}({\omega}^4) \right] \label{g0}\\
f_\pi(\omega) &=& \left(1+({\omega'\omega}/{M^2})\right)^{1/2} 
[-({e^2}/{4 \pi M})q^2 +       
\omega\omega'({\alpha} - {\beta}) 
+{\cal O}({\omega}^2{{\omega}'}^2)] \label{fpi}\\
g_\pi(\omega) &=& \sqrt{\omega\omega'}[
({e^2}/{8 \pi M^2})  
( {\kappa}^2 + 4q
{\kappa} + 2q^2)
+ \omega\omega'
{{\gamma_\pi}} + {\cal O}
({\omega}^2 {{\omega}'}^2)]  \label{gpi}
\end{eqnarray}
where $q e$ is the electric charge ($e^2/4\pi=1/137.04$), 
$\kappa$ the anomalous magnetic
moment of the nucleon and $\omega'=\omega/(1+\frac{2\omega}{M})$.

In the relations for $f_0(\omega)$ and $f_\pi(\omega)$ the first
nucleon structure dependent coefficients are the photon-helicity non-flip 
$(\alpha+\beta)$ and photon-helicity flip $(\alpha-\beta)$ linear
combinations of the electromagnetic polarizabilities $\alpha$ and
$\beta$. In the relations for $g_0(\omega)$ and $g_\pi(\omega)$
the corresponding coefficients are the spin polarizabilities
$\gamma_0$ and $\gamma_\pi$, respectively.

In principle it is possible to define polarizabilities related to
higher orders in $\omega$. We do not consider these higher-order
polarizabilities  is this paper. For a discussion of these higher-order
polarizabilities we refer to \cite{drechsel03,wissmann04,holstein00,babusci98}
and references therein.

\subsection{Sum-rules for the forward direction}

Sum-rules for the forward direction are derived from
Cauchy's theorem which may be formulated in the form \cite{roman,hoehler83}
\begin{equation}
{\rm Re}F(\omega)=\frac{1}{\pi}{\cal P}\int^{+\infty}_{-\infty}
\frac{{\rm Im}F(\omega')}{\omega'-\omega}d\omega'+ C(\infty),
\label{cauchi}
\end{equation}
where the principal value integral runs along the real $\omega$ axis
and  is   closed by a  contour contribution $C(\infty)$ which in
general is nonzero. However, for the
present consideration we make the assumption that this contour 
(or asymptotic) contribution is equal to zero, i.e. we apply  the 
no-subtraction hypothesis $C(|\omega|\to\infty)\to 0$.
 In case this hypothesis is valid,
(\ref{cauchi}) may be written in the form
\begin{equation}
{\rm Re} F(\omega)=\frac{1}{\pi}\int^\infty_{0}\frac{-{\rm
    Im}F(-\omega')}{\omega'+\omega}d\omega' + \frac{1}{\pi} {\cal P}
\int^\infty_{0}\frac{{\rm
    Im}F(\omega')}{\omega'-\omega}d\omega'.
\label{cauchi-1}
\end{equation}
Crossing symmetry \cite{roman}
\begin{equation}
F(-\omega)=F^*(\omega).
\label{crossingsymmetry}
\end{equation}
implies that
\begin{equation}
{\rm Im}F(-\omega)=- {\rm Im} F(\omega).
\label{optical}
\end{equation}
Inserting (\ref{optical}) into (\ref{cauchi-1}) we arrive at
\begin{equation}
{\rm Re} F(\omega)=\frac{2}{\pi}{\cal P}\int^{\infty}_{0}
\frac{\omega'\,{\rm Im}F(\omega')}{\omega'^2-\omega^2}d\omega'
= \frac{1}{2\pi^2}{\cal
  P}\int^{\infty}_{\omega_0}\frac{\omega'^2\sigma_{\rm tot}(\omega')}
{\omega'^2-\omega^2}d\omega',
\label{cauchi-2}
\end{equation}
where $\omega_0$ is the threshold energy below which the total
photo-absorption cross section is equal to zero.
Applying the same procedure to the function $F_1(\omega)$  defined through
\begin{equation}
\omega^2 F_1(\omega)=f_0(\omega)+(e^2/4\pi M)q^2 = \omega^2 (\alpha+\beta)+ 
{\cal O}(\omega^4) 
\label{cauchi-3}
\end{equation}
and also make use of the crossing symmetry (\ref{crossingsymmetry}) for
$F_1(\omega)$,
we arrive at
\begin{equation}
{\rm Re}f_0(\omega)+(e^2/4\pi M)q^2 =  \frac{\omega^2}{2\pi^2}{\cal
  P}\int^{\infty}_{\omega_0}\frac{\sigma_{\rm tot}(\omega')}
{\omega'^2-\omega^2}d\omega'
\label{cauchi-4}
\end{equation}
and 
\begin{equation}
\alpha+\beta= \frac{1}{2\pi^2}
\int^{\infty}_{\omega_0}\frac{\sigma_{\rm tot}(\omega')}
{\omega'^2}d\omega'.
\label{cauchi-5}
\end{equation}
Similarly, for the function $F_2(\omega)$ defined through
\begin{equation}
\omega F_2(\omega)=g_0(\omega)=-\omega \frac{e^2 \kappa^2}{8\pi M^2} +{\cal O}
(\omega^3)\label{cauchi-6}
\end{equation}
we arrive at
\begin{equation}
{\rm Re}\,g_0(\omega)=  \frac{\omega}{4\pi^2}{\cal
  P}\int^{\infty}_{\omega_0}\frac{\omega' \Delta\sigma(\omega')}
{\omega'^2-\omega^2}d\omega'
\label{cauchi-7}
\end{equation}
and 
\begin{equation}
 \frac{2\pi^2\alpha_e\kappa^2}{M^2}=
\int^{\infty}_{\omega_0}
\frac{ \sigma_{3/2}(\omega)- \sigma_{1/2}(\omega)   }
{\omega}d\omega;\quad \alpha_e=1/137.04.
\label{cauchi-8}
\end{equation}
Applying a Taylor expansion to  (\ref{cauchi-7}) and making  use
of (\ref{g0}) we arrive at
\begin{equation}
\gamma_0=-\frac{1}{4\pi^2}\int^\infty_{\omega_0}
\frac{\sigma_{3/2}(\omega)-\sigma_{1/2}(\omega) }
{\omega^3}d\omega.
\label{cauchi-9}
\end{equation}
It should be noted that in (\ref{cauchi-3}) to  (\ref{cauchi-9})   the 
no-subtraction hypothesis applies  to the function $F_1(\omega)$
and $F_2(\omega)$.

The sum rule given in (\ref{cauchi-5}) was first derived by Baldin
\cite{baldin60}
and later discussed in more detail  by Lapidus
\cite{lapidus62}. It, therefore, appears justified \cite{lvov93} to call it the
``BL'' sum rule.
The sum rule
(\ref{cauchi-8})
has independently been derived by Gerasimov \cite{gerasimov66}
and Drell and Hearn \cite{drell66}. A further  independent derivation in terms
of current algebra has been given by Hosada \cite{hosada66}. It has
become customary to call it the ``GDH'' sum rule.

\subsection{Properties of amplitudes for the  backward
direction}

In order to arrive at general properties of scattering amplitudes for
the  backward direction we make use of the fact that 
in the forward and backward directions Compton scattering takes place
via photon-helicity non-flip and photon-helicity flip amplitudes,
respectively. Denoting the corresponding scattering amplitudes by 
$T_{1,1}$ and $T_{1,-1}$, respectively, 
we apply the usual partial wave expansion of the helicity
amplitudes \cite{huett00,jacob59,hara72} 
\begin{equation}
T_{1,\pm 1}(\omega,\theta)=\sum^\infty_{L=1} T^L_{1,\pm 1 }(\omega)
d^L_{1,\pm 1}(\theta)
\label{mutipoledecomposition}
\end{equation}
with $d^L_{1,\pm 1}(\theta)$
being the $d$-functions, having the special values
\begin{eqnarray}
&& d^L_{1,+1}(\theta=0)=1,\\
&& d^L_{1,- 1}(\theta=\pi)=(-1)^{L-1}.
\label{d-funkt}
\end{eqnarray}
The partial waves $T^L_{1,\pm 1}(\omega)$ contain electric $(T^{EL})$
and magnetic $(T^{ML})$ multipoles:
\begin{equation}
T^L_{1,\pm 1}(\omega) = T^{EL}(\omega) \pm  T^{ML}(\omega).
\label{EjML}
\end{equation}
This leads to the equations
\begin{eqnarray}
T_{1,+1}(\omega,\theta=0)&=&\sum^\infty_{L=1}\left[ T^{EL}(\omega)
+ T^{ML}(\omega)\right], \label{optical-11}\\
T_{1,- 1}(\omega,\theta=\pi)&=&\sum^\infty_{L=1}(-1)^{L-1}
\left[T^{EL}(\omega)- T^{ML}(\omega) \right],\nonumber\\
&=&T^{\Delta P={\rm yes}}(\omega)- T^{\Delta P={\rm no}}
(\omega),
\label{optical-22}
\end{eqnarray}
where $\Delta  P={\rm yes}$ and $\Delta P={\rm no}$ denote excitation
processes with parity change 
($\Delta  P={\rm yes}:E1,M2,E3,\cdots$) and parity 
nonchange ($\Delta P={\rm no}:M1,E2,M3,\cdots$), respectively.
Applying  the optical theorem
in the generic form\footnote{This is a formal use of the optical theorem. 
In a physical sense the optical theorem is only valid in the forward 
direction.}
${\rm Im}F(\omega)=(\omega/4\pi)\sigma(\omega)$
these properties can be used to supplement the relations
(\ref{optical-3}) and (\ref{optical-4}) by 
\begin{eqnarray}
{\rm Im}f_\pi(\omega)&=&\frac{\omega}{4\pi}\left[\sigma(\omega,
\Delta P={\rm yes}) - \sigma(\omega,\Delta P={\rm no})\right],
\label{optical-5}\\
{\rm Im}g_\pi(\omega)&=&\frac{\omega}{8\pi}\left[\Delta\sigma(\omega,
\Delta P={\rm yes}) - \Delta\sigma(\omega,\Delta P={\rm no})\right],
\nonumber\\
&\equiv&\frac{\omega}{8\pi}\{ [\sigma_{1/2}(\omega,\Delta P={\rm
yes})-  \sigma_{1/2}(\omega,\Delta P={\rm no}) ],\nonumber\\&&\,\,\,\,\,\,-
[\sigma_{3/2}(\omega,\Delta P={\rm
yes})-  \sigma_{3/2}(\omega,\Delta P={\rm no}) ],\}\nonumber\\
&\equiv& \frac{\omega}{8\pi}\sum_n P_n[\sigma^n_{3/2}(\omega)
-\sigma^n_{1/2}(\omega)]\label{optical-7}
\end{eqnarray}
with $P_n=+1$ for $\Delta P=$ no and  $P_n=+1$ for $\Delta P=$ yes.  
Eq. (\ref{optical-5}) has been used to derive a sum rule 
for $\alpha - \beta$. Its first formulation was published  by 
Bernabeu, Ericson and Ferro Fontan \cite{bernabeu74}, its final form 
was first given by Bernabeu and Tarrach \cite{bernabeu77}. It, therefore,
appears justified to call  it the ``BEFT'' sum rule.
Eq. (\ref{optical-7}) was used by L'vov and Nathan \cite{lvov99} to
derive a sum rule for $\gamma_\pi$, which may be called the ``LN'' 
sum rule.

We will see later that these four sum rules, 
e.g. ``BL'', ``GDH'', ``BEFT'', and ``LN'', contain very important
information for an understanding of the polarizability of the nucleon.
The following parts of this section will be devoted to a description
of phenomena which are essential for these sum rules.  For the
derivation of the ``BEFT'', and ``LN'' sum rules   dispersion theories for
nonzero angle are needed which will be discussed in  Section 3.

\subsection{Mandelstam variables}

The conservation of energy and momentum in nucleon Compton scattering
\begin{equation}
\gamma(k,\lambda) + N(p) \to \gamma'(k',\lambda')+N'(p')
\label{s-channel}
\end{equation} 
 is given by 
\begin{equation}
k+p=k'+p'\label{enermomen}
\end{equation}
where the four-momenta may be expressed through the energies and
three-momenta in the form  $k=(\omega,{\bf k})$, $k'=(\omega',{\bf k}')$, 
$p=(E,{\bf p})$  and    $p'=(E',{\bf p}')$. 
The metric is chosen such that $k^2=\omega^2-{\bf k}^2=0$, 
$p^2=E^2-{\bf p}^2=M^2$ which analogous expressions for the
final states. 

Let us introduce the three  Mandelstam 
variables
\begin{eqnarray}
s=(k+p)^2=(k'+p')^2, \label{s}\\
t=(k-k')^2=(p'-p)^2, \label{t}\\
u=(k-p')^2=(k'-p)^2. \label{u}
\end{eqnarray}
They are constrained by the relation
\begin{equation}
s+t+u=2M^2. \label{constraint}
\end{equation}
which follows from the energy-momentum relations given above.

\begin{figure}[h]\centering
\includegraphics[scale=0.6]{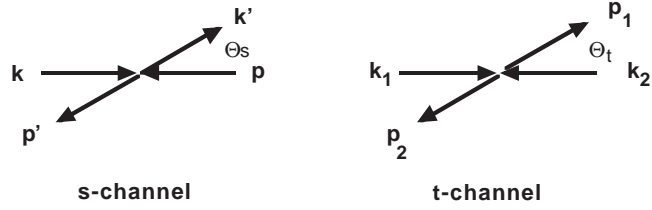}
\caption{Left panel: Definition of the c.m. scattering angle 
of Compton scattering ($s$-channel). Right panel:  
Definition of the c.m. scattering angle of  a two-photon fusion process
($t$-channel).  The quantities  ${\bf k}$
are  three-momenta of photons, the quantities  ${\bf p}$ three-momenta of
nucleons.  }
\label{fig:s-t-channel}
\end{figure}

It is useful to
introduce one other variable
\begin{equation}
\nu=\frac{s-u}{4M} \label{nu}
\end{equation}
which may replace $s$ and $u$ via
\begin{eqnarray}
\nu&=&\frac{s-M^2+t/2}{2M}, \label{nus}\\
-\nu&=&\frac{u-M^2+t/2}{2M}. \label{nuu}
\end{eqnarray}
Furthermore, we have
\begin{eqnarray}
&&\nu=E_\gamma+\frac{t}{4M}=\frac12(E_\gamma+E_{\gamma'}), \label{nuEgamma}\\
&&\sin^2\frac{\theta_{s}}{2}=-\frac{st}{(s-M^2)^2}
\label{scatteringangle}
\label{EnergyAngle}
\end{eqnarray}
where $E_\gamma$ and $E_{\gamma'}$ are the energies of the initial 
and final photon in the laboratory and $\theta_{s}$ the c.m.
scattering angle of Compton scattering ($s$-channel). The definition
of $\theta_s$ is illustrated in the left panel of Figure 
\ref{fig:s-t-channel}. It should be noted that the definition given in
(\ref{scatteringangle}) may lead to scattering angles outside
the ``physical range'' $-1 \leq \cos \theta_s \leq +1$ 
\cite{lvov97}. Unphysical ranges may be of importance for the
completeness of dispersion integrals and the definition given in
(\ref{scatteringangle}) provides the tool for the analytical continuation.

The $t$-channel
corresponds to the fusion of two photons with four-momenta
$k_1$ and $k_2$ and helicities $\lambda_1$ and $\lambda_2$ to form
a $t$-channel intermediate state $|t\rangle$ from which -- in  a
second step -- a proton-antiproton pair is created.  
The corresponding reaction may be formulated in the 
form
\begin{equation}
\gamma(k_1,\lambda_1) + \gamma(k_2,\lambda_2)  \to  \bar{N}(p_1)  +N(p_2).
\label{t-channel}
\end{equation}
For Compton scattering
the related $N\bar{N}$ pair creation-process is only virtual,
i.e. the energy is too low to put the proton-antiproton pair on the
mass shell. This means that we have to treat the process described in
(\ref{t-channel})  in the unphysical region. 
The $s$-channel to $t$-channel transition may be summarized as follows
\begin{eqnarray}
&& k \to k_1 \quad \quad   \veceps \to \veceps_1, \label{transformation1}\\
&& k' \to -k_2 \quad  \veceps'^* \to \veceps_2,\label{transformation2} \\
&&p\to -p_1 \quad \,\, \,\, u(p) \to v(p_1), \label{transformation3}\\
&& p' \to p_2 \quad  \quad u(p') \to u(p_2),
\label{transformation4}
\end{eqnarray}
where $u(p)$, $u(p')$ and  $u(p_2)$ denote nucleon states and $v(p_1)$
an antinucleon state. The sign changes in (\ref{transformation2})
and (\ref{transformation3})  mean that instead of an
outgoing photon $\gamma'$ there is a second incoming photon $\gamma_2$
and instead of an incoming nucleon $N$ there is an outgoing antinucleon 
$\bar{N}$.

The sign change in (\ref{transformation2}) when
replacing the four-momentum of the
outgoing scattered photon by the four-momentum of  a second incoming
photon leads to the following relations between photon-helicities
\begin{eqnarray}
&&\lambda \to \lambda_1,  \label{helicity1}\\
&&\lambda' \to - \lambda_2, \label{helicity2}\\
&&\Delta\lambda^s_{\gamma\gamma}\equiv |\lambda-\lambda'|=0 \to 
\Delta\lambda^t_{\gamma\gamma}\equiv |\lambda_1-\lambda_2|=2,
\label{helicity3}\\
&&\Delta\lambda^s_{\gamma\gamma}\equiv |\lambda-\lambda'|=2 \to 
\Delta\lambda^t_{\gamma\gamma}\equiv |\lambda_1-\lambda_2|=0.
\label{helicity4}
\end{eqnarray}
The case described in (\ref{helicity3}) is given in  forward Compton
scattering ($\theta_s=0$) , the case described in (\ref{helicity4}) in
backward Compton  scattering ($\theta_s=\pi$). Forward Compton
scattering may be related to a two-photon fusion process where the helicity
difference of the two photons is equal to 2. Backward Compton
scattering, on the other hand, may be related with a two-photon fusion
process where the helicity difference of the two photons is 0.
 For sake of completeness we give the following relation for
the $t$-channel scattering angle
\begin{equation}
\cos \theta_t=\frac{s+\frac12 t -M^2}{\frac12 \sqrt{t(t-4M^2)}}
\equiv\frac{4M\nu}{\sqrt{t(t-4M^2)}}.
\label{eq:cos-theta-t}
\end{equation}
For $\theta_s=\pi$ we have $16 M^2 \nu^2 = t(t-4M^2)$ and, 
therefore\footnote{There should be a 
$\pm$ sign in front of $\cos \theta_t$ indicating  that the photons
$\gamma_1$ and $\gamma_2$ are indistinguishable so that $\theta_t$ and
 $\pi-\theta_t$ are physically equivalent.
We will follow the convention to use the $-$ sign.}, $\cos \theta_t =-1$. 
 As in case
of $\cos \theta_s$ also $\cos \theta_t$ is the essential variable for
analytical continuations. 
In the c.m. system of the two photons $k_1$ and $k_2$ the quantity $t$
has the meaning of a total energy squared:
\begin{equation}
t=(k_1 + k_2)^2 = (\omega_1 + \omega_2)^2 = (W^t)^2.
\label{t-interpretation}
 \end{equation}

\subsection{Selection rules and photon-polarization
correlations  of t-channel intermediate states}

For the discussion of properties of the $t$-channel it is useful to
study the selection rules for spins, parities and photon-polarization
correlations of mesons decaying into two photons. 

Following early work of Landau \cite{landau48} the problem 
of selection rules and photon-polarization correlation was treated 
by C.N. Yang \cite{yang50} in its general form. In  \cite{yang50}
two-photon decay  of $J^{PC}=3^{++},5^{++},\cdots$ mesons
was discussed without a definite statement about the correlation of linear
polarizations.  The necessary completion was possible using the formulae 
given in  \cite{yang50}.
The results obtained in this way are summarized in the following.

The polarization correlation can be formulated in terms circular and
linear polarization. In terms of circular polarization the states are
denoted by $\psi^{RR}$ ,$\psi^{LL}$, $\psi^{RL}$  and $\psi^{LR}$, 
where, e.g.,  $RR$ means that the two photons going into $+z$ and $-z$ 
direction, respectively, 
both are righthanded circular, i.e. they have positive helicity. 
The linear combinations $\psi_1$ to $\psi_4$ of these states describe  
two-photon states with definite linear polarization-correlation, 
where symmetric 
linear combinations correspond to parallel linear polarization 
of the two photons and 
antimetric linear combinations to perpendicular linear polarization.
It is easy to see that $\psi_2=\psi^{RR}- \psi^{LL}$ has 
negative parity whereas the other states, $\psi_1$, $\psi_3$ and  $\psi_4$
have positive parity. The states $\psi^{RR}$ and $\psi^{LL}$
correspond to $t$-channel helicity differences of 
$\Delta\lambda^t_{\gamma\gamma}=0$, the states $\psi^{RL}$ and 
$\psi^{LR}$ to $\Delta\lambda^t_{\gamma\gamma}=2$. Furthermore,
two-photon decay implies that the $C$-parity quantum number is $C=+1$.
Collecting all pieces of
information we arrive at Table
\ref{yang}.
\begin{center}
\begin{table}[h]
\caption{Selection rules and photon-polarization correlations
of meson decaying into two photons.}
\vspace{0.3cm}
\begin{center}
\begin{tabular}{||l|l|l|l|l|l||}
\hline
\hline
$\gamma\gamma$-state&$\gamma\gamma$-polarization&
$\Delta\lambda^t_{\gamma\gamma}$&
$\Delta\lambda^s_{\gamma\gamma}$&quantum numbers&\\
\hline
$\psi_1=\psi^{RR}+\psi^{LL}$& parallel& 0&2& $J^{PC}=0^{++},2^{++},
4^{++}, 
\cdots$&$f_\pi$\\
$\psi_2=\psi^{RR}-\psi^{LL}$& perpendicular&0 &2&  $J^{PC}=0^{-+},2^{-+},
4^{-+}, \cdots$&$g_\pi$\\
$\psi_3=\psi^{RL}+\psi^{LR}$& parallel &2&0& $J^{PC}=2^{++},4^{++},
6^{++}, 
\cdots$&$f_0$\\
$\psi_4=\psi^{RL}-\psi^{LR}$& perpendicular&2&0&
$J^{PC}=3^{++},5^{++}, 
7^{++}, \cdots$&$g_0$\\
\hline
 \hline
\end{tabular}
\end{center}
\label{yang}
\end{table}
\end{center}
The helicity differences $\Delta^t_{\gamma\gamma}$ and
$\Delta^s_{\gamma\gamma}$ (Eqs. (\ref{helicity3}) and 
(\ref{helicity4}))
and the $\gamma\gamma$-polarization
correlations corresponding  to the four states $\Psi^1,..,\Psi^4$
coincide with the properties of  the four scattering amplitudes 
$f_\pi,g_\pi,f_0,g_0$, respectively. This means that the respective  
quantum numbers 
$J^{PC}$ ($J$=spin, $P$=parity, $C=C$-parity) given in Table \ref{yang}
may be attributed to the $t$-channel intermediate states of these amplitudes.
Making use of the reasonable assumption that the lowest spin quantum
number $J$ is the most relevant one  we arrive at
the $t$-channel intermediate states given in Table \ref{fg-0}. 
\begin{center}
\begin{table}[h]
\caption{$t$-channel intermediate states.}
\vspace{3mm}
\begin{center}
\begin{tabular}{|l|l|l|}
\hline
&$J^{PC}$&exchanged particles\\
\hline
$f_\pi(\omega)$ & $0^{++}$&
 $\sigma(600), f_0(980), a_0(980)$ \\
$g_\pi(\omega) $ & $0^{-+}$& $\pi^0(135)$, 
$\eta(547)$, $\eta'(958)$\\
$f_0(\omega)$ & $2^{++}$ &  Pomeron I\!\!P, $f_2(1270)$, 
$a_2(1320)$\\
$g_0(\omega):$ & $3^{++}$ &   mesons not known\\
\hline
\end{tabular}
\end{center}
\label{fg-0}
\end{table}
\end{center}

\subsection{Properties of scalar  mesons}

It may be expected that the dominating $t$-channel contributions to 
the amplitudes $f_\pi$ and $g_\pi$ result from the mesons with lowest
angular momenta, i.e. $J^{PC}=0^{++}$  and  $J^{PC}=0^{-+}$,
respectively, as shown in Table \ref{fg-0}.
In contrast to the pseudoscalar mesons $\pi^0, \eta, \eta'$ which 
have comparatively small widths,
the scalar counterparts $f_0(600), a_0(980)$ and $f_0(980)$ have open
two-particle hadronic channels and, therefore, have large widths. 
 For the pseudoscalar mesons the
$|q\bar{q}\rangle$ structure in the $^1S_0$ spectroscopic state 
is the generally adopted  description of the internal structure. For the
scalar mesons the $^3P_0$ $|q\bar{q}\rangle$ content of the wave-function
may be of minor importance, whereas the major part of the
wave-function may be due to the two-meson configurations, also showing 
up in the decay-channel and the two-meson configurations with
thresholds  above the resonance energies.  

An up-to-date
description of the properties of scalar  mesons is given by the
Particle Data Group \cite{eidelman04}.  
In connection with the
interpretation of Compton scattering the $f_0(600)$ meson is the most
relevant one in  the scalar sector. We, therefore, restrict the
following discussion to the  $f_0(600)$ meson.

\subsubsection{Historical introduction to the $f_0(600)$ or $\sigma$}

The iso-singlet (isoscalar $I=0$) scalar ($J=0$) $\sigma$ meson  was introduced
theoretically by Schwinger in 1957 \cite{schwinger57} as a supplement of
the $\pi$ meson. The $\sigma$ meson was predicted
\cite{schwinger57} to have strong 
interactions  and, therefore, should immediately disintegrate into
two $\pi$ mesons 
thus making the experimental observation difficult.
This difficulty was considered as the explanation for the fact that the
$\sigma$ meson had not been found in experiments. On the other hand
the theoretical arguments were considered firm enough to be convinced  
that this particle should  exist. Extending the arguments of Schwinger, 
Gell-Mann and Levi \cite{gellmann60} developed the  $\sigma$ model
in its linear and nonlinear versions. 

Later, the  existence of the $\sigma$ meson 
was suggested in the one-boson-exchange potential model
of nuclear forces \cite{taketani67}, where it was
introduced to supply the attraction between nucleons at intermediate 
distances. Fits of the $NN$ partial-wave phase-shift led to a mass of
$m_\sigma \approx 500 - 600$ MeV.

The $\sigma$ meson was also introduced in relation 
\cite{nambu61}
with the dynamical symmetry breaking of QCD based on the
Nambu-Jona-Lasinio model, making  a prediction for the $\sigma$ 
mass of  $m_\sigma\approx 600 -700 $ MeV.

In spite of the  very strong demands on the side of the theory for the
existence of the $\sigma$ meson  the experimental verification
was considered  uncertain for a long time, mainly  based on negative
results \cite{morgan87} of $\pi\pi$ scattering phase-shift analyses. 
This failure was explained later \cite{ishida96} to be a consequence 
of not considering  the cancellation mechanism between the resonant
(Breit-Wigner type) part of the $\sigma$ meson $\pi\pi$ phase-shift 
$\delta^0_0$ and a nonresonant 
(repulsive core type) $\pi\pi$ phase-shift, also contributing to
 $\delta^0_0$. This will be discussed 
in more detail in  Section 2.6.2.

In the 2002 edition, 
the PDG group \cite{hagiwara02}
has given the $\sigma$ meson the status of a particle, 
$f_0(600)$,  with a  mass 
assignment\footnote{In the foregoing editions only a mass range, 
{\it viz.} 400--1200 MeV, was given.}. This  mass assignment may be understood
as a nominal value, the meaning of which will be explained in 
Sections 2.6.2. and 2.6.3.
Of the large number of the 
experimental aspects of the
$\sigma$ meson we only discuss those which are of importance for our purposes.

\subsubsection{The $f_0(600)$ or $\sigma$ in the 
$\pi + N  \to N \pi\pi$ reaction}
\begin{figure}[h]
\begin{center}
\includegraphics[scale=0.55]{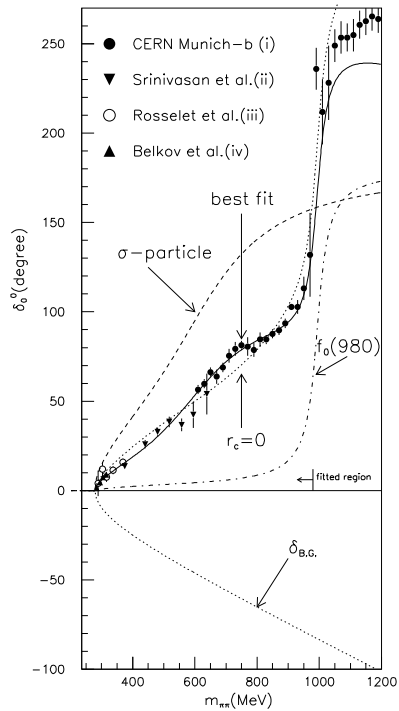}
\end{center}
\caption{Fit to $I=0$ S-wave $\pi\pi$ scattering phase-shift
 $\delta^0_0$  according to Ishida \cite{ishida03}. The fit takes into account
a hard-core (seagull type)  component $\delta_{\rm BG}$ shown as a
  dotted curve in the lower part of the figure  and a resonant
(Breit-Wigner type) component shown in the upper part of the figure 
as a dashed curve  labeled $\sigma$-particle
  }
\label{delta-0}
\end{figure}
Of the more recent analyses \cite{achasov94} of the reaction 
$\pi + N  \to N \pi\pi$ 
in terms of $\pi\pi$ scattering we want to inspect the ones given in
\cite{colangelo01} and \cite{ishida03} in more detail. In
\cite{colangelo01} the analysis of the $\pi\pi$ scattering amplitude is 
based on chiral perturbation theory starting from the observation that
chiral symmetry determines the low energy behavior of the $\pi\pi$
scattering amplitude to within very small uncertainties.  According to
this analysis the $\sigma$ meson $\pi\pi$ phase-shift $\delta^0_0$ 
consists of a
superposition  of a resonant (pole) part and a nonresonant part
stemming from  a repulsive core. The destructive interference of the two
parts leads to a $90^\circ$ crossing of the scalar phase at 
$(844\pm13)$ MeV. This result is summarized in the second line of Table
\ref{SigmaParameters}.
 \begin{table}[h]
\caption{Position of the $\sigma(600)$ pole, and $90^\circ$ crossing 
of the scalar phase. Supplement a) is an estimate provided by the 
present author.} 
\begin{center}
\begin{tabular}{|c|c|c|}
\hline
$\sqrt{s}({\rm pole}) [MeV]$&$\sqrt{s}(\delta_S=90^\circ)$&reference\\
\hline
$(470\pm 30) -i(295\pm 20)$& $844\pm13$ MeV&\cite{colangelo01}\\
$(585\pm 20)-i (193\pm 35)$ &$\sim$900 MeV&\cite{ishida03}\\
\hline
PDG summary$^{a)}$&& {\it recommended}\\
$(500\pm 40) -i (250\pm 40)$&& average \cite{eidelman04}\\
\hline
\end{tabular}
\end{center}
\hspace{4cm}{\footnotesize a) supplemented by estimated errors.}
\label{SigmaParameters}
\end{table}

In the second approach \cite{ishida03}
the analysis of the phase-shift $\delta^0_0$
of the $\pi\pi$ scattering amplitude is based on the 
Interfering Amplitude method, where the total phase-shift $\delta^0_0$
below $m_{\pi\pi}\simeq 1 $ GeV is represented by the sum of the
component phase-shifts, 
\begin{equation}
\delta^0_0=\delta_\sigma+\delta_{\rm BG}+\delta_{f_0}.
\label{deltaScalar}
\end{equation}
This analysis is shown in  Figure \ref{delta-0}.
The term $\delta_\sigma$ is from the attractive resonant  (Breit-Wigner
type) 
part of the $\pi\pi$ scattering amplitude of  the $\sigma$ meson
and 
$\delta_{\rm BG}$  from the background non-resonant repulsive
$\pi\pi$ amplitude which is taken phenomenologically to be of the 
hard-core type.
The term $\delta_{f_0}$ is from the $f_0(980)$ Breit-Wigner amplitude with
comparatively narrow width. This part has to be separately determined  and  
eliminated by subtraction.
The combined phase-shift 
$\delta_\sigma+\delta_{\rm BG}$ passes $90^\circ$ at about $\sim$ 900 MeV
as listed in the third line of Table  \ref{SigmaParameters}.

\subsubsection{ The $f_0(600)$ or $\sigma$ in the 
$\gamma+\gamma \to \pi^+\pi^-$ reaction}

\begin{figure}[h]
\begin{center}
\includegraphics[scale=0.5]{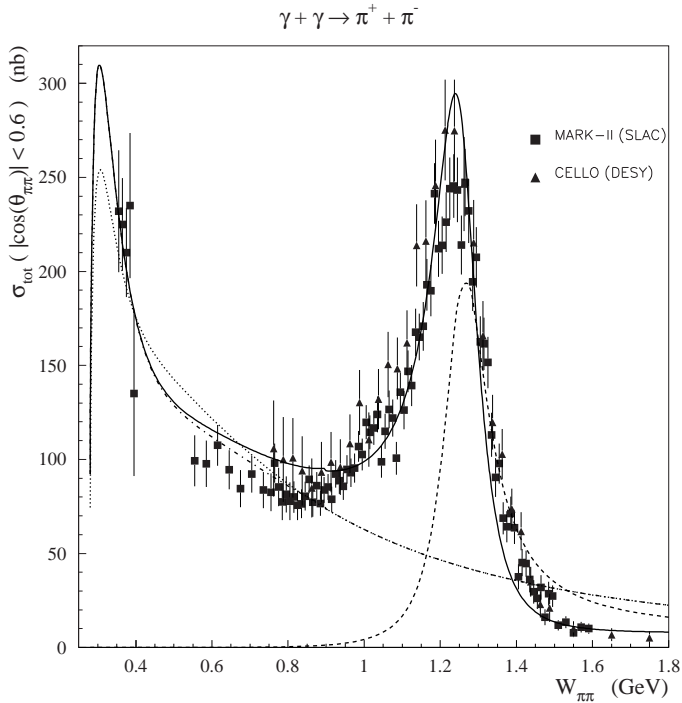}
\end{center}
\caption{Total cross section for the $\gamma\gamma\to \pi^+ \pi^-$
process as function of the c.m. energy and the analysis given by 
Drechsel et al.
\cite{drechsel99}: Born terms (dotted line),
Born amplitude with unitarized $s$-wave (dashed-dotted line),
$f_2(1270)$ resonance contribution (dashed line), and total
cross section  (full line). }
\label{gammagammacross}
\end{figure}
Figure \ref{gammagammacross} shows experimental data on  
the two-photon
fusion reaction $\gamma+\gamma \to \pi^+ + \pi^-$ and their
decomposition into  components. 
At threshold the main part of the cross 
section is due to the Born term which may be viewed as the production of two
uncorrelated point-like pions. The Born term is superimposed  by  a 
pion-structure dependent term which -- at threshold -- may be
expressed through the polarizabilities of the pion. It should be noted
that there is no clear indication of a bump-like structure at 600 MeV
which could be attributed to a $f_0(600)$ particle. But there is a
prominent bump at 1270 MeV corresponding to the tensor meson    
$f_2(1270)$. According to Tables \ref{yang} and \ref{fg-0} this
tensor-meson may be related to the forward Compton scattering
amplitude $f_0$ and its extension to larger angles.  This will be
discussed in  Section 2.7.1.

\subsubsection{The $f_0(600)$ or $\sigma$ in the 
$\gamma\gamma  \to \pi^0\pi^0$ reaction}

Unknown parameters of the $S$-wave $\pi\pi$ interaction, represented
as a broad Breit-Wigner resonance, have recently been  determined
\cite{filkov99} from fits to the experimental data \cite{marsiske90}
for the $\gamma\gamma\to\pi^0\pi^0$ process using dispersion relations. 
For the reaction $\gamma\gamma\to \pi^0\pi^0$ the Born term is equal
to zero and the main contribution is determined by the $S$-wave of
$\pi\pi$ interaction. For the fit to the experimental data there are
five free parameters: the mass, the full width and the decay width
of the $\sigma$ meson
into $\gamma\gamma$  and the sum and the
difference of $\pi^0$ meson polarizabilities. The adopted set
of parameters \cite{ahrens04} obtained
in this way are given in Table \ref{filkovtable}. 
\begin{table}[h]
\caption{Parameters of the $\sigma$ meson obtained from 
fits \cite{filkov99} to the experimental
data \cite{marsiske90} of the process $\gamma\gamma\to \pi^0\pi^0$}
\vspace{3mm}
\begin{center}
\begin{tabular}{ccccc}
\hline
$m_\sigma$ (MeV)&$\Gamma_\sigma$ (MeV)&
$\Gamma_{\sigma\to\gamma\gamma}$ (keV)& $(\alpha+\beta)_{\pi^0}$&
$(\alpha-\beta)_{\pi^0}$\\
\hline
547$\pm$ 45 & 1204 $\pm$ 362 & 0.62 $\pm$ 0.19&
0.98$\pm$0.03&$-1.6\pm 2.2$\\  
\hline
\end{tabular}
\end{center}
\label{filkovtable}
\end{table}

\subsection{Regge phenomenology and Compton scattering}

Regge phenomenology has been introduced in connection with hadronic
reactions but is also widely used in connection with photo-absorption and
Compton scattering. Though attempts have been made to replace Regge
phenomenology by QCD, its  apparent successes require a coverage
of this topic in connection with the present review.

\subsubsection{The $\rho-\omega$ and Pomeron trajectories}

\begin{figure}[h]
\includegraphics[scale=0.4]{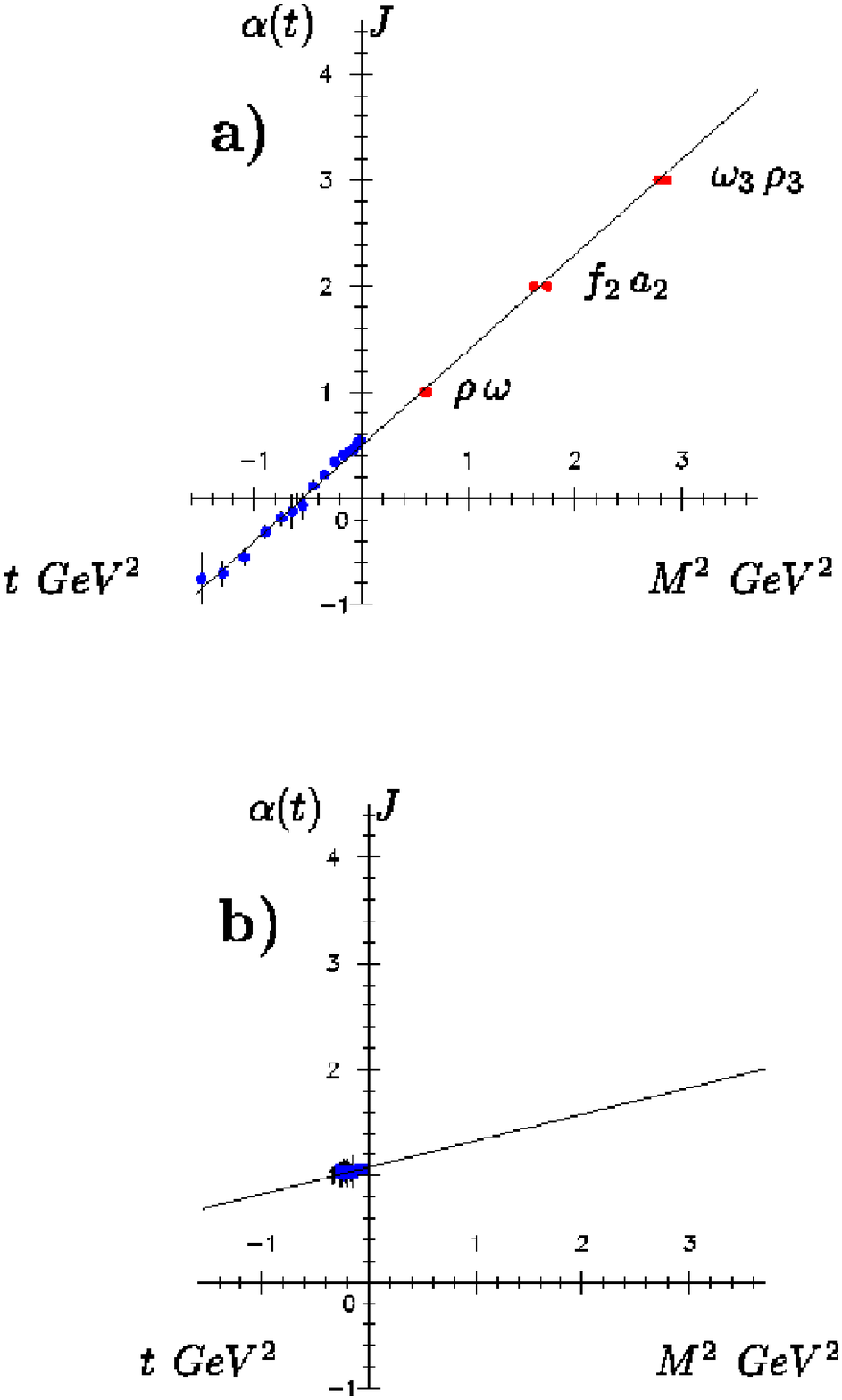}
\includegraphics[scale=0.5]{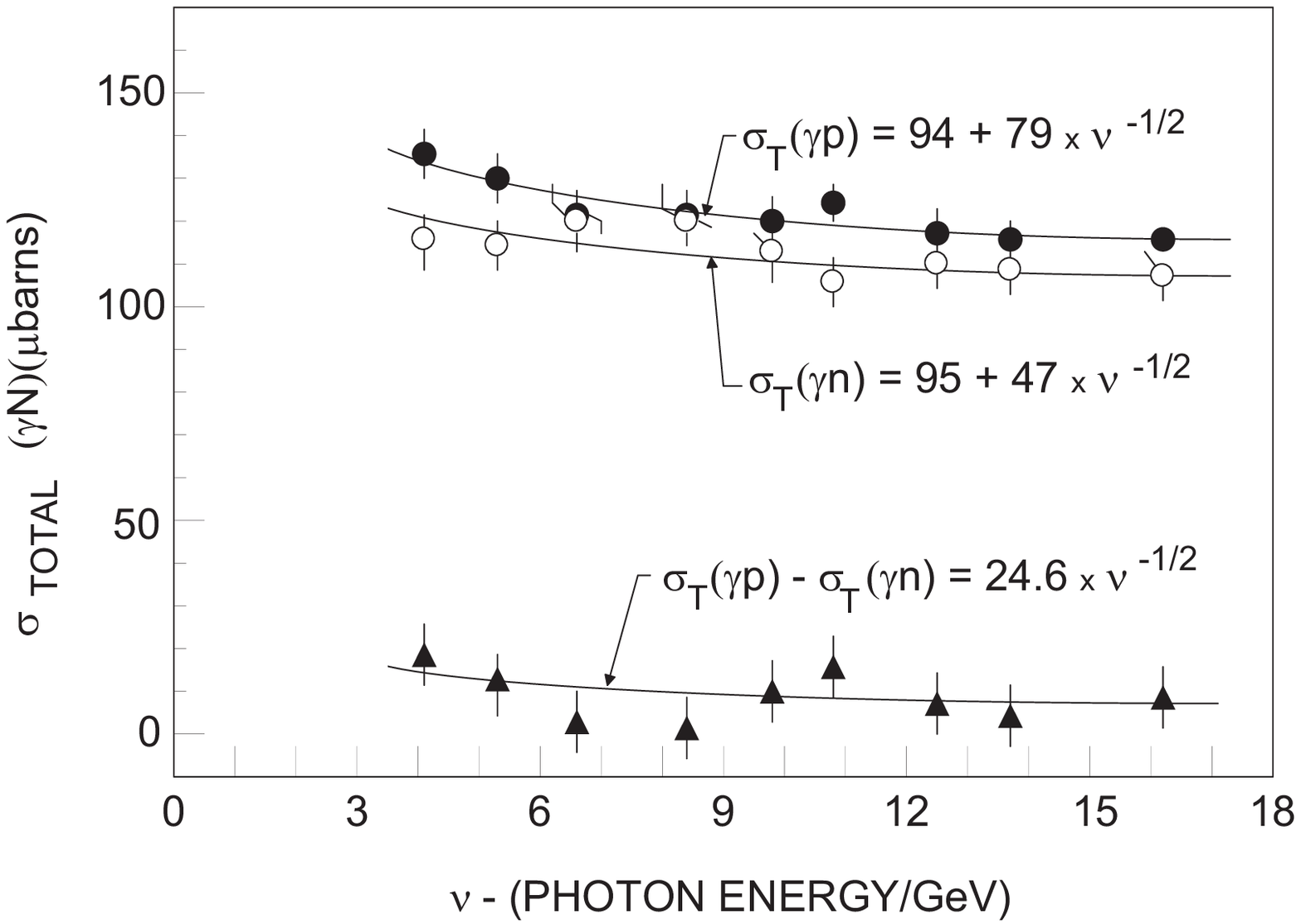}
\caption{Left panel: Regge trajectory for the charge exchange a) and elastic
 scattering b) reaction \cite{rostovtsev01}.
Right panel: Comparison of the proton, neutron, and proton-neutron
  difference data of photo-absorption cross sections with $\nu$-fits
as given in  \cite{hesse70}.}
\label{FigureRegge}
\end{figure}
The left panel  \cite{rostovtsev01} 
of Figure~\ref{FigureRegge} shows a) the Regge
trajectories   of the meson exchange  reaction 
observed through 
\begin{equation}
\pi^- + p \to \pi^0 +n,
\label{ChargeExchange}
\end{equation}
and b) the Regge trajectory for an elastic
scattering reaction, observed through
\begin{equation}
p+p \to p+p.
\label{ProtonElastic}
\end{equation}
Regge phenomenology says, that the differential cross-sections of 
these reactions are given by the function
\begin{equation}
\frac{d\sigma}{dt}\sim s^{2\alpha(t)-2},
\label{differentialRegge}
\end{equation}
 where $s$ and $t$ are the
total energy and momentum transfer squared, respectively.
The experimental data obtained for the meson-exchange reaction
at negative $t$ can be
parameterized by a linear trajectory 
\begin{equation}
\alpha(t) = \alpha(0)+\alpha' t.
\label{trajectory}
\end{equation} 
The extrapolation of the trajectory to positive values 
$t\equiv M^2$ goes through the $\rho$ point and through other 
resonances which belong to this trajectory including the $f_2(1270)$
and the $a_2(1320)$.  The trajectory for the meson-exchange reaction
may be represented by 
\begin{equation}
\alpha_\rho(t)=0.56+ 0.97 {\rm GeV}^{-2} t.
\label{rhotrajectory}
\end{equation}

The  trajectory for the elastic reaction has parameters which differ from
those of the $\rho$-trajectory, namely an intercept 
$\alpha_{\!I\!\!P}(0) \approx 1.1$ and a slope 
$\alpha'_{\!I\!\!P} \approx 0.25 {\rm GeV}^{-2}$. This trajectory is
called Pomeron trajectory and the object exchanged in the elastic
reaction (\ref{ProtonElastic}) is called Pomeron. It carries vacuum quantum
numbers ($P=C=+1$) and is thought to be a composite system of gluons. 

     
For diffractive processes where the scattering amplitude only has an
imaginary part the optical theorem provides us with a total cross
section in the form \cite{donnachie92}
\begin{equation}
\sigma^{\rm tot} \sim s^{\alpha(0)-1}.
\label{totalcross}
\end{equation}
 The validity of this relation for the photo-absorption cross section
has already shown in the 1970's where the  total
photo-absorption cross sections for the proton and the neutron
in the 3 -- 16 GeV range could be represented in the form
\cite{hesse70}
\begin{equation}
\sigma^{\rm tot} = X   + Y \nu^{-1/2}.
\label{totalcross-1}
\end{equation}
In (\ref{totalcross-1}) the variable $s$ has been replaced by $\nu$
for convenience.
The constant term corresponds to the Pomeron exchange 
where the approximation $\alpha_P(0)=1$ has been used, 
whereas the $\nu^{-1/2}$ dependent
term corresponds to $f_2(1270)$ and $a_2(1320)$ exchanges where the 
approximation $\alpha_\rho(0)=0.5$ has been used. The decrease of the cross
section with photon energy due to the second term in (\ref{totalcross-1})
and the isovector component due to $a_2(1320)$ exchange are clearly seen in the
right  panel of Figure \ref{FigureRegge}.

The conclusion we can draw from this finding is that in the forward
direction the $f_2(1270)$ and $a_2(1320)$ meson exchanges are part of
the photo-absorption cross section. The reason for this is that these
mesons  interact with the  photons via intermediate vector mesons
thus making  photo-absorption  and  Compton scattering effectively   
hadronic reactions. These hadronic reactions take place at the
periphery of the nucleon and consequently can be observed in
small-angle Compton scattering only. For larger
Compton scattering angles the $f_2(1270)$ and the   $a_2(1320)$
mesons undergo a direct coupling to the two photons and, therefore,
have to be treated as genuine $t$-channel exchanges.
This point will be discussed in Section 6.1 in more detail
in connection with the validity of the Baldin sum rule.

\subsubsection{The $f_1$ -- $a_1$ trajectories}

According to Table \ref{yang} the $t$-channel exchanges contributing
to the spin independent amplitude $f_0$ are of natural parity whereas 
the $t$-channel exchanges contributing to the spin dependent 
amplitude $g_0$ are of unnatural parity.
We, therefore,  expect that the 
 $f_1$ -- $a_1$ trajectory 
is the relevant one for the spin-dependent amplitude $g_0$,
where the meson $a_1(1260)$ has the quantum
numbers $I^G(J^{PC})=1^-(1^{++})$ and the meson $f_1(1285)$ the
quantum numbers  $I^G(J^{PC})=0^+(1^{++})$. The first meson on the 
trajectory which is capable of  coupling to two photons has the
quantum number $J^{PC}=3^{++}$. Such a meson is not known up to the present.

The $f_1$ and $a_1$ exchanges have been discussed \cite{bianchi99} 
in connection with a parameterization of the cross section difference
$\Delta\sigma=\sigma_{1/2} -\sigma_{3/2}$ investigated with virtual photons,
leading to  a nonresonant
contribution to the GDH sum rule when extrapolated to the photon point
($Q^2\to 0$). The isovector contribution to
$\Delta\sigma$ is described by the $a_1(1260)$ meson trajectory,
$\Delta\sigma_V\sim s^{\alpha^0_{a_1}-1}$, the isoscalar contribution
to $\Delta\sigma$ by the $f_1(1285)$ meson trajectory, 
$\Delta\sigma_I\sim s^{\alpha^0_{f_1}-1}$.  The trajectory slope
is  $\alpha'\simeq0.8-0.9\,\, {\rm GeV}^{-2}$ and the intercepts are 
$\alpha^0_{a_1}\approx-0.3$
and $\alpha^0_{f_1}= -0.4\pm 0.1$, respectively. 
The intercept $\alpha^0_{a_1}$ shows a rather strong model dependence
whereas  the intercept $\alpha^0_{f_1}$ turns out to be more stable.

\section{Scattering Amplitudes and Dispersion Relations}

In the following we discuss the theoretical tools for the description
of Compton scattering and the determination of polarizabilities. 
Two versions of dispersion theories are presented, the nonsubtracted
fixed-$t$ and fixed-$\theta$ dispersion theories. When applied
appropriately, these versions are capable of providing predictions
in the first and second resonance range of the nucleon, whereas the 
subtracted fixed-$t$ dispersion theory has only been successful in 
the lower part of the $\Delta$ range \cite{drechsel03,drechsel99}.
Though fixed-$\theta$ dispersion theory has the advantage
of guaranteeing that the use of $s$- and $t$-channel contributions does 
not lead
to problems like double counting of empirical input, it runs into technical
difficulties \cite{drechsel03} when applied at small scattering angles. 
On the other hand fixed-$t$ dispersion theory is constructed in  formal
analogy to forward-direction dispersion theory, where the $s$-channel 
contribution is replaced by the largely  equivalent integral part 
\cite{lvov97} and the $t$-channel contribution by a contour integral,
which already was discussed in Section 2.2 in connection with 
forward-direction dispersion theory. For the interpretation of the 
contour integral 
arguments were developed stemming from Regge theory 
\cite{petrunkin81,baranov00}. In this work we use arguments from Regge theory
only in connection with forward-direction $t$-channel exchanges as discussed
in Section 2.7.

\subsection{S-matrix and invariant amplitudes}

The amplitude $T_{fi}$ for Compton scattering
\begin{equation}
\gamma(k) N(p) \to \gamma'(k') N(p')
\label{D1}
\end{equation}
is related to the $S$-matrix of the reaction through the relation
\begin{equation}
\langle f|S-1|i \rangle= i (2\pi)^4 \delta^4(k+p-k'-p')T_{fi}.
\label{D2}
\end{equation}
The quantities $k=(\omega,{\bf k})$, $k'=(\omega',{\bf k}')$,
$p=(E,{\bf p})$, $p'=(E',{\bf p'})$ are the four momenta of the photon
and the nucleon in the initial and final states, respectively,
related to the Mandelstam variables via (see also Section 2.4)
\begin{equation}
s=(k+p)^2, \quad t=(k-k')^2, \quad u=(k-p')^2.
\end{equation}

The scattering amplitude $T_{fi}$ may be expressed on an
orthogonal basis suggested by Prange \cite{prange58} by 
means of six invariant amplitudes $T_k(\nu,t)$ leading to the general
form \cite{lvov97} 
\begin{eqnarray}
T_{fi}&=\bar{u'} e'{}^{*\mu} \Big[
-\frac{P'_\mu P'_\nu}{P'{}^2}(T_1+(\gamma \cdot K)T_2) - 
\frac{N_\mu N_\nu}{N^2}
(T_3+(\gamma \cdot K)T_4)\nonumber\\ &+i \frac{P'_\mu N_\nu -P'_\nu
  N_\mu}{P'{}^2K^2}\gamma_5
T_5 + i\frac{P'_\mu N_\nu +P'_\nu N_\mu}{P'^2K^2}
\gamma_5 (\gamma \cdot K) T_6\Big] e^\nu u. \label{Tif}
\end{eqnarray}
In (\ref{Tif}) $u'$ and $u$ are the Dirac spinors of the final and the
initial nucleon, $e'$  and $e$  are the
polarization 4-vectors of the final and the initial photon, and 
$\gamma_5=-i\gamma_0\gamma_1\gamma_2\gamma_3$.  The 4-vectors
$P'$, $K$ and $N$ together with the vector $Q$ are orthogonal and
are expressed in terms 
of the 4-momenta $p'$, $k'$
and $p$, $k$ of the final  and initial nucleon and photon,
respectively,
by
\begin{eqnarray}
&&K=\frac12 (k+k'),\quad P'=P-\frac{K(P\cdot K)}{K^2},\quad 
N_\mu=\epsilon_{\mu\nu\lambda\sigma}  P'{}^\nu Q^\lambda
K^\sigma,\nonumber\\
&&P=\frac12 (p+p'),\quad Q=\frac12 (k'-k)=\frac12 (p-p')\label{KPN}
\end{eqnarray}
where $\epsilon_{\mu\nu\lambda\sigma}$ is the antisymmetric tensor with
$\epsilon_{0123}=1$. The amplitudes  $T_k$ are functions of the two
variables $\nu$ and $t$ and $M$ is the mass of the nucleon.
The normalization of the amplitude $T_{fi}$ is determined by
\begin{equation}
\bar{u}u=2M, \quad \frac{d\sigma}{d\Omega}=\frac{1}{64 \pi^2
  s}\sum_{\rm spins}|T_{fi}|^2.
\label{NORM}
\end{equation}
It follows from crossing symmetry of $T_{fi}$, that $T_{1,3,5,6}$ and
$T_{2,4}$
are even and odd functions of $\nu$, respectively.

The amplitudes $T_k(\nu,t)$  do not have
kinematic singularities, but there are kinematic constraints, which
arise from the vanishing of $P'{}^2$, $N^2$ and $P'{}^2K^2$ in the
denominators of the decomposition (\ref{Tif}) at certain values of $\nu$
and $t$. The kinematic constraints can be removed by introducing
linear combinations of the amplitudes $T_k(\nu,t)$. This problem has first
been satisfactorily solved 
 by Bardeen and Tung \cite{bardeen68}. However, the Bardeen and Tung 
amplitudes contain the inconvenience that part of them are even functions of
$\nu$ and part of them odd functions. This inconvenience has been removed by
L'vov \cite{lvov81} so that these latter amplitudes now have become standard.

The linear combinations  introduced by 
L'vov \cite{lvov81} are 
\begin{eqnarray}
&&A_1=\frac{1}{t}[T_1+T_3+\nu (T_2+T_4) ], \nonumber \\
&&A_2=\frac{1}{t}[2T_5 +\nu (T_2+T_4)], \nonumber\\
&&A_3=\frac{M^2}{M^4-su}\left[T_1-T_3 - \frac{t}{4\nu}(T_2-T_4)\right],
\nonumber\\
&&A_4=\frac{M^2}{M^4-su}\left[2M
  T_6-\frac{t}{4\nu}(T_2-T_4)\right],\nonumber\\
&&A_5=\frac{1}{4\nu}[T_2+T_4],\nonumber\\
&&A_6=\frac{1}{4\nu }[T_2-T_4]. \label{T1-6}
\end{eqnarray}
The amplitudes $A_i(\nu,t)$ are even functions of $\nu$ and have no
kinematic singularities or kinematic constraints. They
have poles at zero energy because of contributions of the nucleon in
the intermediate state. These poles are contained in two Born diagrams
with the pole propagator $(\gamma\cdot p -M)^{-1}$ and on-shell
vertices $ \Gamma_\mu(p+k,p)=\gamma_\mu+[\gamma\cdot
  k,\gamma_\mu]\kappa/4M$,
where $\kappa=1.793 q- 1.913(1-q)$ is the nucleon anomalous magnetic
moment. Here the electric charge of the nucleon,
$q=\frac12 (1+\tau_3)=1$ or $0$ is introduced. The Born contributions to the
amplitudes $A_i$ have a pure pole form
\begin{equation}
A^{\rm B}_i(\nu,t)=\frac{M e^2 r_i(t)}{(s-M^2)(u-M^2)},
\label{Born}
\end{equation}
where e is the elementary electric charge ($e^2/4\pi = 1/137.04$)
and 
\begin{eqnarray}
&&r_1=-2q+(\kappa^2+2q\kappa)\frac{t}{4M^2},\quad
r_2=2q\kappa+2q+(\kappa^2+2q\kappa)\frac{t}{4M^2},\nonumber\\
&&r_3=r_5=\kappa^2+2q\kappa,\quad r_4=\kappa^2,\quad
r_6=-\kappa^2-2q\kappa-2q^2.
\label{residue}
\end{eqnarray}

\subsection{Lorentz invariant definition of polarizabilities}

The relations between 
the amplitudes $f$ and $g$ introduced in  Section 2.1
and the invariant amplitudes $A_i$ \cite{babusci98,lvov97}
 are
\begin{eqnarray}
&&f_0(\omega)= -\frac{\omega^2}{2\pi}\left[A_3(\nu,t)+ A_6(\nu,t)
\right],\quad\quad\quad
g_0(\omega)=\frac{\omega^3}{2\pi M}A_4(\nu,t), \label{T3}\\
&&f_\pi(\omega)=-\frac{\omega\omega'}{2\pi}\left(1+\frac{\omega\omega'}
{M^2}\right)^{1/2}\left[ 
A_1(\nu,t) - \frac{t}{4 M^2}A_5(\nu,t)\right],\label{T4}\\
&&g_\pi(\omega)=-\frac{\omega\omega'}{2\pi M}\sqrt{\omega\omega'}
\left[
A_2(\nu,t)+ \left(1-\frac{t}{4 M^2}\right)A_5(\nu,t)\right]
,\label{T5}\\
&&\omega'(\theta=\pi)=\frac{\omega}{1+2\frac{\omega}{M}},\,\,
\nu=\frac12 (\omega+\omega'),\,\, t(\theta=0)=0,
\,\,t(\theta=\pi)=-4\omega\omega.'
\label{T6}
\end{eqnarray}
For the electric, $\alpha$, and magnetic, $\beta$,  polarizabilities 
and the spin polarizabilities $\gamma_0$ and $\gamma_\pi$ for the
forward and backward directions, respectively, 
we obtain the relations
\begin{eqnarray}
&&\alpha+\beta = -\frac{1}{2\pi}\left[A^{\rm nB}_3(0,0)+ 
A^{\rm nB}_6(0,0)\right], \quad 
\alpha-\beta = -\frac{1}{2\pi}
\left[A^{\rm nB}_1(0,0)\right], \nonumber\\ 
&&\gamma_0= \frac{1}{2\pi M}\left[A^{\rm nB}_4(0,0)
\right], \quad\quad\quad\quad \quad\quad\quad\,\,\,
\gamma_\pi = -\frac{1}{2\pi M}
\left[A^{\rm nB}_2(0,0)+A^{\rm nB}_5(0,0) \right],
\label{T7}
\end{eqnarray}
where $A_i^{\rm nB}$ are the non-Born parts of the invariant amplitudes.

According to Eqs. (\ref{T3}) to  (\ref{T7}) the following linear
combinations of invariant amplitudes are of special importance because 
they contain the physics of the four fundamental sum rules, { \it viz.}
the BEFT, LN, BL and GDH sum rules, respectively:
\begin{eqnarray}
&& {\tilde A}_1(\nu,t)\equiv A_1(\nu,t)-\frac{t}{4M^2}A_5(\nu,t),\label{T8}\\
&& {\tilde A}_2(\nu,t)\equiv A_2(\nu,t)+\left(1-\frac{t}{4M^2}\right)
A_5(\nu,t),\label{T9}\\
&& {\tilde A}_3(\nu,t)  \equiv A_{3+6}(\nu,t)\equiv A_3(\nu,t)+ 
A_6(\nu,t), \label{T10}\\
&& {\tilde A}_4(\nu,t)   \equiv A_4(\nu,t).
\label{T11}
\end{eqnarray}

\subsection{Properties of the $s$- and $t$-channel}
 
For the general discussion of dispersion theories
we shall consider the invariant amplitudes $A_i(s,t,\tilde{u})$ 
as  functions of two {\it complex
variables} ($s$ and $t$) and establish their analytical properties in the
``topological product''  of the $s$- and $t$-planes \cite{roman}. 
For sake of completeness and for further applications we introduce the
further variable $u$ which is constrained by the two others,
what frequently is denoted by the tilde \cite{hoehler83}, {\it i.e.},
in the present case $\tilde{u}$.
 
\begin{figure}[h]
\begin{center}
\includegraphics[width=0.5\linewidth]{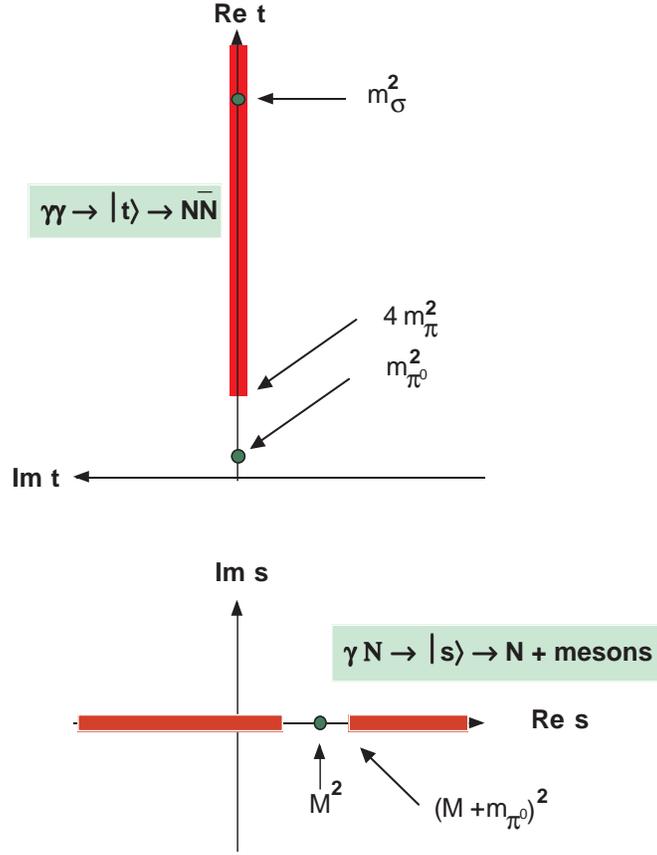}
\end{center}
\caption{Singularities in the $s$ and $t$ planes. The $s$-plane
contains the nucleon pole at $s=M^2$ and a right hand cut
corresponding to the $s$-channel and a left-hand cut corresponding to the 
$u$-channel. The $t$-channel contains a pole at $t=m^2_{\pi^0}$ and a
cut starting at $t_{\rm thr}=4 m^2_{\pi}$. In the range of the  
$t$-channel cut 
there are further poles at $t=m^2_{\eta}$ and
$t=m^2_{\eta'}$. Furthermore,  there is a
pole-like phase substructure  defining a ``mass'' at $t=m^2_{\sigma}$. }
\label{s-t-plane}
 \end{figure}

The amplitudes $A_i(s,t,\tilde{u})$ are chosen such that they 
do not have kinematical
singularities or  constraints. But there are {\it physical}
singularities which are the basis for constructing the amplitudes.
In the $s$-plane there are  singularities on the real axis,
a pole at $s=u=M^2$ and two cuts, representing the $s$- and the $u$-channel.
This is illustrated in  Figure
\ref{s-t-plane}.
The pole at $s=u=M^2$ represents the Born term, i.e. Thomson
scattering without  excitation of internal degrees of freedom of the nucleon.  
The main singularities in the $t$-plane are poles on the positive 
real $t$ axis corresponding   
to the pseudoscalar mesons $\pi^0$, $\eta$ and $\eta'$, and a cut
starting $t=4 m^2_\pi$. This cut consists of a 
$\pi\pi$ channel with  a  $\pi\pi$ phase-substructure which may be
interpreted in terms of the   scalar isoscalar $f_0(600)$ or $\sigma$ 
particle ( see Section 2.6.2).
The singularities in the two planes are taken into account by
imaginary parts of the amplitudes, i.e.  
${\rm Im}_s A_i(s,t,\tilde{u})$ for the $s$-plane and 
${\rm Im}_tA(s,t,\tilde{u})$ for the $t$-plane. 

\subsection{Properties of the Mandelstam plane}

\begin{figure}[t]
\centering\includegraphics[width=0.6\linewidth]{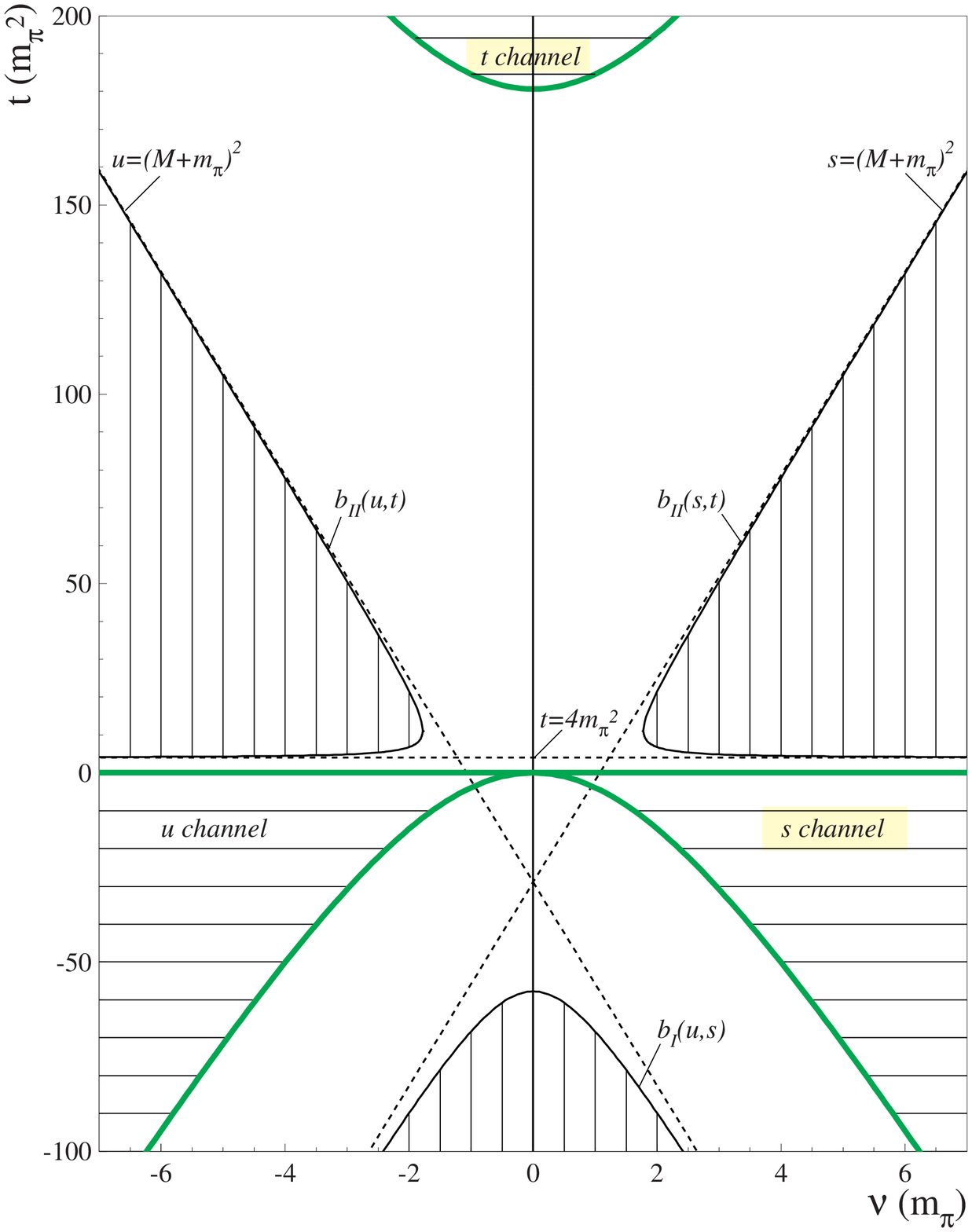}
\caption{The Mandelstam plane for Compton scattering according to
\cite{drechsel03}. The physical
regions are horizontally hatched. The spectral regions (with
boundaries $b_I$ and $b_{II}$) are vertically hatched.}
\label{MandelstamPlane}
\end{figure}

For illustration it is useful to merge the complex $s$ and $t$ planes
into one plane by choosing  the real axes as abscissa and
ordinate. Furthermore, we may replace  $s$ by $\nu$. This leads to the
Mandelstam plane shown in Figure   \ref{MandelstamPlane}.

First we want to identify the region  in the Mandelstam plane where
Compton scattering may take place and will name that region  the $s$-channel
{\it physical region}. There are also physical regions for the $u$
and $t$ channel as shown in Figure  \ref{MandelstamPlane}.
From a mathematical point of view the physical regions  are
of minor importance because the scattering amplitudes in principle are
defined in the whole {\it topological product} of the complex $s$ and $t$
planes.  As stated before, the appropriate tools for the analytical
continuations are the generalized angles $\theta_s$ and $\theta_t$ given in
Eqs. (\ref{scatteringangle}) and (\ref{eq:cos-theta-t}). 

The boundary of the $s$-channel physical region (Figure \ref{s-t-plane})
is  defined through the
threshold $s_0=(M+m_\pi)^2$ for photo-absorption. In the Mandelstam plane
(Figure \ref{MandelstamPlane})
this physical region is given by the  interval
between $\theta=0$ and $\theta=\pi$ of the {\it physical} scattering
angle. The boundary $\theta=0$ is equivalent with $t=0$. The boundary
corresponding to $\theta=\pi$ may be obtained through the following
consideration. Let $s_\pi(t)$, $u_\pi(t)$ and $\nu_\pi(t)$ be the 
Mandelstam variables corresponding to $\theta=\pi$. Then with
\begin{eqnarray}
&& s_\pi(t) u_\pi(t) =M^4, \nonumber \\
&& s_\pi(t) + t + u_\pi(t)=2M^2, \nonumber\\
&& \nu_\pi(t) =  \left( s_\pi(t) - u_\pi(t)\right)/(4M), \label{VariablesAtPi}
\end{eqnarray}
we arrive at
\begin{eqnarray}
&&s_\pi(t)=\frac12 \left( 2M^2-t + \sqrt{t(t-4M^2)} \right), \label{s-pi}\\
&&u_\pi(t)=\frac12 \left( 2M^2-t - \sqrt{t(t-4M^2)} \right), \label{u-pi}\\
&&\nu_\pi(t)= \frac{1}{4M}\sqrt{t(t-4M^2)}. 
\label{nu-pi}
 \end{eqnarray}
In Eqs. (\ref{s-pi}) to (\ref{nu-pi}) the signs in front of the square
roots have been chosen such that $\nu_\pi(t)$ is positive for negative
$t$. In the Mandelstam
plane of Figure \ref{MandelstamPlane}, Eq. (\ref{nu-pi}) describes the
lower borderline of the physical region of the 
$s$-channel for positive $\nu$ and   the 
lower borderline of the $u$-channel for negative  $\nu$ if  $t\leq 0$.
For  $t \geq 4M^2$ Eq. (\ref{nu-pi}) describes the lower borderline of
the physical region of the $t$-channel. The $t$-channel
singularities of interest
in connection with Compton scattering are positioned at
$t=m^2_{\pi^0}$
and at $4M^2\geq t \geq 4m^2_\pi$ i.e. outside the physical region
of the $t$-channel for the reaction $\gamma\gamma\to N\bar{N}$. 
For the $t$-channel 
this means that the two photon fusion process $\gamma\gamma\to\pi\pi$
leading to the $|\pi\pi\rangle$ intermediate state 
or some other resonant or nonresonant intermediate state $|t\rangle$, 
takes place
as a real (on-shell) process whereas the subsequent 
process $\pi\pi\to N\bar{N}$
takes place virtually, i.e. below threshold for $N\bar{N}$
pair production. This corresponds to low-energy Compton scattering processes
proceeding through $|\pi^0\rangle$, $|\pi\pi\rangle$,
etc. $t$-channel exchanges in the intermediate state 
with no excitation of the 
constituent-quark-meson structure of the nucleon.

In Figure \ref{MandelstamPlane}
the vertically hatched areas are the regions of nonvanishing double
spectral functions \cite{drechsel03,hoehler83,drechsel99}. 
These spectral regions 
are those regions in the Mandelstam plane where two of the three
variables
$s$, $t$ and $u$ take on values that correspond with a physical 
(i.e., on-shell) intermediate state. The boundaries of these regions 
follow from unitarity. 
The double spectral functions are of importance in connection with the
foundation of dispersion theories \cite{hoehler83}. No use is made
of them in the following  discussion of fixed-$\theta$ and
fixed-$t$ dispersion theories.

\subsection{Discussion of Compton scattering in the complex $s$-plane}

The $\theta =\pi$ constraint given in the first line of
Eq. (\ref{VariablesAtPi}) may be generalized  for other angles using the 
$s-u$ crossing symmetric hyperbolic integration paths 
\cite{hoehler83,drechsel03}
\begin{equation}
(s-a)(u-a)=b, \quad b=(a-M^2)^2,
\label{hyperbolicPaths}
\end{equation}
where $a$ is in one-to-one correspondence with the lab and 
  c.m.
scattering angles as:
\begin{equation}
a=-M^2\frac{1+\cos \theta_{\rm lab}}{1- \cos \theta_{\rm lab}},
\quad\quad  
a=-s\frac{1+\cos \theta_{ s}}{1- \cos \theta_{s}},
\label{a-s-relation}
\end{equation}
where $\theta_{\rm lab}$ is the scattering angle in the laboratory
system and 
$\theta_{s}$ the c.m. scattering angle in the $s$-channel. 
Eqs. (\ref{hyperbolicPaths}) and (\ref{a-s-relation}) contain the
main ingredients of the kinematics of fixed-$\theta$ dispersion theory.

The equations given above allow to project the $t$-channel singularities into
the complex  $s$-plane as first discussed by Hearn and Leader in 1962
\cite{hearn62} and later worked out in more detail by K\"oberle in 1968
\cite{koeberle68}. 
\begin{figure}[h]
\centering\includegraphics[width=0.5\linewidth]{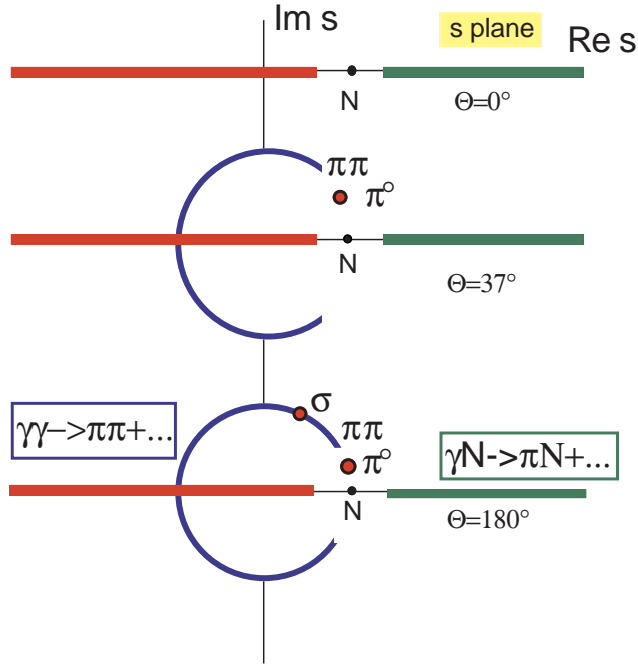}
\caption{Singularities of Compton scattering amplitudes depicted in
the complex $s$-plane according to Hearn and Leader
\cite{hearn62}. The right thick line on the real axis
corresponds to the $s$-channel, the circular line and $\pi^0$-pole 
to the ``unphysical'' part of the  $t$-channel and the  left thick line
on the horizontal axis partly to the $u$-channel and partly to the
``physical'' part of the $t$-channel. For $\theta=180^\circ$ the lower
limit of the $u$-channel cut is at $s=0$.
}
\label{complexSplane}
\end{figure}
The results  of these considerations are very important
because they prove that in fixed-$\theta$ dispersion theory double counting
of physical input as  coming from the $s$-channel and the $t$-channel 
is avoided.
Figure \ref{complexSplane} reproduces the result obtained in \cite{hearn62},
where the singularities contributing to the imaginary parts of the
Compton scattering amplitudes are depicted for three scattering angles. 
The $s$-channel is represented by the right-hand cut on the real $s$-axis.
The circular part contains the $\pi^0$ pole, the $\pi\pi$ cut and 
the other structures located in the unphysical
region of the $t$-channel, i.e. below the $N\bar{N}$ threshold
(see Figure \ref{s-t-plane}).  The left-hand cut on the real $s$-axis
is shared by the $u$-channel and the physical part of the $t$-channel,
i.e. the range above the $N\bar{N}$ threshold. 
It is of
interest that the circular part corresponding to the ``unphysical''
part of the $t$-channel does only contribute when the scattering angle
is nonzero and is most pronounced for $\theta=180^\circ$.

\subsection{Fixed-$\theta$ dispersion relations}

Along a path at fixed $a$ defined in (\ref{hyperbolicPaths}) and 
(\ref{a-s-relation}), one can write down \cite{drechsel03,hoehler83}
a dispersion integral as
\begin{eqnarray}
{\rm Re}A_i(s,t,a)&=& A^{\rm B}_i(s,t,a)+A^{t-{\rm
    pole}}_i(s,t,a)\nonumber\\
&+& \frac{1}{\pi}\int^\infty_{(M+m_\pi)^2} ds'{\rm
    Im}_s A_i(s',\tilde{t},a)
\left[\frac{1}{s'-s}+\frac{1}{s'-u}-\frac{1}{s'-a}\right]\nonumber\\
&+&\frac{1}{\pi} \int^\infty_{4m^2_\pi} dt' \frac{{\rm Im}_t
 A_i(\tilde{s},t',a)}{t'-t},
\label{Drechsel(185)}
\end{eqnarray}
where  ${\rm Im}_s A(s',\tilde{t},a)$
is evaluated along the hyperbola given by
\begin{equation}
(s'-a)(u'-a)=b, \quad s'+\tilde{t}+u'=2M^2,
\label{Drechsel(186)}
\end{equation}
and ${\rm Im}_t A_i(\tilde{s},t',a)$  runs along the path defined by
the hyperbola
\begin{equation}
(\tilde{s}-a)(\tilde{u}-a)=b, \quad \tilde{s}+t'+\tilde{u}=2M^2.
\label{Drechsel(187)}
\end{equation}
The amplitude $A^{t-{\rm pole}}_i(s,t,a)$ entering into
(\ref{Drechsel(185)})
describes the contribution of pseudoscalar mesons to the scattering
amplitude
and may be written in the form
\begin{equation}
A^{{\pi^0}+\eta+\eta'}_2(t)=\frac{g_{{\pi^0} NN}F_{\pi^0\gamma\gamma}}
{t-m^2_{\pi^0}}\tau_3+\frac{g_{{\eta} NN}F_{\eta\gamma\gamma}}
{t-m^2_{\eta}}+\frac{g_{{\eta'} NN}F_{\eta'\gamma\gamma}}
{t-m^2_{\eta'}},
\label{pseudoscalarPole}
\end{equation}
where the quantities $g_{\pi^0 NN}$, etc. are the meson-nucleon coupling
constants and the quantities $F_{\pi^0 \gamma\gamma}$, etc. the 
two-photon decay amplitudes. 
The last term in (\ref{Drechsel(185)}) represents the contribution
of the scalar
$t$-channel to the scattering amplitude.
The integration in the upper half plane (for $t>0$) runs through the 
unphysical region $4m^2_\pi \leq t < 4M^2$. In the latter region the
$t$-channel partial wave expansion of ${\rm Im}_t A_i$ converges
if $101^\circ \leq \theta_{\rm lab}  \leq 180^\circ$
\cite{drechsel03}. This statement shows that
fixed-$\theta$ dispersion theory cannot be applied at all angles
with equally good precision. At $\theta=180^\circ$ the application is
rather straightforward  so that backward-angle sum rules can be based
on this approach. This will be discussed in Section 6. 

\subsection{Fixed-$t$ dispersion relations}

The natural
extension of forward-angle dispersion theory to larger angles
is given by fixed-$t$ dispersion theory. 
Fixed-$t$ dispersion theory has been proven to be successful
in a wide   range of Compton scattering angles 
and for energies up to about 1 GeV with high
precision (see Section 5). A formal drawback, however, is that in 
fixed-$t$ dispersion theory  a
strict separation of the $s$- and $t$-channel contributions is not
possible but we may assume an approximate separation. In order 
to distinguish between the strict separation in case of
fixed-$\theta$ dispersion theory and the approximate
separation in case of fixed-$t$ dispersion theory we use the term 
``integral''  amplitude instead
of ``$s$''-channel amplitude  and ``asymptotic'' amplitude instead
of ``$t$-channel''  amplitude.

The  ansatz for a fixed-$t$ dispersion relation may be given
in the following form \cite{hoehler83}:
\begin{eqnarray}
{\rm Re}A_i(s,t,{\tilde u})=A^B_i(s,t,{\tilde u}) &+& \frac{1}{\pi}
{\cal P} \int^{s_{\rm max}(t)}_{s_0}\left( \frac{1}{s'-s} + \frac{1}{s' -
  \tilde{u}}
\right) 
\times {\rm Im}_s A_i(s',t,\tilde{u})ds'\nonumber\\
&+& \frac{1}{\pi}
{\cal P} \int^\infty_{s_{\rm max}(t)}\left( \frac{1}{s'-s} + \frac{1}{s' -
  \tilde{u}}
\right)\times {\rm Im}_s A_i(s',t,\tilde{u})ds' \nonumber\\
&+& A^{t-{\rm pole}}_i(t)+ A^{\rm scalar}_i(t)
\label{fixed-t}
\end{eqnarray}
with the  constraint
\begin{equation}
s'+ \tilde{u}+t=2M^2
\label{constraints}
\end{equation}
where use has been made of the crossing symmetry \cite{hoehler83}
\begin{equation}
{\rm Im}_sA_i(s,t,u)={\rm Im}_uA_i(u,t,s).
\label{crossing}
\end{equation}
The terms in the first two lines in Eq. (\ref{fixed-t}) represent  
the standard fixed-$t$
dispersion relation \cite{hoehler83}. 
The $s$-channel integral has been split up into one integral
extending from $s_0$ to $s_{\rm max}(t)$ and an other extending from
$s_{\rm max}(t)$ to $\infty$. This precaution
has been taken in order to avoid  the use of  divergent $s$-channel
integrals. The quantity $s_{\rm max}(t)$ in (\ref{fixed-t})  
has been proposed \cite{lvov97} to correspond to the excitation energy
$E_{\rm max}=(s_{\rm max}(t)-M^2)/2M= 1.5\,\, {\rm GeV}$.
The fourth  term $A^{t-{\rm pole}}_i(t)$ is the
analogue of the corresponding term in (\ref{Drechsel(185)}). The term
$A^{\rm scalar}_i(t)$ replaces the $t$-channel integral in
(\ref{Drechsel(185)}). It is interesting to mention that $A^{\rm scalar}_i(t)$
cannot be expressed through a $t$-channel integral in a strict sense, 
as in case of
the fixed-$\theta$ dispersion theory. The reason for this is that in
case of fixed-$t$ dispersion theory parts of the $t$-channel strength
is already contained in the integral amplitude.
This is especially true for contributions which -- in a diagrammatic
representation -- have a $t$-channel cut as well as a $s$-channel cut
\cite{lvov04}.
However, there are reasons to assume that the term 
$A^{\rm scalar}_i$ in (\ref{fixed-t}) and the $t$-channel integral 
in (\ref{Drechsel(185)}) are approximately equivalent.

For the present case of a fixed-$t$ dispersion relation
it is useful to work with the 
variable $\nu=(s-u)/(4M)$, because the crossing operation
$s\leftrightarrow u$ at fixed $t$ is then simply a reflection 
$\nu \leftrightarrow -\nu$. In terms of the variable $\nu$, 
the first term of the  $s$-channel part of the 
fixed-$t$ dispersion relation reads
\begin{eqnarray}
{\rm Re}A^{\rm NB}_i(\nu,t)&=&\frac{1}{\pi}{\cal P}
\int^{\nu_{\rm max}(t)}_{\nu_{\rm thr}(t)}
d\nu'{\rm Im}A_i(\nu',t)
\left[\frac{1}{\nu'-\nu}+\frac{1}{\nu'+\nu}\right]\\
&=&\frac{2}{\pi}{\cal P}\int^{\nu_{\rm max}(t)}_{\nu_{\rm thr}(t)}
d\nu'\frac{ \nu' {\rm Im}A_i(\nu',t)}{{\nu'}^2-\nu^2}.
\label{variable-nu}
\end{eqnarray}
where NB denotes non-Born.
The integral given in Eq. (\ref{variable-nu}) starts at the threshold
\begin{equation}
\nu_{\rm
  thr}(t)=\nu\left( s=(M+m_\pi)^2,t\right)=m_\pi+\frac{1}{2M}
\left(m^2_\pi+\frac{t}{2}
\right). \label{nu-th}
\end{equation}
This function is shown in Figure \ref{MandelstamPlane}  as a dotted line
labeled $s=(M+m_\pi)^2$. It is easy to see that the threshold of the
integral (\ref{variable-nu}) is located in the physical region of the 
$s$-channel only for very small negative $t$. Outside the physical region
$ \nu_{\rm thr}(t)$ remains positive up to $t=-28.8\,\, m^2_\pi$ and then
becomes negative. This unwanted property will be discussed in connection with
the evaluation of the fixed-$t$ dispersion integrals (see Section 4.5).
 
The second integral  and the two terms in the last line
of   in (\ref{fixed-t})              are dependent on $t$ but not
on $s$. Therefore, they can be added for each amplitude $A_i$ 
to  give one term which
traditionally is termed the ``asymptotic'' contribution 
$A^{\rm as}_i(\nu,t)$
where we keep the $\nu$ dependence for sake of completeness.  
In a formal way this ``asymptotic'' contribution can be incorporated
into the fixed-$t$ dispersion theory formulated in a  complex 
$\nu$ plane,  by using
a loop of finite size (a closed semicircle of radius
$\nu_{\rm max}$) to close the Cauchy integral.  This leads to the 
representation \cite{lvov97}:
\begin{equation}
{\rm Re}A_i(\nu,t)=A^B_i(\nu,t)+A^{\rm int}_i(\nu,t)+ A^{\rm as}_i(\nu,t),
\label{threeA}
\end{equation}
with 
\begin{eqnarray}
&&A^{\rm B}_i(\nu,t)=\frac{a_i(t)}{\nu^2-t^2/16M^2}\,\,,\label{threeA-1}\\
&&A^{\rm int}(\nu,t)=\frac{2}{\pi}{\cal P}
\int^{\nu_{\rm max}(t)}_{\nu_{\rm thr(t)}} {\rm Im}A_i(\nu',t)
\frac{\nu'd\nu'}{\nu'{}^2-\nu^2}\,\,,\label{threeA-2}\\
&&A^{\rm as}_i(\nu,t)=\frac{1}{\pi}{\rm Im}
\int_{\nu'=\nu_{\rm max}(t)e^{i\phi},0<\phi<\pi}
A_i(\nu',t)\frac{\nu'd\nu'}{\nu'{}^2-\nu^2}\,\,.
\label{threeA-3}
\end{eqnarray}
Without loss of generality we may use $\nu_{\rm max}(t)\to \infty$, and
this will be done in the following unless something else is stated.

In the foregoing,  the non-Born (NB)
amplitude has been partitioned into an integral part along the real
axis of the complex $\nu$ plane and a contour integral in the upper
half-plane. This partitioning is a formal procedure in the sense that 
only the integral part can be calculated, using the formula given
in (\ref{threeA-2}). The asymptotic (contour integral) part  requires
special considerations. The usual way to take it into account is to
use the ansatz of a phenomenological  effective $\sigma$-pole for the amplitude
$A^{\rm as}_1(\nu,t)$ and poles due to pseudoscalar mesons for the
amplitude $A^{\rm as}_2(\nu,t)$ as given in Eq. (\ref{pseudoscalarPole}):
\begin{equation}
A^{\rm as}_1(\nu,t)=A^{\sigma}_1(t)=\frac{g_{\sigma NN}
F{\sigma }\gamma\gamma}{t-m^2_{\sigma }},
\quad A^{\rm as}_2(\nu,t) \equiv A^{\pi^0+\eta+\eta'}_2(t).
\label{sigmaPole}
\end{equation}
Experimental tests  of this ansatz are described in Sections 5 and 6.

\subsection{Propagator of  the $\sigma$ meson}

For the practical application of the $\sigma$ pole ansatz in the data 
analysis we
assume that  the $\sigma$ meson  has a composite structure with a 
$|q{\bar q}\rangle$ core and a $|\pi\pi\rangle$ cloud.  Due to the
coupling to the $\pi\pi$ hadronic exit channel the $\sigma$ has a
large width which modifies the structure of the corresponding
pole term. This modification can be calculated by standard techniques
\cite{milstein03} leading to the result that the  mass $m_\sigma$
has to be replaced by  an effective mass 
$m_{\rm eff}=(m^2_\sigma +{\cal P}(t))^{1/2}$.
\begin{figure}[h]
\begin{center}
\includegraphics[scale=0.8]{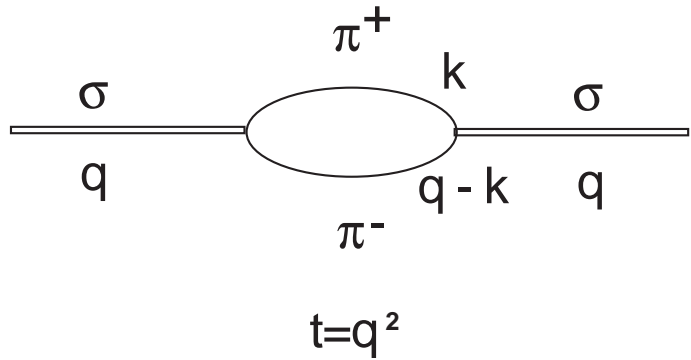}
\end{center}
\caption{Propagator of a $\sigma$ meson in the presence of a $\pi^+\pi^-$ loop.
In addition, the $\pi^0\pi^0$ loop has to be taken into consideration.}
\label{propagator}
\end{figure}
The width of the $\sigma$ meson is taken into account via a 
$\pi\pi$ loop as depicted in Figure \ref{propagator}, leading to 
\begin{equation}
\frac{F_{\sigma\gamma\gamma}g_{\sigma NN}}
{t-m^2_\sigma} \longrightarrow \frac{F_{\sigma\gamma\gamma}g_{\sigma NN}
}{t-m^2_\sigma- {\cal P}(t)}
\label{modifiedpropagator}
\end{equation}
with
\begin{eqnarray}
&&{\cal P}(t)=\frac{m_\sigma\Gamma_\sigma}{\pi\sqrt{1-4m^2_\pi/m^2_\sigma}}
\left[-2+\sqrt{\frac{1+y}{y}}\ln \left( \frac{\sqrt{\frac{1+y}{y}}+1}
{\sqrt{\frac{1+y}{y}}-1} \right)\right], \\
&&y=\frac{-t}{4m^2_\pi}.\nonumber
\label{P-t}
\end{eqnarray}
This means that the width $\Gamma_\sigma$ of the $\sigma$ mesons 
leads to a shift of the pole parameter $m^2_\sigma$. For the data analysis
we replace the $t$ dependent effective mass
$m_{\rm eff}=(m^2_\sigma+{\cal P}(t))^{1/2}$ in the
denomiator of the r.h.s of Eq. (\ref{modifiedpropagator}) by a constant
and make a tentative prediction by assuming
$m_\sigma=500$ MeV, $\Gamma_\sigma=
500$ MeV (see Section 2.6.2) for the $\sigma$ meson and 
$\theta_{\rm cm}=125^\circ$ 
and $E_\gamma= 600$ MeV  for the kinematics of the experiment. Then we 
arrive at the prediction  $m_{\rm eff}= 630$ MeV.

For the 
data analysis we use the relations
\begin{equation}
2\pi(\alpha-\beta)+A^{\rm int}_1(0,0)= -A^{\rm as}_1(0)=
\frac{g_{\sigma NN}F_{\sigma \gamma\gamma}}{m^2_{\rm eff}},\quad 
A^{\rm as}_1(\nu,t)= 
\frac{g_{\sigma NN}F_{\sigma \gamma\gamma}}{t-m^2_{\rm eff}}.
\label{alpha-beta-sigma}
\end{equation}
In (\ref{alpha-beta-sigma}) $(\alpha-\beta)$ is the experimental value for the
difference of electromagnetic polarizabilities measured at low energies 
where the ${\cal O}(\omega^2)$ approximation is valid, $A^{\rm int}_1(0,0)$
the integral part calculated from photomeson data (see Section 4) and  
$m_{\rm eff}$ an adjustable parameter to be determined from experimental
Compton differential cross sections in the second resonance region and a range
of large scattering angles. The fit parameter obtained for a
given parameterization of photomeson amplitudes (see Section 5.5) 
is  $(589\pm 12)$ MeV. The error $\pm 12$ only takes
into account the uncertainties of the fit. The  additional error due to
the uncertainties  in the parameterizations of photomeson amplitudes
can only be estimated with a large
margin of uncertainty. In view of this, the predicted and the experimental 
effective mass  may be considered as being in good agreement with each other.

\section{The imaginary parts of Compton amplitudes}

In the following we discuss the method through which the imaginary
parts   ${\rm Im}A_i(\nu,t) $ of the amplitudes $A^{\rm int}(\nu,t)$
given in Eq. (\ref{variable-nu}) and (\ref{threeA-2}) are obtained. 
The main input are the single pion 
photoproduction data which are available with improving precision.

\subsection{The unitarity relation}

The integral contributions $A^{\rm int}_i$ are determined 
by the imaginary parts of the Compton scattering amplitudes which follow
from the unitarity relation of the generic form \cite{landau90}
\begin{equation}
T_{fi}-T^*_{if}=2 {\rm Im} T_{fi} 
= \sum_n (2\pi)^4 \delta^4 (P_n-P_i) T^*_{nf}T_{ni}.
\label{GenericForm} 
\end{equation}
For the special case of forward Compton scattering where $f\equiv i$, 
we obtain
\begin{equation}
{\rm Im}T_{ii}= \frac12 (2\pi)^4 \sum_n|T_{in}|^2\delta^4(P_i-P_n)=
\frac{\omega}{4\pi}\sigma_{tot}(\omega).
\label{ForwardOptical}
\end{equation}
The quantities
$P_i$ and $P_n$ are the four-momenta before and after the absorption
of the incident photon by the nucleon. The quantity 
$\delta^4 (P_n-P_i)$ guarantees 
the energy and momentum
conservation during the absorption process. In order to measure 
$\sigma_{tot}(\omega)$ all exit channels of the exited nucleon state 
have to be 
registered. This includes all particle channels as well as all photon
decay channels. For the nucleon, in general only the elastic 
photon decay channel into the ground state is available in addition to
meson photoproduction. This exit channel is
termed elastic  radiative decay and, 
of course, is not identical with the solid-angle integrated 
differential-cross section for Compton scattering. 
For the $\Delta$ resonance region it can
be estimated that  elastic radiative decay amounts to 
$\lesssim  1 \%$ of the
total decay.  Below particle threshold  we have $\sigma_{tot}(\omega)=0$.

\subsection{The CGLN amplitudes}

The most precise data have been obtained for single-pion
photoproduction for unpolarized photons and target nucleons.  
Based on the angular distribution of the emitted
pions,  multipole analyses have been carried out. The necessary tools
for this  have been worked out by Chew, Goldberger, Low and Nambu
\cite{chew57}:
 
Let the  photoproduction amplitude be denoted by $\cal{F}$,
such that the differential cross section for meson production in the 
c.m. system is
\begin{equation}
\frac{d\sigma}{d\Omega}=\frac{q}{k}|\langle \pi N|{\cal F}|\gamma N\rangle|^2,
\label{pioncross}
\end{equation}
where the matrix element  is taken between initial and final
Pauli (not Dirac) spinors. For a given isospin  it
is then possible to write $\cal{F}$ as follows
\begin{equation}
{\cal F} = i  {\bs \sigma}\cdot {\bs \epsilon}{\cal F}_1
+ \frac{  {\bs \sigma}\cdot{\bf q}\,{\bs \sigma}\cdot({\bf k}\times
{\bs \epsilon})}{q k}{\cal F}_2
+ \frac{ i {\bs \sigma}\cdot{\bf k}\,{\bf q }\cdot {\bs \epsilon}}{q k}
{\cal F}_3
+ \frac{i {\bs \sigma}\cdot{\bf q}\,{\bf q}\cdot{\bs \epsilon}}{q^2}
{\cal F}_4,
\label{calF}
\end{equation}
where ${\cal F}_1 \cdots {\cal F}_4$ are functions of energy and angle
in the c.m. system and {\bf q} and {\bf k} are the meson and
photon three-momenta.
The angular dependence may be made explicit through a multipole expansion
involving derivatives of Legendre polynomials:
\begin{eqnarray}
&&{\cal F}_1 = \sum^\infty_{l=0}[l M_{l+} + E_{l+}] P'_{l+1}(x)
+ [(l+1)M_{l-}+ E_{l-}]P'_{l-1}(x),\label{F1}\\
&&{\cal F}_2 = \sum^\infty_{l=1}[(l+1) M_{l+} + l M_{l-}]P'_{l}(x),\label{F2}\\
&&{\cal F}_3 = \sum^\infty_{l=1}[E_{l+} - M_{l+}] P''_{l+1}(x)
+[E_{l-}+M_{l-}]P''_{l-1}(x),\label{F3}\\
&&{\cal F}_4 = \sum^\infty_{l=1}[M_{l+}-E_{l+}-M_{l-}-E_{l-}]P''_l(x).
\label{F4}
\end{eqnarray}
Here $x$ is the cosine of the angle of emission in the c.m.
system.

The energy-dependent amplitudes $M_{l\pm}$ and $E_{l\pm}$ refer to
transitions initiated by magnetic and electric radiation,
respectively, leading to final states of orbital angular momentum $l$
and total angular momentum $l\pm\frac12$. Superscripts may be added
to each amplitude in formulas (\ref{F1}) to (\ref{F4}) in order to
designate the isospin of the transition.

The most prominent CGLN photomeson amplitudes 
at low and intermediate energies 
are the nonresonant
$E_{0+}$ amplitude where the pion is emitted as an $s$-wave, and the
resonant 
$M^{3/2}_{1+}$ and $E^{3/2}_{1+}$ amplitudes leading to the $I=3/2$
component of the $\Delta$ resonance via magnetic dipole or electric
quadrupole transitions, respectively, whereafter the pion is emitted
as a $p$-wave.

\subsection{Helicity amplitudes}

Walker \cite{walker69,{karliner73}} writes the integral cross section for pion
photoproduction in terms of helicity ``elements'' (or amplitudes)  
$A_{l\pm}$ and 
$B_{l\pm}$, given in the form
\begin{eqnarray}
&&\sigma_T= \frac12 \left(\sigma_{1/2} +\sigma_{3/2}\right),\nonumber\\
&&\sigma_{1/2}=\frac{8\pi q}{k}\sum^\infty_{l=0}(l+1)(|A_{l+}|^2
+|A_{(l+1)-}|^2),\nonumber\\
&&\sigma_{3/2}=\frac{8\pi q}{k}\sum^\infty_{l=0}\frac14
[l(l+1)(l+2)](|B_{l+}|^2
+|B_{(l+1)-}|^2),
\label{helicitycrosssections}
\end{eqnarray}
where the subscript notation of the $A'$s and $B'$s is the same as that 
of CGLN \cite{chew57}; e.g. $B_{l\pm}$ refers to a state with pion
orbital angular momentum $l$ and total angular momentum $j=l\pm\frac12$.
The $A'$s and $B'$s differ in the absolute values of the
helicities $\Lambda=|\lambda|$ of the initial states
given by $\lambda=\lambda_\gamma-\lambda_p$, where $\lambda_\gamma$ is the
helicity of the incident photon and $\lambda_p$ the helicity of
the nucleon in the initial state. For the $A'$s the helicity is
$\Lambda=1/2$, for the $B'$s $\Lambda=3/2$. Walker \cite{walker69}
finds the following relations between the helicity elements and the
CGLN multipole coefficients
\begin{equation}
E_{0+}= A_{0+}, \quad M_{1-}=A_{1-},
\label{coefficients-0}
\end{equation}
and for $l\geq1$,
\begin{eqnarray}
&&E_{l+} = (l+1)^{-1}(A_{l+} +\frac12 l B_{l+}),\nonumber\\
&&M_{l+}=(l+1)^{-1}[A_{l+}-\frac12 (l+2)B_{l+}],\nonumber\\
&&E_{(l+1)-}=-(l+1)^{-1}[A_{(l+1)-}-\frac12 (l+2) B_{(l+1)-}],\\
&&M_{(l+1)-}=(l+1)^{-1}(A_{(l+1)-}+\frac12 l B_{(l+1)-}).\nonumber
\label{coefficients-l}
\end{eqnarray}
Equivalently we may write
\begin{eqnarray}
&&A_{k+}=\frac12[(k+2)E_{k+}+kM_{k+}],\quad
  B_{k+}=E_{k+}-M_{k+},\nonumber\\
&&A_{(k+1)-}=\frac12[-kE_{(k+1)-}+(k+2)M_{(k+1)-}],\nonumber\\
&&B_{(k+1)-}=E_{(k+1)-}+M_{(k+1)-}.
\label{coefficient-2}
\end{eqnarray}
For the three main multipoles this corresponds to
\begin{eqnarray}
&&E1:\quad\quad E_{0+}=A_{0+}, \nonumber\\
&&M1:\quad\quad M_{1+}=\frac12(A_{1+}-\frac32B_{1+}),\\
&&E2:\quad\quad E_{1+}=\frac12(A_{1+}+\frac12 B_{1+}).\nonumber
\label{lowestmultipoles}
\end{eqnarray}

\subsection{Imaginary parts of Compton amplitudes for $\pi N$
intermediate states}

Following L'vov  \cite{lvov97} we express the invariant amplitudes
$A_i(s,t)$ for Compton scattering through reduced helicity 
amplitudes $\tau_i$ for Compton
scattering:
\begin{eqnarray}
&&A_1=\frac{1}{(s-M^2)^2}\Big[-\frac{s}{M}\left(1-\sigma\frac{s+M^2}{2s}
\right)\tau_4  -\frac{\sqrt{s}}{2}(\tau_5+\sigma\tau_6)   \Big],          
\nonumber\\
&&A_2=\frac{1}{(s-M^2)^3}\Big[-\frac{s}{M}(s+M^2)\left(
1-\sigma\frac{s-M^2}{2s}\right)\tau_4
-\frac{\sqrt{s}}{2}(s-M^2)\tau_5+2s\sqrt{s}\left(1-\sigma\frac{s-M^2}{4s}
\right)\tau_6     \Big],                                 
\nonumber\\
&&A_3=\frac{1}{(s-M^2)^2(s-M^2+t/2)}\Big[M^3[\tau_1+(1-\sigma)\tau_2]
-2M^2\sqrt{s}\left(1-\sigma\frac{s+M^2}{2s}\right)\tau_3   
\Big],                              
\nonumber\\
&&A_4=\frac{1}{(s-M^2)^2(s-M^2+t/2)}\Big[M^3\tau_1-M^3
\left(1+\sigma\frac{M^2}{s}\right)\tau_2
+\frac{2M^4}{\sqrt{s}}\sigma\tau_3  \Big],  \nonumber\\
&&A_5=\frac{1}{(s-M^2)^2(s-M^2+t/2)}\Big[M(s+M^2)\sigma\tau_4 
-M^2\sqrt{s}(\tau_5+\sigma\tau_6) \Big],                                    
\nonumber\\
&&A_6=\frac{1}{(s-M^2)^2(s-M^2+t/2)}\Big[-\frac{M}{2}(s+M^2)[\tau_1+
(1-\sigma)\tau_2]+  
2M^2\sqrt{s}(1-\sigma)\tau_3 \Big],                                    
\label{invarianthelicity}
\end{eqnarray} 
where
\begin{equation}
\sigma=\sin^2\frac{\theta_s}{2}=-\frac{st}{(s-M^2)^2}
\label{defsigma}
\end{equation}
and $\theta_s$ the c.m. scattering angle.

The imaginary parts of the reduced helicity amplitudes $\tau_i$ may be
written in the form \cite{lvov97}
\begin{eqnarray}
&&{\rm Im}[\tau_1]^{(1\pi)}=8\pi q\sqrt{s}\,\sum^\infty_{k=0}
(2k+2)(|A_{k+}|^2+|A_{(k+1)-}|^2)
F(-k,k+2,1,\sigma),
\nonumber\\
&&{\rm Im}[\tau_2]^{(1\pi)}=8\pi q\sqrt{s}\,\sum^\infty_{k=1}
\frac{k(k+1)(k+2)}{2}(|B_{k+}|^2+|B_{(k+1)-}|^2
F(-k+1,k+3,1,\sigma),
\nonumber\\
&&{\rm Im}[\tau_3]^{(1\pi)}=8\pi q\sqrt{s}\,\sum^\infty_{k=1}
k(k+1)(k+2)(-A_{k+}B^*_{k+}-A_{(k+1)-}B^*_{(k+1)-})F(-k+1,k+3,2,\sigma),
\nonumber\\
&&{\rm Im}[\tau_4]^{(1\pi)}=8\pi q\sqrt{s}\,\sum^\infty_{k=1}
\frac{k(k+1)^2(k+2)}{2}(A_{k+}B^*_{k+}-A_{(k+1)-}B^*_{(k+1)-})
F(-k+1,k+3,3,\sigma),
\nonumber\\
&&{\rm Im}[\tau_5]^{(1\pi)}=8\pi q\sqrt{s}\,\sum^\infty_{k=0}
2(k+1)^2(|A_{k+}|^2-|A_{(k+1)-}|^2)F(-k,k+2,2,\sigma),
\nonumber\\
&&{\rm Im}[\tau_6]^{(1\pi)}=8\pi q\sqrt{s}\,\sum^\infty_{k=1}
\frac{k^2(k+1)^2(k+2)^2}{12}(-|B_{k+}|^2+|B_{(k+1)-}|^2 )F(-k+1,k+3,4,\sigma),
\label{imaginarytau}
\end{eqnarray}
where $q$ is the pion momentum and the sum over different isotopic
channels is implied. $F$ is a hypergeometric polynomial of
$\sigma=\sin^2(\theta_s/2)$:
\begin{equation}
F(a,b,c,x)=1+\frac{ab}{c}\frac{x}{1!}+\frac{a(a+1)b(b+1)}{c(c+1)}
\frac{x^2}{2!}+\cdots\,.
\label{hypergeometrical}
\end{equation}
For angular momenta $j\leq j_{\rm max} = 7/2$
predictions of imaginary parts related to single-pion photoproduction
have been based  on phenomenological analyses of photopion
experimental data.  Higher multipoles, $j >  7/2$, are assumed
to be given by one-pion-exchange (OPE) diagrams. The corresponding
formulae are given in \cite{lvov97}. 
Further information  concerning the $\pi N$ intermediate state
and information on the $\pi\pi N$, the
$\pi\Delta$ and the $\rho^0 N$ intermediate states may be found 
in the appendices  of \cite{lvov97}.

\subsection{The integral parts of the amplitudes 
calculated in fixed-$t$ dispersion theory for the unphysical region}

As noted  in connection with the Mandelstam plane 
(Figure \ref{MandelstamPlane}), part of the 
integral amplitude of  fixed-$t$ dispersion theory has to be determined
in the unphysical region between $\nu_{\rm thr}(t)$ and the borderline
$\nu_\pi(t)$ of the physical region. As discussed in \cite{lvov97}
the unphysical region corresponds to very small $\nu'$
 and very high $-t$ which corresponds to an unphysical  $z'=
\cos\theta_s$ of the photon scattering angle arising in the integrand  
when taking into account Compton scattering at high energies and
backward angles. For small angular momenta $j$ the functions
${\rm Im} A_i(\nu',t)$ do not show any pathological behavior in the
unphysical region. This, however, may not be the case when partial
waves with high $j$ are taken  from experimental fits. 
For this case special procedures have been developed \cite{lvov97}
involving model calculations which take care of the problem in a reliable way.
In conclusion, the unphysical region did not provide problems in the range
of the present investigation which is restricted to
$E_\gamma < 1 \,\,{\rm GeV}$ and $\theta_{\rm lab} < 150^\circ$.

\subsection{Application of photomeson amplitudes in Compton
  scattering}
In general the available  pion photoproduction amplitudes have 
been applied without modification, but selections between different
versions have been made in order to achieve optimum fits to the
Compton differential cross sections. The sensitivity of this selection
procedure has been demonstrated at an early stage of our
investigations where it was shown that the resonance part of the
$M_{1+}$ amplitude which was available at that time 
\cite{arndt96,arndt95} had to be scaled down
by 2.8\% \cite{peise96}.

Updated versions of photomeson amplitudes are published regularly in
the SAID and  MAID data bases 
\cite{MAID}. Though large improvements
have been made in recent years, the application of the photomeson
amplitudes to the calculation of Compton scattering amplitudes
requires a very sophisticated  procedure. One example is given in
Section 5.3.3 where the selection procedure is described to find
an optimal representation of differential cross sections  for Compton
scattering by the neutron. In the course of our work it also was
noticed that updates of photomeson amplitudes may lead to less good
agreement with Compton scattering data than the versions published
before. 
This shows that Compton scattering is well suited to make independent 
constraints on these amplitudes.

\section{Recent experiments on Compton scattering and
polarizabilities of the proton and the neutron}

\subsection{Electromagnetic polarizabilities of the proton}

A common feature of experiments carried out in the 1950's to 1970's
has been the use of continuous-energy bremsstrahlung photon beams and
photon detectors having poor energy resolution. These factors made it 
difficult to determine the incident photon flux accurately. With the
advent of high duty factor electron accelerators and large-volume
NaI(Tl) detectors these problems were largely overcome. The first
experiment making use of these new achievements was carried out 
at the University of Illinois at Urbana-Champaign
\cite{federspiel91}. Photons were
produced via the bremsstrahlung process in a thin Al foil. The 
post-bremsstrahlung electrons were momentum analyzed in a magnetic 
spectrometer and detected in an array of plastic scintillators,
thereby tagging the associated photons through their time-coincidence
with the scattered photons. In this way quasi-monochromatic photons
were obtained with an energy resolution of few MeV. The advantage
of the quasi-monochromatic photons is paid for by low  intensity 
which is limited to $10^5 - 10^6 s^{-1}$ per channel as 
defined through  a  plastic scintillator. The result obtained was
$\alpha_p=10.9\pm 2.2\pm 1.3$, $\beta_p= 3.3 \mp 2.2 \mp 1.3$ using the 
constraint $\alpha_p+\beta_p =14.2$ calculated in an early
evaluation \cite{lvov79} from photo-absorption
data assuming the validity of Baldin's sum rule.

A different method was used \cite{zieger92} at the low 
duty-cycle linear accelerator
at Mainz where brems\-strah\-lung was used for Compton scattering by the
proton through $\theta^{\rm lab}=180^\circ$.  The forward scattered
protons were deflected by a magnetic spectrometer with 12 detector 
channels to measure the energies of the recoil protons. Rather thin
targets (5 --10 mm lq. H$_2$) had to be used so that the recoil
protons had to be separated from protons produced in the thin windows
of the target. The result obtained was $\alpha_p=10.62\pm 1.22 \pm
1.05$, $\beta_p=3.58 \mp 1.22 \mp 1.05$, using the constraint 
$\alpha_p+\beta_p= 14.2$.

Two experiments were carried out in the SAL laboratory in Saskatoon.
In the 1993 experiment \cite{hallin93} a high duty-factor
bremsstrahlung beam  and a high-resolution NaI(Tl) detector
was used to measure Compton scattering using the end-point technique.
This method has the disadvantage that the shape of the bremsstrahlung 
spectrum  at the endpoint has to be known. Furthermore, the energies
(150 - 300 MeV) were far outside the range of validity of the
low-energy approximation, leading to possible large uncertainties in
the polarizabilities due to model-dependent effects. The result
obtained was $\alpha_p=9.8\pm 0.4\pm 1.1$,  $\beta_p=4.4\mp 0.4\mp
1.1$, using the constraint $\alpha_p+\beta_p=14.2$. In the 1995
experiment \cite{macgibbon95} tagged photons with energies from 70 to
100 MeV and un-tagged photons with energies from 100 to 148 MeV were
used.
The entire energy region from 70 to 148 MeV  was measured simultaneously. The
result
obtained is $\alpha_p= 12.5 \pm 0.6 \pm 0.7 \pm 0.5$,
$\beta_p= 1.7 \mp 0.6 \mp 0.7 \mp 0.5$, using the constraint 
$\alpha_p+\beta_p= 14.2$. Here a model dependent error is given in
addition to the statistical and systematic error.

Recently a re-evaluation of the  experiments on the electromagnetic 
polarizabilities of the proton has been carried out using the data
cited above and using also all the other data from the 
data base of the  1950's -- 1990's. The overall (global) average obtained
from these data is \cite{baranov00}
\begin{eqnarray}
&&\alpha_p= 11.7 \pm 0.8({\rm stat+syst}) \pm 0.7({\rm model}), 
 \nonumber \\
&& \beta_p=2.3 \pm 0.9({\rm stat+syst}) \pm 0.7({\rm model})
\label{baranovreevaluation}
\end{eqnarray}
where no use has been made of the Baldin sum rule result 
for $\alpha_p+\beta_p$ as obtained from photo-absorption data. The sum of 
polarizabilities following from (\ref{baranovreevaluation}) is 
\begin{equation}
\alpha_p + \beta_p 
= 14.0 \pm 1.3 ({\rm stat+syst)} \pm 0.6 ({\rm model}).
\label{baranovreevaluation-2}
\end{equation}

\subsubsection{Electromagnetic polarizabilities of the proton measured at
MAMI (Mainz)}

\begin{figure}[h]
\includegraphics*[width=0.4\linewidth]{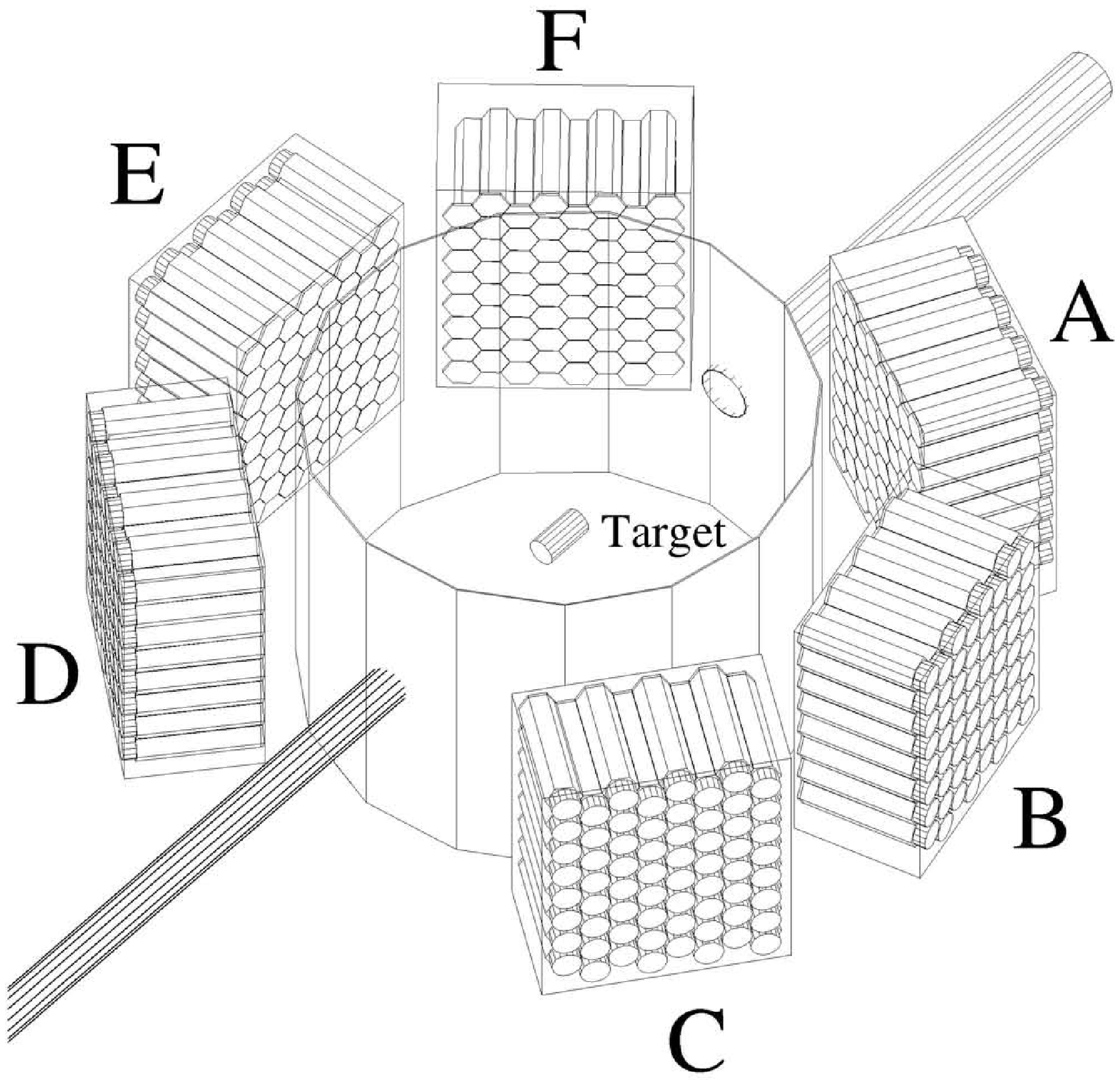}
\includegraphics[width=0.5\linewidth]{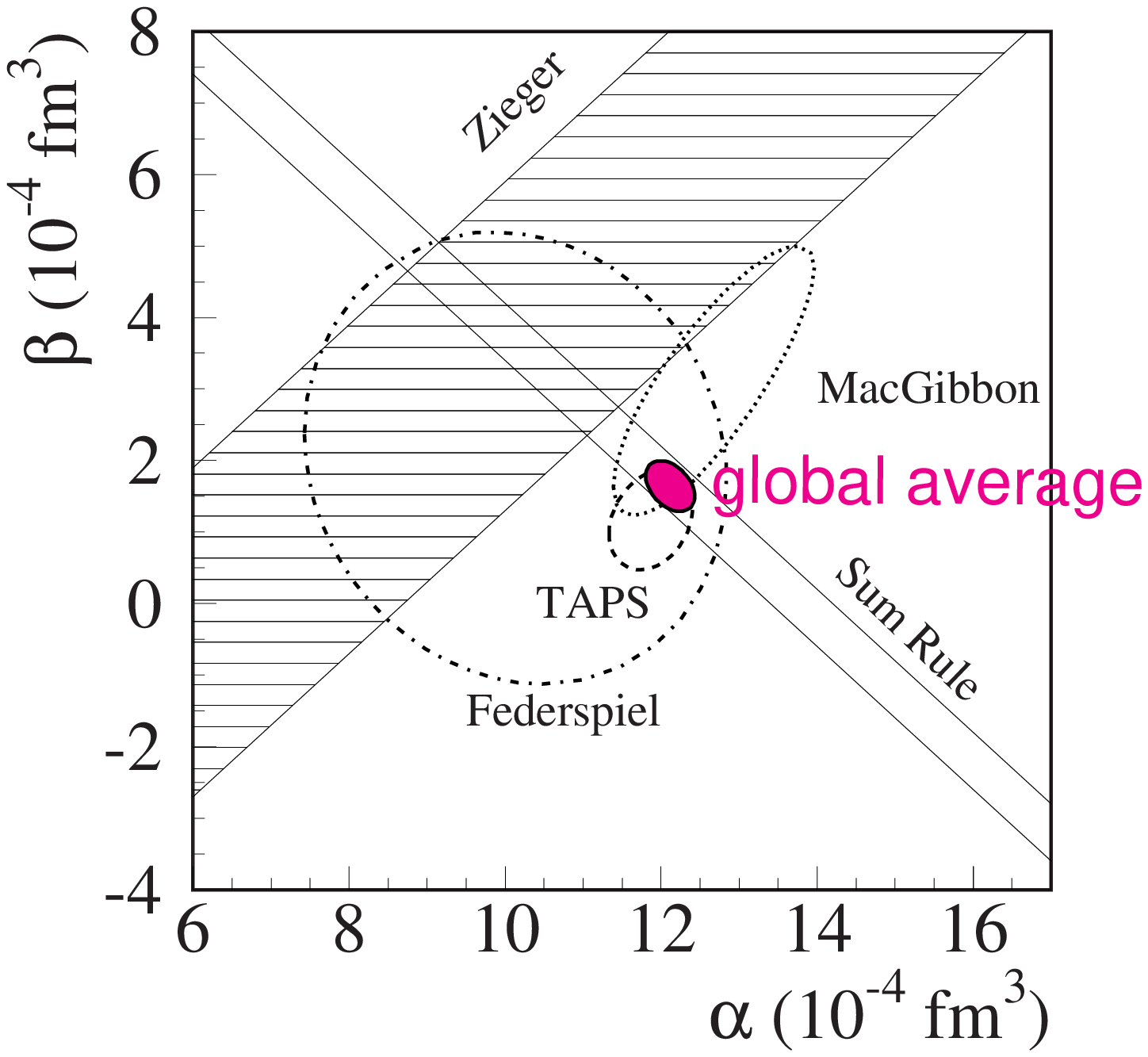}
\caption{Left panel: The TAPS arrangement at MAMI. The photon beam enters
  the setup between the blocks A and F. Also shown are the target
  cell and the scattering chamber. Right panel: Error contour plot 
$\chi^2(\alpha,\beta)=\chi^2_{\rm min}+1$ for which the errors are
  taken as the statistical ones only.
Also shown are the constraint form the Baldin
sum rule  and the value $\alpha -\beta$ as follows from the experiment
by Zieger et al. \cite{zieger92}. The grey ellipse corresponds to the
  global fit of \cite{olmos01,wissmann04}.} 
\label{fig:TapsExperiment}
\end{figure}

A new precise determination of the electromagnetic polarizabilities 
of  the proton $\alpha_p$ and $\beta_p$  has been performed using the 
TAPS apparatus set up at the  tagged-bremsstrahlung  facility at MAMI
(Mainz) \cite{olmos01,wissmann04}. This facility provided
quasi-monochromatic photons with an energy resolution of
$\Delta E_\gamma \approx 2 $ MeV at energies in the range between
$ E_\gamma \approx 60$ MeV and $160$ MeV.
At these low energies the  low-energy approximation is valid to a
large extent so that model dependencies arising from higher-oder terms
in the photon energy are small in a major part of this interval. 
This experiment was the first
which covered a large angular interval with one experimental setting
ranging from $\theta_{\rm lab}= 59^\circ$ to $155^\circ$. The set-up
is depicted in the left panel of Figure \ref{fig:TapsExperiment}.
The coverage of this large angular range was achieved by using 6
blocks A to F of the TAPS detector, each consisting of 64 BaF$_2$
detectors. The procedure used for the data analysis was to  take 
$\alpha_p$ and $\beta_p$ as free parameters and to make no use of the
constraints provided by Baldin's sum rule.
The result obtained in this way was
\begin{eqnarray}
\alpha_p&=& 11.9 \pm 0.5 ({\rm stat}) \mp 1.3 ({\rm syst}),\nonumber\\
\beta_p&=& 1.2 \pm 0.7 ({\rm stat}) \mp 0.3 ({\rm syst}).
\label{olmosresult}
\end{eqnarray}
The sum  of electromagnetic  polarizabilities obtained from these data is
\begin{equation}
\alpha_p+\beta_p=13.1\pm 0.9 ({\rm stat}) \mp 1.0 ({\rm syst}).
\label{olmos-baldin-sum}
\end{equation}
Comparing the results of the TAPS experiment as given in Eqs. 
(\ref{olmosresult}) and (\ref{olmos-baldin-sum}) with the average of all
results obtained in previous experiments as given in Eqs. 
(\ref{baranovreevaluation}) and  (\ref{baranovreevaluation-2}) we
recognize that  the precision   achieved in the TAPS experiment is  
comparable with the precision of all the previous experiments
taken together. This finding is illustrated in the right panel of
Figure  \ref{fig:TapsExperiment}.

The right panel of Figure \ref{fig:TapsExperiment} shows a 
comparison of the result of the recent TAPS experiment \cite{olmos01} 
with  other experiments carried out in the 1990's. Also shown
is  the constraint given by 
Baldin's  sum rule. The results  are shown  
in a two-dimensional diagram $\alpha_p$ versus
$\beta_p$. The contours correspond to $\chi^2(\alpha_p,\beta_p)
= \chi^2_{\rm min}+1$. 
The result obtained for $\alpha_p-\beta_p$ when taking into account
the constraints of Baldin's sum rule (taken to be 
$\alpha_p+\beta_p= 13.8\pm 0.4$ \cite{olmos01}) is represented by 
the small grey ellipse corresponding to
\begin{equation}
\quad\quad
\alpha_p-\beta_p=10.5\pm 0.9({\rm stat+syst})\pm 0.7({\rm model}).
\end{equation}

From the numbers given in Eqs. (\ref{baranovreevaluation-2}) and
(\ref{olmos-baldin-sum}) we obtain as the global --  i.e. all existing
data for low-energy Compton scattering containing --  result 
\begin{equation}
\alpha_p+\beta_p= 13.6\pm 0.8({\rm stat+syst}) \pm 0.5({\rm model}).
\label{globalglobal}
\end{equation}

As a {\it recommended} final result of the experimental
electromagnetic
polarizabilities of the proton we propose to used the weighted average
of the data obtained from low-energy Compton scattering by the proton
given in Eqs. (\ref{baranovreevaluation}) and (\ref{olmosresult})    
and from the adopted result for $\alpha_p+\beta_p$ obtained from  
photo-absorption
data via the Baldin sum rule.  
The corresponding numbers are given 
Table \ref{recommended-alpha+beta}.
\begin{table}[h]
\caption{{\it Recommended} values for the experimental electromagnetic
polarizabilities of the proton. The values are constrained by Baldin's sum
rule using $\alpha_p+\beta_p=13.9\pm 0.3$ 
(see Table \ref{alpha+beta-absorption}).
The unit is $10^{-4}{\rm fm}^3$.}
\begin{center}
\begin{tabular}{||c|c||}
\hline\hline
$\alpha_p$ & $\beta_p$\\
\hline
$12.0\pm 0.6$ & $1.9\mp 0.6$\\
\hline\hline
\end{tabular}
\end{center}
\label{recommended-alpha+beta}
\end{table}

\subsection{Spin polarizability of the proton}

The spin polarizability of the proton for the backward direction 
has first been extracted from experimental data by Tonnison et al. (LEGS)
\cite{tonnison98} leading to  $\gamma^{(p)}_\pi=-[27.1\pm2.2({\rm stat+syst})
^{+2.8}_{-2.4}({\rm model})]\times 10^{-4}{\rm fm}^4$. This result
received great attention because it differed considerably
from the prediction based on dispersion theory with the asymptotic
contribution calculated from the $\pi^0$,  $\eta$ and $\eta'$ poles.
One possible interpretation of this finding was that 
a new contribution to the spin structure of the proton was
discovered. 

However, a Compton scattering experiment
carried out at MAMI using the 48 cm $\oslash\times$ 64 cm NaI(Tl) detector
\cite{wissmann99} confirmed the predicted value of $\gamma^{(p)}_\pi=-37.6$
\cite{wissmann99}. This latter result has recently been confirmed in two
further experiments carried out at MAMI using the large acceptance
arrangement LARA \cite{galler01,wolf01} and the TAPS detector
\cite{olmos01}. The
result obtained with LARA through fits to the experimental
differential cross sections \cite{galler01,wolf01} is
\begin{equation}
\gamma^{(p)}_\pi=-37.1\pm 0.6({\rm stat+ syst}) \pm 3.0 ({\rm model})
\label{gammapi-1}
\end{equation}
when using the SAID-SM99K parameterization \cite{arndt96,arndt95} as a basis in
the un-subtracted fixed-$t$ dispersion theory \cite{lvov97} and 
\begin{equation}
\gamma^{(p)}_\pi=-40.9 \pm 0.6({\rm stat+ syst}) \pm 2.2 ({\rm model})
\label{gammapi-2}
\end{equation}
when using the MAID2000 parameterization \cite{drechsel99a}. The
difference between the results given in (\ref{gammapi-1})
and (\ref{gammapi-2}) may be considered as a typical example of a
model dependence introduced by inconsistencies in the  parameterizations of 
photomeson amplitudes. In the present case these parameterizations are
used unmodified, i.e.  without adjustments to the Compton scattering 
data whereas the asymptotic part of the spin-polarizability is 
treated as an adjustable parameter. 
The result 
obtained from low energy Compton scattering using the TAPS detector 
\cite{olmos01} is
\begin{equation}
\gamma^{(p)}_\pi=-35.9 \pm 2.3({\rm stat+ syst})
\label{gammapi-3}
\end{equation}
where the model error has been  included into the systematic error.

\begin{figure}[h]
\includegraphics[width=0.45\linewidth]{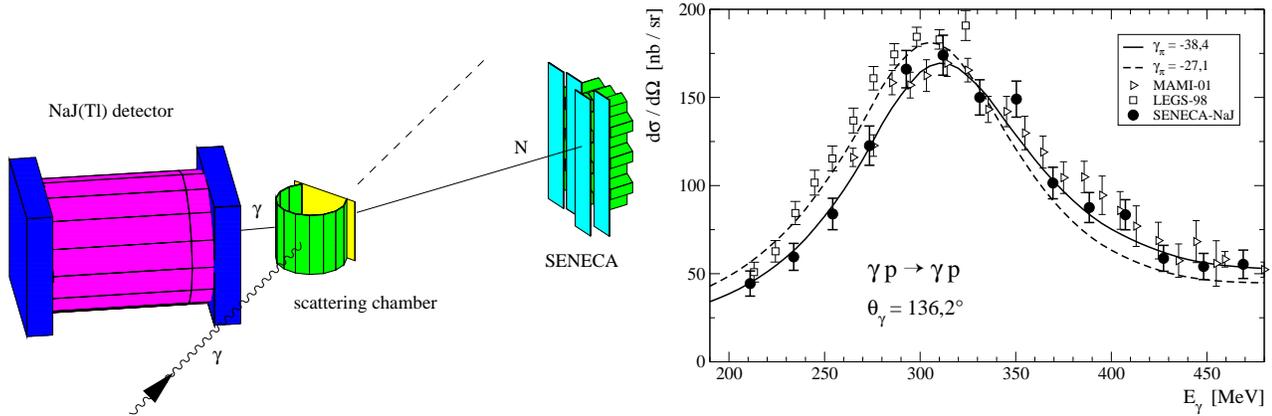}
\includegraphics[width=0.45\linewidth]{fig/pggp-27.eps}
\caption{Left panel: Experimental arrangement used for Compton scattering by
  the proton \cite{camen02}. Compton scattering events are identified 
through coincidences between the Mainz 48 cm $\oslash\times$ 64 cm
  NaI(Tl) photon detector positioned under 136$^\circ$ and the
  G\"ottingen segmented recoil counter SENECA positioned under
  18$^\circ$.
Right panel: Experimental differential cross sections for Compton scattering
  by the proton at $\theta^{\rm lab}_\gamma=136.2^\circ$. Closed
circles: SENECA-NaI (left figure) data \cite{camen02}. Open triangles
Mainz-LARA data \cite{galler01,wolf01}. Open squares: LEGS data 
\cite{tonnison98,blanpied01}}
\label{fig:ProtonSpinPol}
\end{figure}

It has been shown \cite{wissmann99,galler01,wolf01} that these
discrepancies between MAMI and LEGS can be traced back to a
discrepancy in the experimental differential cross sections 
for Compton scattering by the proton obtained in the $\Delta$
resonance region. This finding has recently been confirmed
\cite{camen02} at MAMI
using the apparatus shown in the left panel of
Figure \ref{fig:ProtonSpinPol}. 
This
apparatus consists of the Mainz 48 cm $\varnothing$ $\times$ 64 cm
NaI(Tl) photon detector positioned under 136$^\circ$ to the incoming
photon beam and the G\"ottingen segmented recoil spectrometer  SENECA
positioned under 18$^\circ$. This apparatus was set up for an
experiment to measure the electromagnetic polarizabilities  of the
neutron via quasi-free Compton scattering by neutrons bound in the
deuteron above $\pi$ photoproduction threshold. As a test of the
apparatus the target cell was filled with liquid hydrogen so that Compton
differential cross sections for the proton were measured. The 
experimental result is shown in the right  panel of Figure
\ref{fig:ProtonSpinPol} together with the LEGS data
 \cite{tonnison98,blanpied01}. 
The backward spin polarizability
extracted from the MAMI data of Figure \ref{fig:ProtonSpinPol} is
\begin{eqnarray}
&&\gamma^{(p)}_\pi=-36.5 \pm 1.6({\rm stat})\pm 0.6 ({\rm syst}) 
\pm 1.8 ({\rm model})\quad \mbox{(SAID-SM99K)},
\label{gammapi-4}\\
&&\gamma^{(p)}_\pi=-39.1 \pm 1.2({\rm stat})\pm 0.8 ({\rm syst}) 
\pm 1.5 ({\rm model})\quad \mbox{(MAID2000)}.
\label{gammapi-5}
\end{eqnarray} 
From the Mainz values of Eqs. (\ref{gammapi-1}) - (\ref{gammapi-5})
given above we obtain \cite{camen02} a the weighted average given in
Table \ref{gammapi-6}.
\begin{table}[h]
\caption{{\it Recommended} value for the experimental backward 
spin polarizability of the proton (MAMI weighted average). The unit is 
$10^{-4}{\rm fm}^4$.}
\begin{center}
\begin{tabular}{||c||}
\hline\hline
$\gamma^{(p)}_\pi=-38.7 \pm 1.8$ \\
\hline\hline
\end{tabular}
\end{center}
\label{gammapi-6}
\end{table}
We recommend to consider  the weighted average given in Table \ref{gammapi-6}
as the accepted   result for the experimental backward 
spin-polarizability 
of the proton.

\subsection{Electromagnetic polarizabilities  of the neutron}

For a long time the determination of the electromagnetic
polarizabilities of the neutron via Compton scattering was considered as
impossible.
Therefore, the application of electromagnetic scattering
of neutrons in the Coulomb field of heavy nuclei was given preference,
as discussed in the introduction. However, these early results
remained unsatisfactory.

\subsubsection{Electromagnetic scattering of neutrons in a Coulomb
  field in the 1990's}

The only  experiment on electromagnetic scattering 
of neutrons in the
Coulomb field of heavy nuclei leading to a value for $\alpha_n$ with a
meaningful precision was carried out in Oak Ridge
\cite{schmiedmayer91}
using a Pb  target
enriched in $^{208}$Pb in order to reduce the uncertainties arising from
nuclear neutron scattering. The result of this experiment was
\begin{equation}
\alpha_n=12.6\pm 1.5 \pm 2.0.
\label{schmiedmayerresult}
\end{equation}
The number given here has been corrected by adding the Schwinger term
\cite{lvov93} $e^2\kappa^2_n/4M^3=0.6$, containing the neutron
anomalous magnetic moment $\kappa_n$  and the neutron mass M. This term
compensates for a missing term (see \cite{lvov93}) in the original 
evaluation of this experiment. After
including the Schwinger term the number is directly comparable with
the one defined through Compton scattering. A later experiment
carried out in Munich \cite{koester95} lead to $\alpha_n=0.6\pm 5$.
This value for $\alpha_n$ appears to us unreasonable and, furthermore,
has a  large error. Therefore, we consider it advisable to disregard 
this result in the following considerations.
The Oak
Ridge experiment \cite{schmiedmayer91} is of very high precision.
However this precision has been questioned by a number of researchers
active in the field of neutron scattering \cite{enic97}. Their
conclusion is that the Oak Ridge result \cite{schmiedmayer91} possibly
might be quoted as $7\leq \alpha_n \leq 19$. (For a more detailed
discussion see \cite{wissmann98}.) Certainly, this criticism
has to be kept in mind. On the other hand we feel obliged to the rule
that authors have to be trusted in a first place and, therefore, 
cite and adopt the result of \cite{schmiedmayer91} as it is given in Eq.
(\ref{schmiedmayerresult}).

\subsubsection{Compton scattering by  neutrons below meson
  photoproduction threshold}

The electromagnetic polarizabilities $\alpha_p$ and $\beta_p$ of the
proton have  been  measured through Compton scattering at energies below
$\pi$-photoproduction threshold. The differential cross section for
Compton scattering is comparatively large in this case because of the sizable
interference term between the Born and the non-Born amplitude. In fact,
it is this interference term which is of essential help in experiments on
the proton.

A corresponding experiment for the neutron is extremely
difficult, because  there is  no Thomson amplitude to
interfere with the non-Born part of the scattering 
amplitude. Therefore, differential cross sections for Compton
scattering by the neutron at energies below $\pi$-photoproduction
threshold are  extremely small. Furthermore, there are no free-neutron
targets with sufficient density, and for neutrons bound in the deuteron
sizable corrections are expected to the quasi-free differential cross section
due to final-state interaction between the outgoing nucleons.
Therefore, any experiment of this kind has to be accompanied 
by a carefully carried out theory \cite{levchuk86}.

The first experiment on  low-energy $\gamma n $ scattering with subsequent
extraction of the  polarizabilities of the neutron has been carried
out by the
G\"ottingen-Mainz group \cite{rose90} which followed an earlier
theoretical suggestion \cite{levchuk86} to exploit  the reaction
$\gamma d \to \gamma np$ in the quasi-free kinematics. The experiment
was carried out using non-tagged bremsstrahlung produced by a
130 MeV 20 $\mu$A c.w. electron beam of MAMI A which was available
for a small period of time during the MAMI construction. Quasi-free Compton
scattering by the neutron was investigated via the 
$^2$H$(\gamma,\gamma'n)^1$H reaction by detecting Compton 
scattered photons in two
25 cm $\varnothing$ $\times$ 36  cm NaI(Tl) detectors positioned at lab
scattering angles of $\theta_{\gamma'}=90^\circ$ and $135^\circ$.  
The energy spectrum and angular distribution of recoiling neutrons
were measured in coincidence with the Compton scattered photons via
time of flight using a plastic-scintillator hodoscope.

This experiment was successful in the sense that the relevant effect,
i.e. coincidence events between Compton scattered photons and recoil
neutrons, was definitely identified. It was possible to extract the
value $\alpha_n=10.7$  for the electric polarizability from the
experimental data with an upper error of $+3.3$. The determination of
a lower error failed because of the rather large lower error of 18\%
of the differential cross section, so that its lower bound  
did not correspond to a possible electromagnetic polarizability. 
In order to avoid this difficulty,
the lower error of the differential cross section should have been 10\%
or smaller. 
The result of
this first and only experiment carried out below the meson photoproduction 
threshold may be quoted as
\begin{equation}
\alpha_n =10.7^{+3.3}_{-10.7}.
\label{roseresult}
\end{equation}

\subsubsection{Quasi-free Compton scattering by neutrons bound in the
  deuteron above meson photoproduction threshold}

\begin{figure}[t]
\vspace{1cm}
\unitlength=1mm
\begin{picture}(180,100)
\linethickness{1mm}
\put(60,78){\framebox(40,30){}}
{\par\centering \resizebox*{10cm}{!}{
\includegraphics{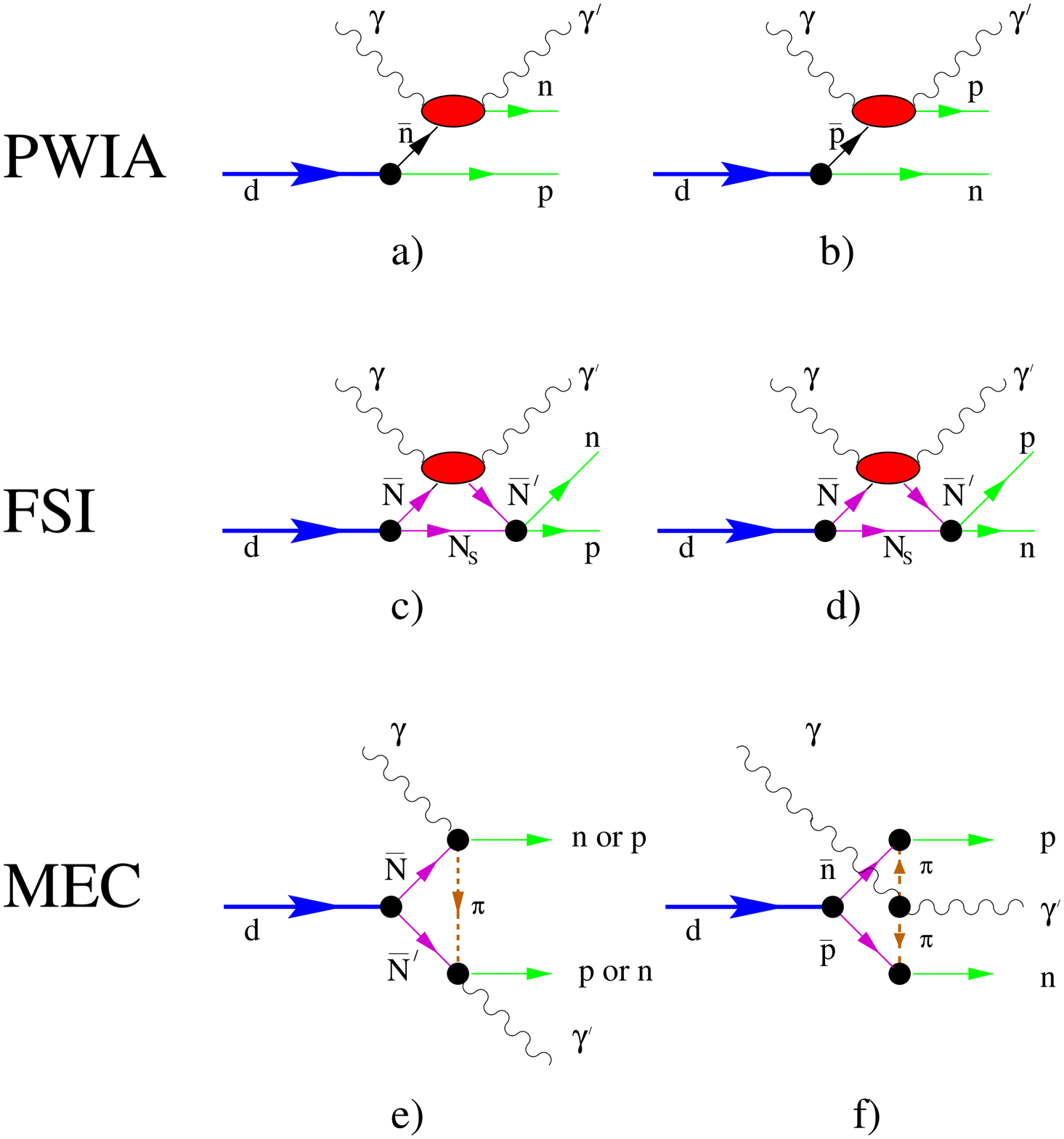}}
\put(0,82){\line(1,0){38.1}}
\put(37.9,82){\line(0,-1){20}}
\put(37.9,61.9){\line(1,1){5}}
\put(37.9,62.){\line(1,1){5}}
\put(37.9,62.1){\line(1,1){5}}
\put(37.9,62.2){\line(1,1){5}}
\put(37.9,62.3){\line(1,1){5}}
\put(37.9,62.4){\line(1,1){5}}
\put(37.9,62.5){\line(1,1){5}}
\put(37.9,62.6){\line(1,1){5}}
\put(37.9,62.7){\line(1,1){5}}
\put(37.9,62.8){\line(1,1){5}}
\put(37.9,62.9){\line(1,1){5}}
\put(37.9,63.){\line(1,1){5}}
\put(37.9,61.9){\line(-1,1){5}}
\put(37.9,62.){\line(-1,1){5}}
\put(37.9,62.1){\line(-1,1){5}}
\put(37.9,62.2){\line(-1,1){5}}
\put(37.9,62.3){\line(-1,1){5}}
\put(37.9,62.4){\line(-1,1){5}}
\put(37.9,62.5){\line(-1,1){5}}
\put(37.9,62.6){\line(-1,1){5}}
\put(37.9,62.7){\line(-1,1){5}}
\put(37.9,62.8){\line(-1,1){5}}
\put(37.9,62.9){\line(-1,1){5}}
\put(37.9,63.){\line(-1,1){5}}
\put(45,76){{\small \(E_\gamma = 250\) MeV}}
\put(45,70){{\small \(\Theta_\gamma = 135^\circ\)}}
\put(45,64){{\small \(\Theta_p = -18^\circ\)}}
\resizebox*{7cm}{!}{
\includegraphics{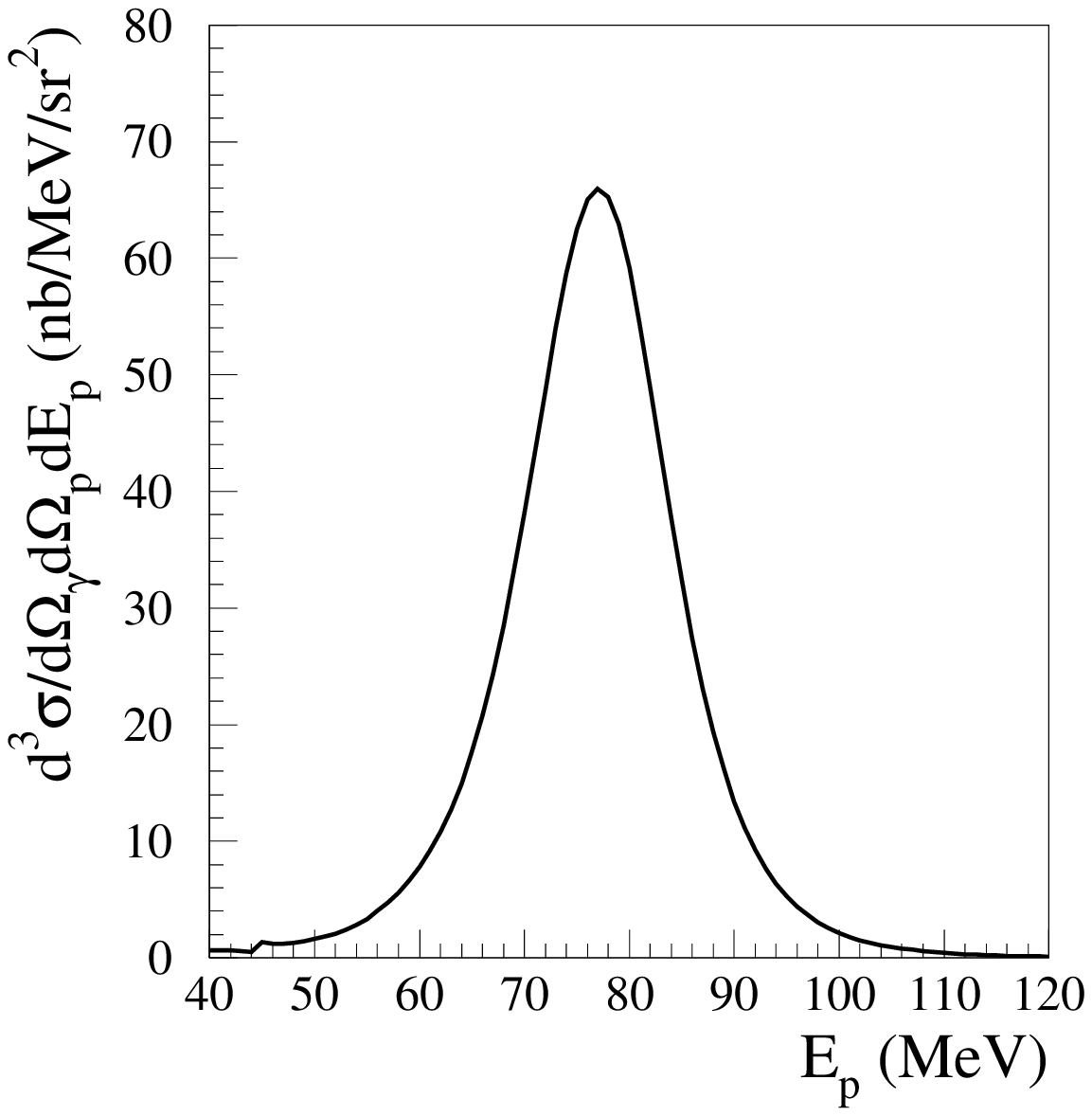}} \par}
\end{picture}
\caption{Left panel: Feynman diagrams for quasi-free Compton scattering
by the proton and neutron bound in the deuteron. a) Compton scattering
by the neutron in  quasi-free kinematics calculated in
plane-wave approximation. b) Same as a) but calculated for the
proton. c) and d) final-state interactions due to the nucleon-nucleon
interaction in the final state. e) and f) meson-exchange corrections. 
Right panel:
Triple differential $d^3\sigma/d\Omega_\gamma d\Omega_p dE_p$
cross section for Compton scattering by the proton in the quasi-free 
peak including final-state interactions and corrections due to
meson-exchange.}
\label{quasifreediagrams}
\end{figure}

Because of the difficulties involved in a Compton scattering
experiment at energies below meson photoproduction threshold it has been
proposed \cite{levchuk86} to measure quasi-free Compton scattering 
by the neutron at energies between pion threshold and the $\Delta$ 
peak  to determine $\alpha_n -\beta_n$. The underlying properties of the
method are depicted in Figure \ref{quasifreediagrams}. Compton
scattering by the nucleon has to be calculated 
under kinematical conditions fulfilled
for Compton scattering by a free nucleon. Under these kinematical
conditions the quasi-free peak is expected. The calculation is
performed in a  diagrammatic approach. 
As a first approximation, the nucleon wave functions in the final state are
treated as plane waves.
Thereafter,
the modification of the wave-functions due to the nucleon-nucleon
potential is taken into account. 
Since the nucleon-nucleon potential is isospin dependent, additional
diagrams taking into account meson exchange currents (MEC) (see
Figure \ref{quasifreediagrams} e) and f)) have to be taken into account.
The result of the calculation is the quasi-free peak either for the
recoiling proton or recoiling neutron having a width which is larger 
than the one expected from Fermi motion alone. This increase of width is large
at energies below meson photoproduction threshold and becomes small at 
higher energies. 
The further steps of the calculation are (i) to relate the area underneath
the quasi-free peak to the triple differential cross section in the center of
the quasi-free peak and (ii) to relate the triple differential cross section
in the center of the quasi-free peak to the differential cross section
for the free nucleon. With this theoretical input, the number of events
found experimentally in the range of recoil energies where the quasi-free peak
is expected can be used (i) to calculate the experimental triple differential
cross section in the center  of the quasi-free peak and (ii) to calculate
the corresponding experimental differential cross section for a ``free''
nucleon obtained by the quasi-free method.

Compton scattering experiments on the free proton and quasi-free
Compton scattering on the proton bound in the deuteron  can be used to
test the validity of the procedure. The first experiment of this kind
has been carried out by Wissmann et al. \cite{wissmann99} at MAMI
(Mainz) using the  Mainz 48 cm$\varnothing$ $\times$ 64 cm
NaI(Tl) spectrometer for the free-proton experiment and the TAPS 
spectrometer for the quasi-free proton experiment.
Differential cross sections sections for Compton scattering by the
free proton were measured in the energy range from 
$E_\gamma$ = 200 MeV to 400 MeV at a scattering angle of
$\theta^{\rm lab}_{\gamma'} = 131^\circ$. Triple differential cross 
sections in the  center of the proton quasi-free peak (PQFP) at 
$\theta^{\rm lab}_{\gamma'} = 149^\circ$ were measured in the energy
range from $E_\gamma=$ 200 MeV to 290 MeV. The results were
encouraging enough   to carry out a dedicated 
experiment on quasi-free Compton scattering by the neutron 
(see also Section 5.2).

\begin{figure}[htb]
\includegraphics[scale=0.35]{fig/fig10.eps}
\includegraphics[scale=0.45]{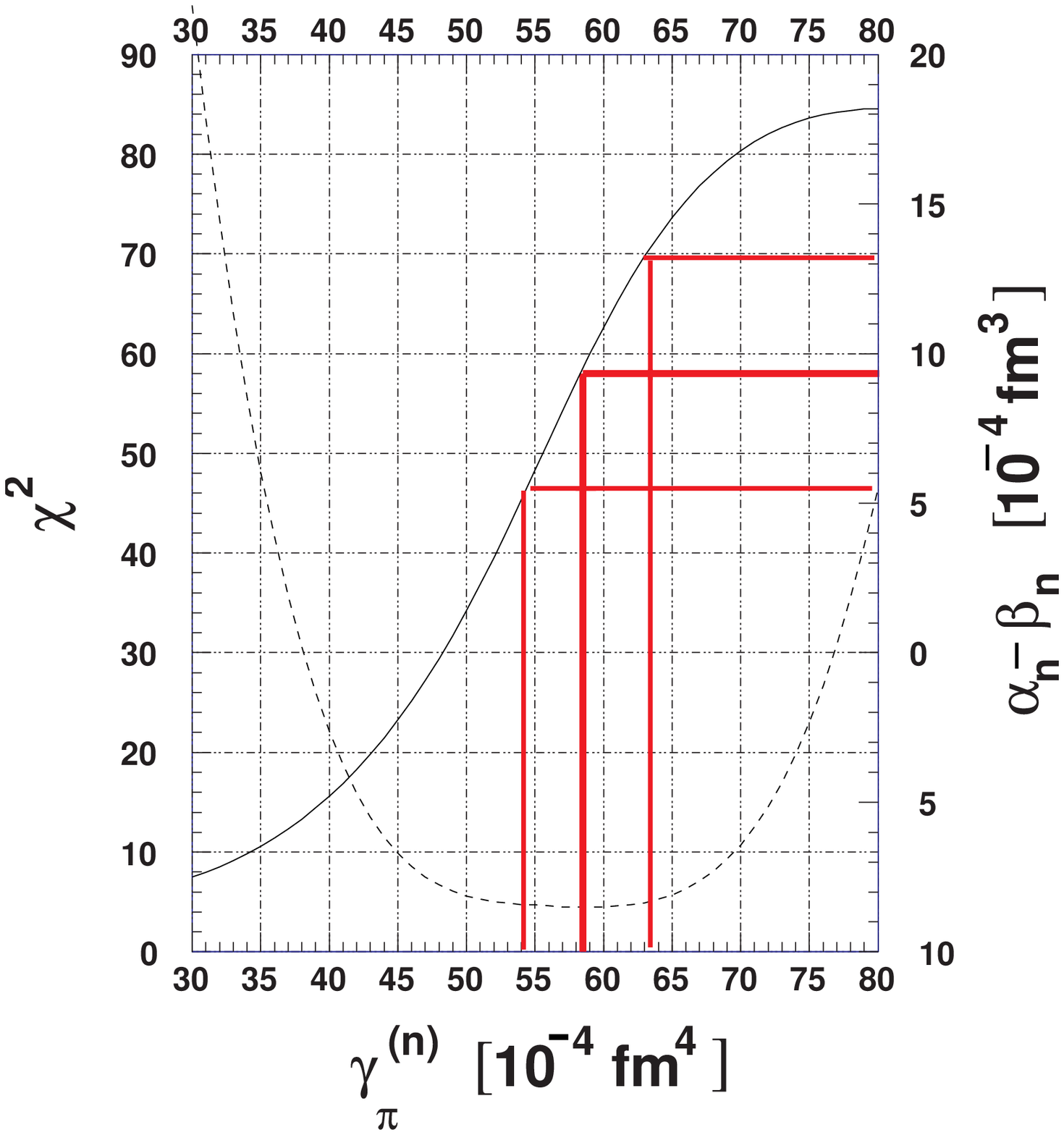}
\caption{Results of the NaI-SENECA experiment on quasi-free Compton scattering 
on neutrons bound in the deuteron. Left panel: Differential
cross-section for the ``free'' neutron extracted from the quasi-free
cross-sections for the bound neutron at $\theta^{\rm lab}_\gamma=
136.2^\circ$
(triangles). The SAL result \cite{kolb00} is shown by a diamond. Only
statistical errors are given. Right panel: The solid curve shows pairs
of values $\alpha_n-\beta_n$ and $\gamma^{(n)}_\pi$ 
used as independent parameters in fits to the experimental differential cross
sections shown in the left panel of this figure.
The corresponding $\chi^2$ is
depicted by the dashed line. The center horizontal and vertical bars
correspond to the best fit obtained for $\gamma^{(n)}_\pi=58.6$. The
outer horizontal and vertical bars correspond to the statistical errors
to be attributed to the best fits.}
\label{SENECAresults}
\end{figure}

This dedicated experiment \cite{kossert02}
was carried out using the experimental
arrangement shown in Figure \ref{fig:ProtonSpinPol}. Though the
foregoing experiments had shown that the basis of the data-evaluation
procedure is sound, this experiment was carried out in a self-testing
way containing the following steps:\\
(i) The target-container was filled with liquid hydrogen so that
Compton scattering by the free proton was measured in an energy
range from $E_\gamma$ = 200 to 400 MeV. The results of this 
experiment led to the following information. The asymptotic part
of the spin-polarizability $ \gamma^{(p)}_\pi$  is in agreement with the 
prediction obtained from the $\pi^0$, $\eta$ and $\eta'$
poles. Therefore, there is a good reason to use the corresponding
prediction to calculate the spin-polarizability for the neutron,
leading to $\gamma^{(n)}_\pi= 58.6$.
This value was used as a fixed input for the evaluation  of $\alpha_n-\beta_n$
from the quasi-free data. The other fixed input is 
$\alpha_n+\beta_n=15.2\pm 0.5$ \cite{levchuk00} obtained
via  Baldin's sum rule from photo-absorption data. 
From the large number of parameterizations of the photomeson 
amplitudes the MAID2000 parameterization led to the best overall
agreement with the experimental differential cross-sections  measured in
this experiment for the free proton. This parameterization, therefore,
served as the leading one in the data analysis.
Other parameterizations which led
to an almost equivalent quality of agreement were used to obtain
information on the model error.\\
(ii) The  target-container was filled with liquid deuterium so that ``free''
differential cross sections for Compton scattering by the proton and
the  neutron were obtained from quasi-free data. 
The ``free'' differential
cross sections obtained for the proton were not in a complete agreement
with the corresponding differential cross section for the free proton.
But the agreement was good enough to justify the application of the
theory of quasi-free Compton scattering in its present form for   
the data analysis and to use the remaining inconsistencies for an 
estimate of the  model error.   The measured ``free'' differential 
cross-sections extracted from  quasi-free data on 
Compton scattering by the neutron is displayed   in the left panel
of Figure \ref{SENECAresults}. There is  consistency  with the one 
existing SAL \cite{kolb00} data point. The fit to the data was
obtained by keeping $\alpha_n+\beta_n=15.2$ \cite{levchuk00}
and $\gamma^{(n)}_\pi=58.6$
and the MAID2000 parameterization fixed and treating $\alpha_n-\beta_n$
as an adjustable parameter. 
The result obtained for the polarizability difference of the neutron 
obtained in this way is
\cite{kossert02}
\begin{equation}
\quad\quad
\alpha_n - \beta_n = 9.8 \pm 3.6({\rm stat}){}^{+2.1}_{-1.1}({\rm syst})
\pm 2.2({\rm model}).\label{alphan}
\end{equation} 
Using  $\alpha_n+\beta_n= 15.2\pm 0.5$
obtained from the photo-absorption cross section \cite{levchuk00} 
as a constraint we obtain
\begin{eqnarray}
 &&\alpha_n=12.5 \pm 1.8({\rm stat})^{+1.1}_{-0.6}({\rm syst})\pm 1.1({\rm
      model}),
\nonumber\\
&& \beta_n= 2.7 \mp  1.8({\rm stat})^{+0.6}_{-1.1}({\rm syst})\mp 1.1({\rm
      model}).
\label{alpha-beta-neutron}
\end{eqnarray}

\subsubsection{Coherent-elastic scattering of photons by deuterons}

In-medium nucleon polarizabilities are an interesting 
field of research. Whereas in complex nuclei the interest
is directed to the question  whether or not the in-medium
electromagnetic polarizabilities are the same as the free
polarizabilities \cite{huett00}, for the deuteron no medium-modifications are
expected because of the weak binding of the  nucleons. Since the 
polarizabilities of the proton are known with comparatively good
precision it may be expected that the neutron polarizabilities
can be extracted from the isospin-averaged nucleon polarizabilities.
The latter can be investigated in elastic Compton scattering experiments
by deuterons.
An advantage of this method is that the effect of the isospin-averaged 
polarizabilities is strongly enhanced in comparison with Compton scattering 
by free neutrons due to interference of the polarizability-dependent term
with the Born term. This situation resembles that of Compton scattering by the
free proton. The price we have to pay for the advantage is the necessity
of taking into account the $np$-interaction in the intermediate state,
meson exchange currents and two-body seagull amplitudes. All these effects
introduce noticeable model dependencies in the extracted values of the
polarizabilities. For details we refer the reader to \cite{levchuk00,lundin03}.

Differential cross sections for Compton scattering from the deuteron
were measured at MAX-Lab (Lund) \cite{lundin03} 
for incident photon energies of 55 and
66 MeV at nominal laboratory angles of 45$^\circ$, 125$^\circ$, and
135$^\circ$. Tagged photons were scattered from liquid deuterium into
three NaI(Tl)  spectrometers. By comparing the data with theoretical
calculations in the framework of a one-boson-exchange model 
\cite{levchuk00}, the sum
and the difference of the isospin-averaged nucleon polarizabilities,
$\alpha_N+\beta_N= 17.4\pm 3.7$ and 
$\alpha_N-\beta_N= 6.4\pm 2.4$ have been determined. By combining the
latter with the global-average value for $\alpha_p-\beta_p$ 
and using the predictions of the Baldin sum rule for the sum of the
nucleon polarizabilities, the following  values for the neutron
electric and magnetic polarizabilities 
\begin{eqnarray}
&&\alpha_n=8.8 \pm 2.4 ({\rm total}) \pm 3.0 ({\rm model}), \nonumber\\
&&\beta_n = 6.5 \mp 2.4 ({\rm total}) \mp 3.0 ({\rm model})
\label{LundPolarizabilities}
\end{eqnarray}
have been obtained \cite{lundin03}.

\subsubsection{Summary on the experimental results for the electromagnetic  
polarizabilities of the neutron}

\begin{table}[h]
\caption{Electromagnetic polarizabilities of the  neutron determined
  by three different methods and their weighted average. The unit is
$10^{-4}{\rm fm}^3$} 
\label{tableMAMI}
\begin{center}
\begin{tabular}{|l|l|}
\hline
experimental method&polarizability\\
\hline
electromagnetic scattering \cite{schmiedmayer91}& $\alpha_n=12.6\pm 2.5$\\
quasi-free Compton scattering \cite{kossert02}& $\alpha_n=12.5\pm 2.3$\\
coherent Compton scattering \cite{lundin03}&$\alpha_n=8.8\pm 3.8$\\
\hline
weighted average&$\alpha_n= 11.9 \pm 1.5$\\
constrained by $\alpha_n+\beta_n=15.2\pm 0.5$&$\beta_n=3.3\mp 1.5$\\
\hline
\end{tabular}
\end{center}
\label{Table:table14}
\end{table}
\begin{table}[h]
\caption{{\it Recommended} results for the electromagnetic polarizabilities 
of the neutron obtained from the Baldin sum rule constraint, from
quasi-free Compton scattering and from electromagnetic neutron
scattering.
The unit is $10^{-4}{\rm fm}^3$.} 
\centering\begin{tabular}{||c|c||}
\hline
\hline
$\alpha_n$&$\beta_n$\\
\hline
$12.5\pm 1.7$&$2.7 \mp 1.8$\\
\hline
\hline
\end{tabular}
\label{NeutronRecommended}
\end{table}
Table \ref{tableMAMI} shows a summary of  the experimental results 
obtained for the electromagnetic polarizabilities of the neutron.
It is satisfactory that the three different methods led to results
for $\alpha_n$
which are in agreement with each other within the given errors. It cannot 
be excluded that a recalculation of the nuclear scattering amplitude of
the deuteron which still is under discussion\footnote{\label{hildebrandt04}
Very recently the data have been re-evaluated in the framework
of chiral effective field theory to next-to-leading order. The
results obtained are $\alpha_n=14.2\pm 2.0 ({\rm stat})\pm 1.9 ({\rm
  syst})$, $\beta_n= 1.8 \pm 2.2 ({\rm stat}) \pm 0.3 ({\rm syst})$
\cite{hildebrandt04}. In \cite{beane04} $1/2(\alpha_p+\alpha_n)
=(13.0\pm1.9({\rm stat}))^{+3.9}_{-1.5}({\rm theory})$ was obtained.
}
may shift the coherent 
Compton scattering result \cite{lundin03} to some extent. Therefore,
without further clarification we recommend to use the Baldin sum-rule
constraint $\alpha_n+\beta_n=15.2\pm 0.5$ and the values for the electric
polarizability from electromagnetic scattering and quasi-free
Compton scattering of Table \ref{tableMAMI} to arrive at 
{\it recommended} values for the electromagnetic polarizabilities of the
neutron as given in Table \ref{NeutronRecommended}.

\subsection{Spin polarizability of the neutron}

 In the frame of the invariant amplitudes and in frame of fixed-$t$
dispersion theory, the backward spin polarizability is given by
\begin{equation}
\gamma_\pi=-\frac{1}{2\pi M}[A^{\rm int}_2(0,0)+A^{\rm as}_2(0,0)
+A^{\rm int}_5(0)]
\label{equ:gammapiintegral}
\end{equation}
with the integral parts  being the smaller contribution. In case of 
fixed-$t$ dispersion theory it is not granted that $A^{\rm as}_2(t)$
is exhausted by the $t$-channel poles due to $\pi^0$, $\eta$ and
$\eta'$ exchanges, though it has been shown for the proton that this is
the case. Nevertheless, it may be of interest to find out whether or
not the adopted quantity $\gamma^{(n)}_\pi=58.6$ is the optimum for  the
data shown in the left panel of Figure \ref{SENECAresults}.
This investigation is carried out in the right panel of Figure
 \ref{SENECAresults}.  The main difference as compared to the analysis
 procedure described so far is that $\alpha_n - \beta_n$ and
$\gamma^{(n)}_\pi$ are treated as free parameters when fitting the
 predicted  Compton differential cross-section to the experimental
 ``free'' differential cross sections. In this way for each pair
of values  $\alpha_n - \beta_n$ and
$\gamma^{(n)}_\pi$ a $\chi^2$ was obtained which is shown by a dashed
line in Figure \ref{SENECAresults}. It is apparent that the adopted
values $\alpha_n-\beta_n=9.8$ and $\gamma^{(n)}_\pi=58.6$ both are in
line with the center of  the rather broad $\chi^2$ distribution. This 
means that this pair of data is supported by the experiment on
quasi-free Compton scattering  as evaluated here. According to  
 Figure \ref{SENECAresults} it is also possible to attribute an
experimental error to $\gamma^{(n)}_\pi$, leading to the number given in
Table \ref{SpinPolNeutron}.
\begin{table}[h]
\caption{{\it Recommended} experimental value for the spin
  polarizability of the neutron. The unit is $10^{-4}{\rm fm}^4$.}
\begin{center}
\begin{tabular}{||c||}
\hline\hline
$\gamma^{(n)}_\pi=58.6 \pm 4.0$\\
\hline\hline
\end{tabular}
\end{center}
\label{SpinPolNeutron}
\end{table}

\subsection{Compton scattering by the proton using the large
  acceptance arrangement LARA}
For a long time dispersion theories for Compton scattering by the
proton existed only for the $\Delta$ range. This restriction was
overcome  in the  1990's \cite{lvov97} when  a
non-subtracted dispersion theory at fixed-$t$ was developed with an 
applicability   extending  through  the second resonance region. This
apparent success led to an experimental program  at MAMI (Mainz)
aimed to carry out a rigorous 
test of the predictions. In order to save beam time is was necessary
to develop an apparatus with which large energy and angular ranges
could be covered with one experimental set up. One essential
difficulty of Compton scattering above $\pi$-threshold is that a very small
rate of Compton scattered photons has to be identified in the presence
of a large rate of photons from $\pi^0$ decay. Though the largest part
\begin{figure}[htb]
\begin{center}
\includegraphics*[scale=.45]{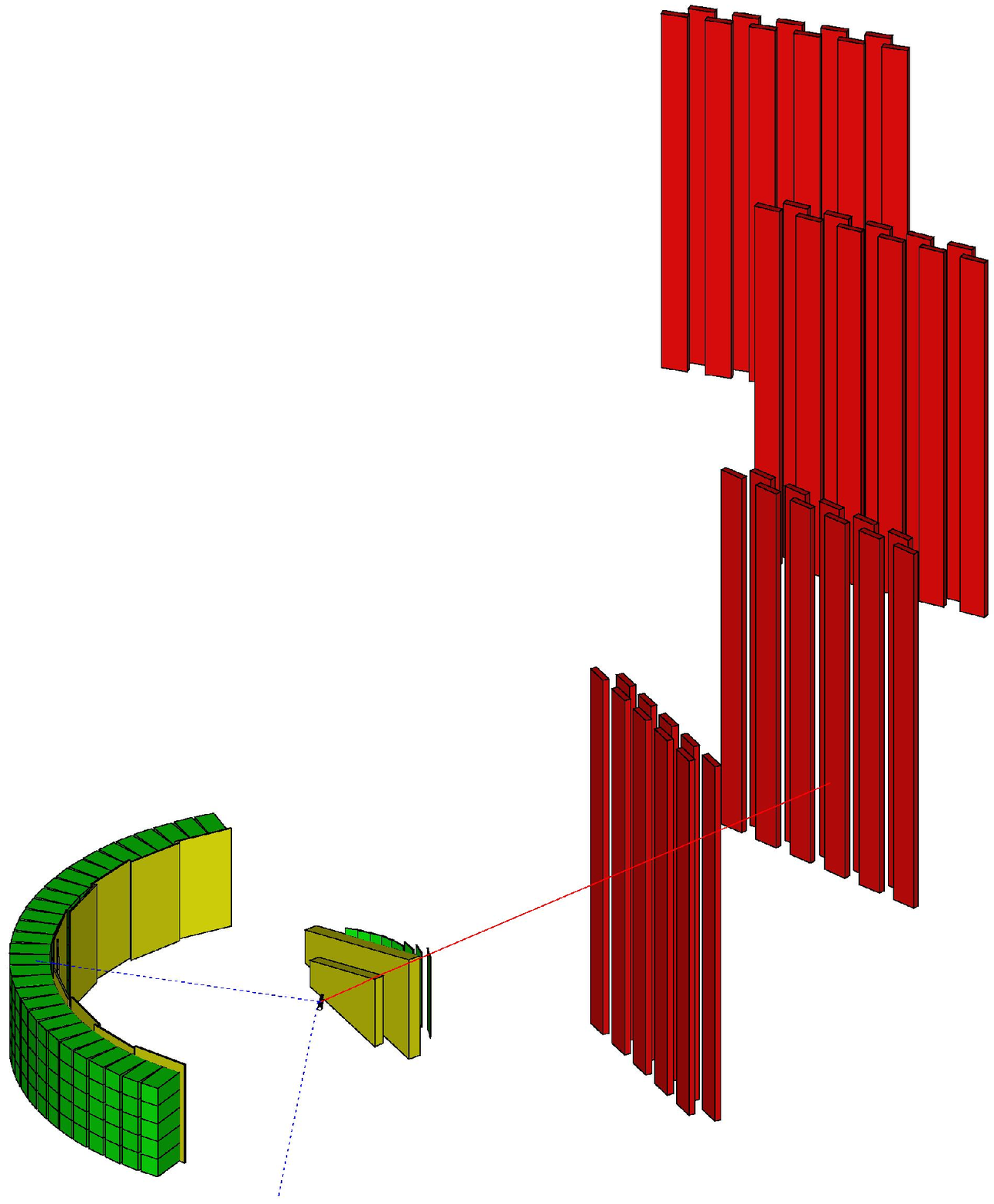}
\includegraphics[scale=.45]{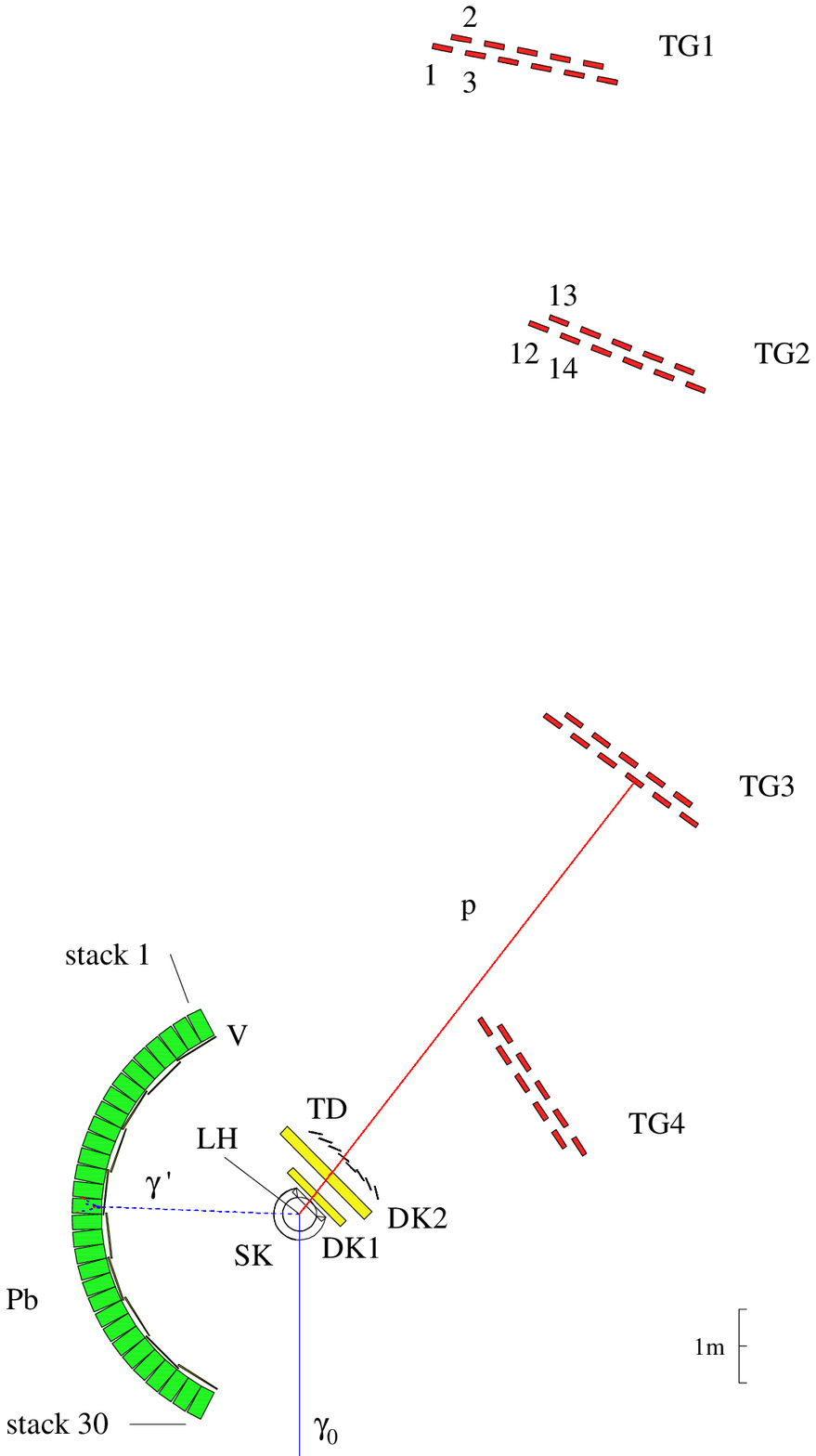}
\end{center}
\caption{
Perspective view (left panel) and vertical projection 
(right panel) of the LARA arrangement. The photon arm consists of 10
  blocks of 3 (horizontal) $\times$ 5 (vertical) lead glass detectors
(Pb),  each block equipped with a 1 cm plastic scintillator (V). The
  proton arm consists of two wire chambers (DK1,DK2) at distances of 25 and
  50 cm from the target center, 8 plastic scintillators serving as
  trigger detectors (TD) and 43 bars of 20 cm $\times$ 300 cm $\times$
5 cm plastic scintillators serving as time-of flight (TOF) stop
detectors. The scattering target consists of liquid H$_2$ contained in
a 3 cm $\varnothing$ $\times$ 20 cm Kapton cylinder. 
}
\label{fig:LARAapparatus}
\end{figure}
of these latter photons have a kinematics which is considerably
different from the Compton kinematics, there is a sizable rate of
photons from $\pi^0$ photoproduction which has an almost identical
kinematics. These are those decays where one of the decay photons is
emitted almost parallel to the direction of the $\pi^0$ velocity. 
In this case  a high-energy photon is emitted with an energy almost
equal to the energy of the  Compton scattered photon. The only way to 
overcome this difficulty is to determined the kinematical
variables of all particles with the highest possible precision. 
Before constructing a large acceptance arrangement  several experiments
were carried out where some of the experimental principles to be
applied later were  tested on a smaller scale
\cite{peise96,molinari96}.

For the construction of the large acceptance arrangement LARA shown in
Figure  \ref{fig:LARAapparatus}
\cite{galler01,wolf01} it had to be
taken into account that one principal  limitation 
is given by the angular straggling of the recoil protons
which limits the angular resolution to about $\pm$0.5$^\circ$ to 
$\pm$1$^\circ$. Due to kinematics the angular spread  of the 
Compton scattered photons
is about twice that of the corresponding recoil protons. Therefore, a wall
of lead glass photon-detectors was built  with an angular
resolution of about  $\pm$2$^\circ$ both in the horizontal
and in the vertical direction. The trajectories of the recoil
\begin{figure}[htb]
 \includegraphics[scale=.45]{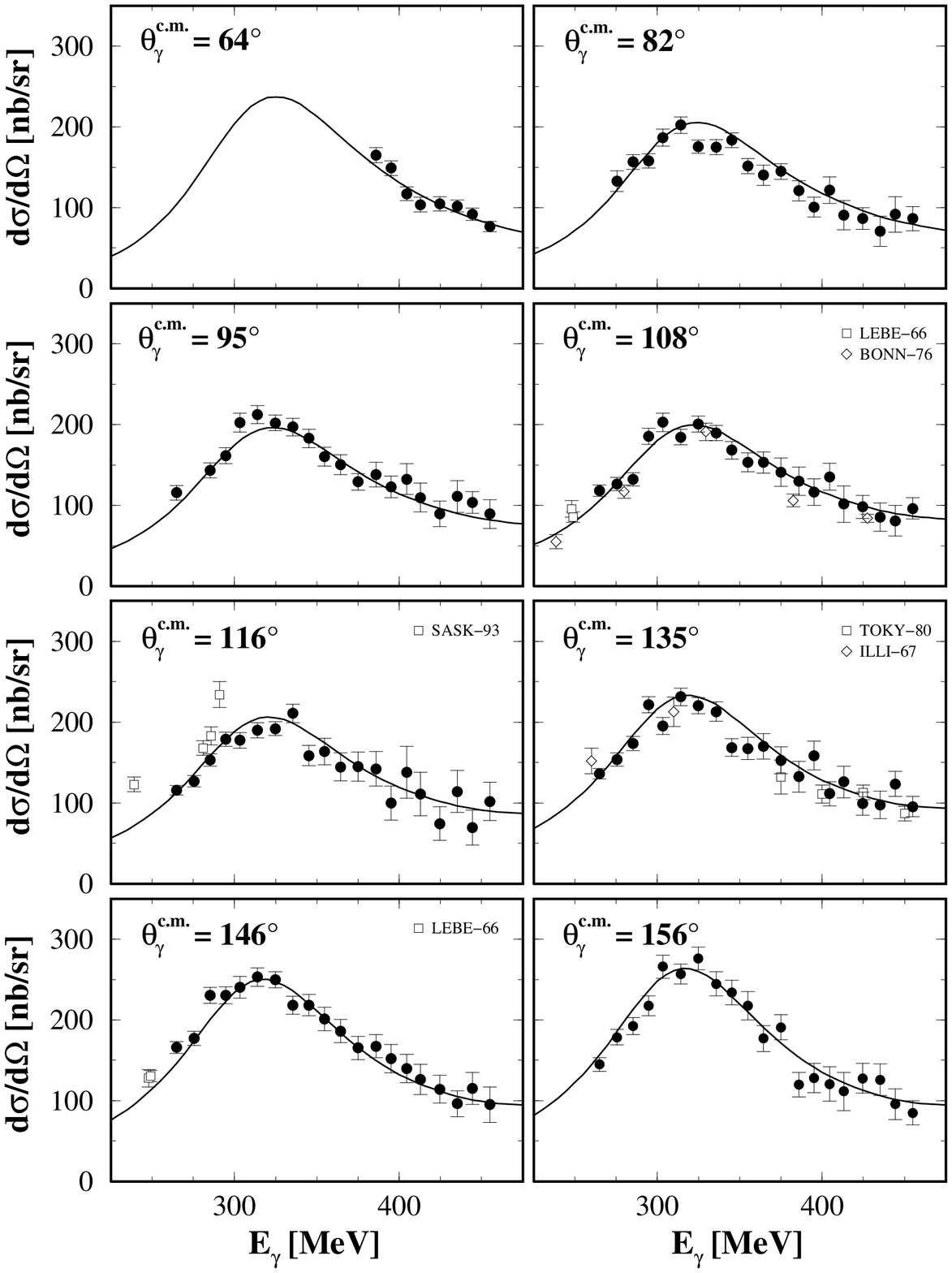}
 \includegraphics[scale=0.4]{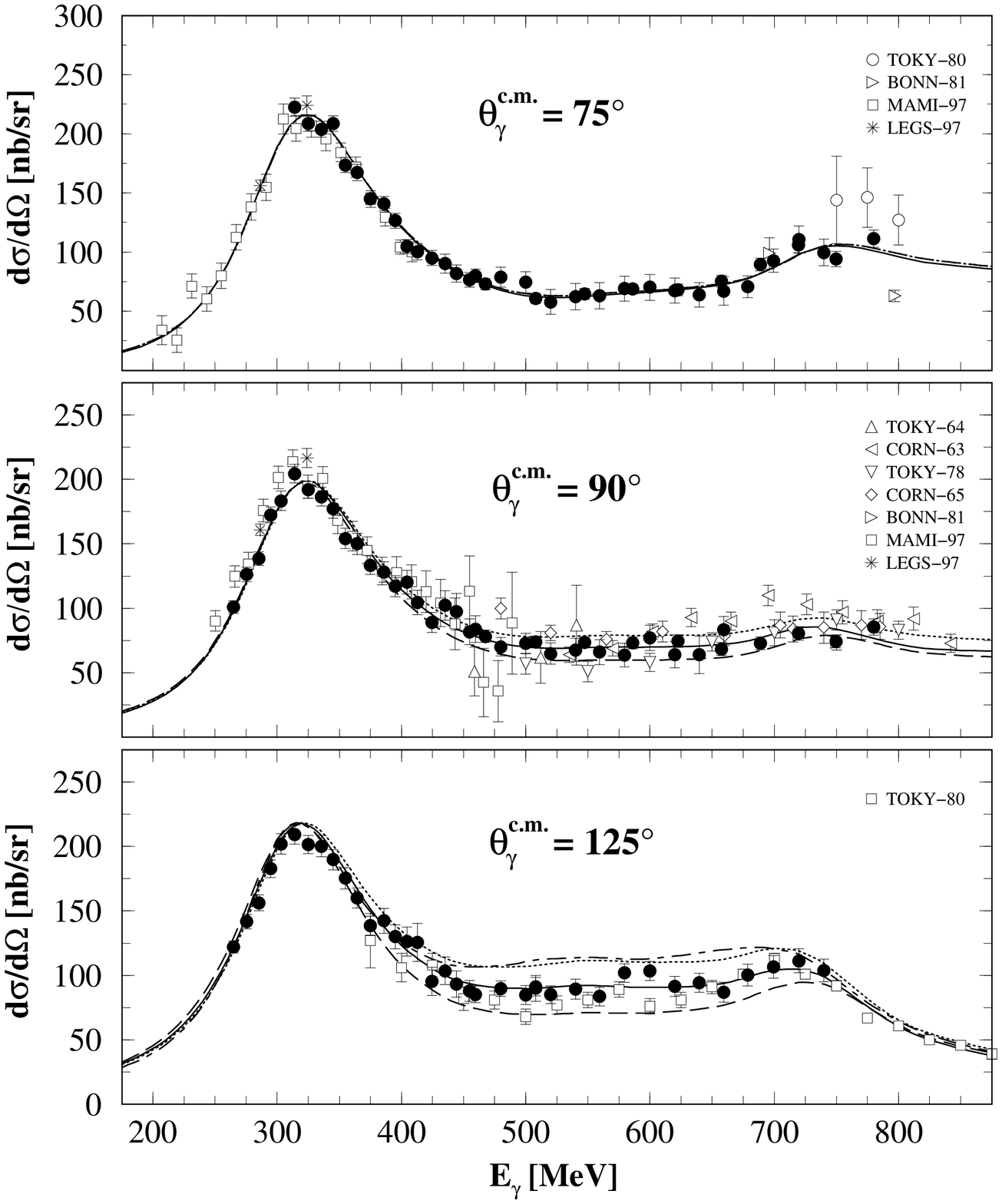}
\caption{
Left panel: Eight out of 24 measured energy distributions from
$\theta^{\rm c.m.}_\gamma$ = 59$^\circ$ to 156$^\circ$ obtained with
the LARA arrangement ($\bullet$) in the first resonance range compared 
with predictions from dispersion theory. The previous data are taken
from \cite{baranov66}(LEBE-66); \cite{gray67} (ILLI-67); 
\cite{genzel76} (BONN-76); \cite{hallin93}
(SASK-93); \cite{ishii80} (TOKYO-80).
Right panel:
Differential cross sections for Compton scattering by the proton
versus photon energy \cite{galler01,wolf01}. The three 
panels contain data corresponding to the
c.m.-angles of 75$^\circ$, 90$^\circ$ and  125$^\circ$. The three curves
are calculated for different mass parameters 
$m_{\rm eff}=800$ MeV (upper), $600$
MeV (center) and $400$ MeV (lower). A good sensitivity to a variation
of $m_{\rm eff}$ is observed at
$\theta^{\rm c.m.}_\gamma$ = 125$^\circ$
with an optimum fit for $m_{\rm eff}=589\pm 12$ MeV.}
\label{fig:LARAdata}
\end{figure}
protons were determined by a pair of wire chambers with an angular
resolution of better than 1$^\circ$. The reaction point in the
scattering target was determined in this way with a precision of 
about $\pm 0.7$
cm. The proton energies were determined via 
time-of-flight with flight paths ranging from 2.6 to 12.0 m,
depending on the energies of the recoil protons. The energies of the 
incoming photons were fixed by tagging to about 2 MeV whereas the
energy resolution of the lead glass detectors was poor.
With this arrangement it was possible to measure differential cross
sections for Compton scattering by the proton in the energy range
$250\lesssim E_\gamma\lesssim 800$ MeV and angular range 
$30^\circ\lesssim \theta^{\rm lab}_\gamma\lesssim 150^\circ$,
 with the
exception of small scattering angles and small photon energies where the
energies of the recoil protons were  too small for an escape from the 
liquid hydrogen target.

Figure \ref{fig:LARAdata} shows a selection of Compton differential 
cross sections obtained by the 
LARA experiment. The experimental data are compared with the predictions of
fixed-$t$ dispersion theory \cite{lvov97} where the
SAID-SM99K parameterization of photomeson amplitudes is used as a
basis and the difference $\alpha_p-\beta_p$ measured at energies
below meson photoproduction threshold is used as an  external input.
It is remarkable to note that in the $\Delta$
resonance range and for $\theta^{\rm c.m.}_\gamma = 75^\circ$ also at
higher energies, an excellent fit to the experimental data is obtained
without any major effect of the scalar $t$-channel -- or, in an other
language  -- the
effective $\sigma$ pole.
This proves that the basic concept of the fixed-$t$ dispersion theory
in the present form is correct.
At $\theta^{\rm c.m.}_\gamma = 90^\circ$ and $125^\circ$ the
differential cross sections show a sensitivity to the effective
mass of the $\sigma$ pole as discussed in Section 3.8. This sensitivity
increases with increasing scattering angle as expected. From the fit
to the experimental differential cross sections it is possible to determine 
the effective mass leading to $m_{\rm eff} = 589\pm 12$ MeV. This value
obtained for the effective mass is in agreement with the prediction we
obtained from our general knowledge about the $\sigma$ meson  
(see Section 3.8). In spite of this apparent success of fixed-$t$ dispersion
theory an independent interpretation in terms of fixed-$\theta$ dispersion
theory appears to us highly desirable.

\section{Sum rules}

\subsection{The Baldin (BL) sum rule}

Using (\ref{T7})  Baldin's sum rule (\ref{cauchi-5})
may be formulated in the form
\begin{equation}
\alpha+\beta=-\frac{1}{2\pi}[A^{\rm int}_{3+6}(0,0)+A^{\rm
    as}_{3+6}(0,0)],
\label{SR-2}
\end{equation}
with
\begin{eqnarray}
A^{\rm int}_{3+6}(0,0)&=&\frac{2}{\pi}
\int^{\nu_{\rm max}}_{\nu_{\rm thr}}{\rm Im}A_{3+6}(\nu',0)
\frac{d\nu'}{\nu'} \label{SR-3}\\
A^{\rm as}_{3+6}(0,0)&=&\frac{2}{\pi}
\int^{\infty}_{\nu_{\rm max}}{\rm Im}A_{3+6}(\nu',0)
\frac{d\nu'}{\nu'},
\label{SR-4}
\end{eqnarray}
where $\nu_{\rm max}$ separates the sum of electromagnetic
polarizabilities into an integral part and into an asymptotic part
\begin{equation}
\alpha+\beta=(\alpha+\beta)^{\rm int}+ (\alpha+\beta)^{\rm as}.
\label{separation}
\end{equation}
Numbers may be found in the recent work of Wissmann \cite{wissmann04}
where the following results have been obtained
\begin{eqnarray}
&&(\alpha_p+\beta_p)^{\rm int}= 12.6, \quad (\alpha_p+\beta_p)^{\rm
    as}= 1.2,\quad \mbox{for}\,\,\, \nu_{\rm max}=1.5\,\, {\rm GeV},
\label{alpha+beta-1.5}\\
&&(\alpha_p+\beta_p)^{\rm int}= 13.1, \quad (\alpha_p+\beta_p)^{\rm
    as}= 0.7,\quad \mbox{for}\,\,\, \nu_{\rm max}=2.0\,\, {\rm GeV},
\label{alpha+beta-2.0}
\end{eqnarray}
using a choice of $\nu_{\rm max}$ as  proposed in \cite{lvov97}.
Apparently, the asymptotic part of $(\alpha+\beta)_p$ is only of the 
order of $5-10\%$ of the integral part. This finding is partly responsible
for the validity of  Baldin's sum rule since deviations are expected only
from the asymptotic part.
An other  essential point in connection with the validity of  Baldin's sum
rule is  that $A^{\rm as}_{3+6}(0,0)$ is given by
an   integral along the real $\nu$-axis and that there is no
additional real part stemming from a non-vanishing contour integral
or -- in the language of fixed-$\theta$ dispersion theory -- from a 
$t$-channel contribution. This follows from  the discussion given in Section
2.7.1. According to this discussion  it may be expected that the expression 
given in Eq. (\ref{SR-4})
\begin{table}[h]
\caption{The Baldin sum rule evaluated from total photo-absorption
  cross sections. The adopted 
value is the weighted average of the
  data of Olmos(01) and Lvov(00).  The unit is $10^{-4}{\rm fm}^3$.}
\begin{center}
\begin{tabular}{|l|c|c|}
\hline
Absorption-Exp.& $\alpha_p+\beta_p$ & $\alpha_n+\beta_n$\\
\hline
Damashek(70) \cite{damashek70}& 14.2 $\pm$ 0.3& \\
Lvov(79) \cite{lvov79}& 14.2 $\pm$ 0.5& 15.8 $\pm$ 0.5\\
Babusci(98) \cite{babusci98}& 13.69 $\pm$ 0.14&  14.40 $\pm $0.66\\
\hline
Olmos(01)\cite{olmos01}&13.8 $\pm $0.4&\\
Lvov(00) \cite{levchuk00}& 14.0 $\pm$ 0.3& 15.2 $\pm $ 0.5\\
\hline\hline
adopted value& $13.9\pm 0.3$&$15.2\pm 0.5$\\
\hline\hline
\end{tabular}
\end{center}
\label{alpha+beta-absorption}
\end{table}
is complete,  since -- in the forward direction -- the candidate for a 
$t$-channel contribution, {\it viz.} 
the $f_2(1270)$ meson, is absorbed into the total photo-absorption
cross section via vector meson dominance (VMD)
together with the  $a_2(1320)$ meson and the Pomeron.
Predictions for $\alpha+\beta$ made on the basis of Eq. (\ref{SR-2})
are shown in Table \ref{alpha+beta-absorption}. Of these, the predictions
made by Olmos et al. \cite{olmos01} and Levchuk et
al. \cite{levchuk00} are based on recent sets of photo-absorption data.
Furthermore they take into account the $\lesssim 1\%$ radiative decay
channel of the total photoabsorption cross section. Therefore, we
consider the averages of these two results  as the adopted values,
listed in the last line of Table  \ref{alpha+beta-absorption}. 

\subsubsection{Experimental tests of the Baldin (BL) sum rule}

Experimentally, the validity
of the Baldin sum rule can be tested by measuring the electromagnetic
polarizabilities $\alpha$ and $\beta$ below the meson photoproduction
threshold. The relevant experiments and the results obtained 
have been discussed in  Section  5.1 and are summarized in Table 
\ref{alpha+beta-scattering}. 
\begin{table}[h]
\caption{Experimental results for $\alpha_p+\beta_p$ obtained by measuring
Compton scattering below meson photoproduction threshold without using the
constraint from Baldin's sum rule in the data evaluation. The adopted 
value 
is equal to the  weighted average. The unit is $10^{-4}{\rm fm}^3$.
}
\begin{center}
\begin{tabular}{|l|c|}
\hline
Scattering-Exp.&$\alpha_p+\beta_p$\\
\hline
1950's--1990's \cite{baranov00}&$14.0 \pm 1.4$\\
Olmos(01) \cite{olmos01}&13.1 $\pm$ 1.3\\
\hline\hline
adopted value& 13.6 $\pm$ 1.0\\
\hline\hline
\end{tabular}
\end{center}
\label{alpha+beta-scattering}
\end{table}
The numbers given as the adopted results in Tables 
\ref{alpha+beta-absorption}  and  \ref{alpha+beta-scattering},
{\it viz.} $(\alpha_p+\beta_p)^{\rm absorption}=13.9\pm 0.3$
and $(\alpha_p+\beta_p)^{\rm scattering}=13.6 \pm 1.0$ deviate from
each other by not more than $30\%$ of the standard deviation.  We may
consider this as a firm verification of the Baldin sum rule.

An interesting  aspect of the Baldin sum rule has been 
discussed by Drechsel et al. \cite{drechsel03} in the frame of
fixed-$\theta$ dispersion theory. The structure of the two dispersion
integrals shows that the integral part of the fixed-$t$  dispersion
theory  at $t=0$  and the $s$-channel part of fixed-$\theta$  
dispersion theory at $\theta_s=\theta_{\rm lab}=0$ coincide. The
number obtained for the $s$-channel part of $\alpha_p+\beta_p$ is
\begin{equation}
(\alpha_p+\beta_p)^s= 11.94,\quad \theta_{\rm lab}=0^\circ \quad  
\mbox{for}\,\,\, \nu_{\rm max}=1.5\,\, {\rm GeV}
\label{alpha+beta-theta-1.5}
\end{equation}
which indeed is not too much different form the fixed-$t$ counterpart
given in Eq. (\ref{alpha+beta-1.5}), but again the Baldin sum rule is not
completely saturated at an energy of $\nu_{max}=1.5$ GeV. 
In fixed-$\theta$ dispersion
theory and at non-forward angles the $f_2(1270)$ meson 
is expected to contribute to $\alpha+\beta$ via the $t$-channel
integral due to its direct coupling to two photons.
Calculations of the $t$-channel integral have been found feasible  
\cite{drechsel03} in the angular
range between $\theta_{\rm lab}=100^\circ$ and $180^\circ$. The
results obtained are summarized in Table \ref{drechsel-tab}.
\begin{table}[h]
\caption{Fixed-$\theta$ dispersion theory prediction for
  $\alpha_p+\beta_p$. $(\alpha+\beta)^s_p$: $s$-channel contribution,
 $(\alpha+\beta)^t_p$: $t$-channel contribution,
$(\alpha+\beta)^{(s+t)}_p$: total prediction,  
$(\alpha+\beta)^{{\rm exp}-(s+t)}_p$: deviation of prediction from the
  experimental value. The unit is $10^{-4}{\rm fm}^3$.}
\begin{center}
\begin{tabular}{|c|c|c|c|c|} 
\hline
$\theta_{\rm lab}$& $(\alpha+\beta)^s_p$ & $(\alpha+\beta)^t_p$&
$(\alpha+\beta)^{(s+t)}_p$ & $(\alpha+\beta)^{{\rm exp}-(s+t)}_p$\\
\hline
$180^\circ$& $7.52$ & $3.28$ & $10.80$ & $3.1$\\
$140^\circ$& $7.65$ & $3.28$ & $10.93$ & $3.0$\\
$100^\circ$& $8.13$ & $3.28$ & $11.41$ & $2.5$\\
\hline
\end{tabular}
\end{center}
\label{drechsel-tab}
\end{table}
Unfortunately, the reason for the large deviation of the prediction
from the experimental value of the 
Baldin sum rule shown in the 5th row of Table \ref{drechsel-tab} 
has only been discussed qualitatively. As a test case for the validity 
of fixed-$\theta$ dispersion theory at large angles it would be highly
desirable to investigate  in what way Baldin's sum rule can be saturated.

\subsection{The GDH sum rule}

The GDH sum rule\cite{gerasimov66,drell66}
relates the anomalous magnetic moment, $\kappa$, and mass, $M$, of the
nucleon, i.e. static properties, to its dynamic observables, like the
total absorption cross sections, $\sigma_{3/2,1/2}$, of 
circularly polarized real
photons on longitudinally polarized nucleons in the two relative spin
configurations, parallel (3/2) and antiparallel (1/2):
\begin{equation}
\int^\infty_{\nu_{\rm thr}}(\sigma_{1/2}(\nu)-\sigma_{3/2}(\nu))
\frac{d\nu}{\nu}=-\frac{2\pi^2\alpha_e}{M^2}\kappa^2=
\begin{cases}
-205\mu b, &\text {proton},\\
-233\mu b, &\text {neutron}.
\end{cases}
\label{GDHsumrule}
\end{equation}
Here  $\alpha_e=1/137.04$ denotes the fine-structure constant and $\nu$ the
photon energy. Since the dispersion theory has been tested through 
Compton scattering experiments to a high level of precision,
the only crucial assumption in the dispersion
theoretic approach applied to the Compton forward amplitude, is the 
 {\it no-subtraction-hypothesis}.  Furthermore, the
energy dependence of the cross section in the two spin configurations 
gives important information for multipole analyses in the
resonance region and for parameters of Regge models in the higher
energy regime. 

In connection with a possible violation of the GDH sum rule a fixed
pole existing in the framework of Regge theory has been discussed 
in the literature. For sake of completeness we wish to summarize the
essential features of these considerations here. Starting from
invariant amplitudes we write the relevant dispersion integral in the following
form
\begin{equation}
{\rm Re}A_4(\nu,0)=-\frac{\alpha_e \kappa^2}{4 M \nu^2}
+\frac{2}{\pi}{\cal P}\int^\infty_{\nu_{\rm thr}}
\frac{\nu'{\rm Im}A_4(\nu',0)}{\nu'^2-\nu^2}d\nu'.
\label{FixedPole}
\end{equation}
The existence of a fixed pole then implies that the l.h.s. of Eq.
(\ref{FixedPole}) converges to an asymptotic contribution in the form
\begin{equation}
A^{\rm as}_4(\nu,t) = a_4(t)\nu^{-2}
\label{FixedPole-2}
\end{equation} 
where $a_4(t)$ is a real function of $t$.
Using (\ref{FixedPole-2}) and multiplying both sides of
(\ref{FixedPole}) with $\nu^2$ we arrive 
at
\begin{equation}
a_4(0)=-\frac{\alpha_e\kappa^2}{4M}-\frac{2}{\pi}\int^\infty_{\nu_{\rm thr}}
\nu {\rm Im} A_4(\nu,0)d\nu.
\label{FixedPole-3}
\end{equation}
in the limit $\nu \to \infty$.
Using the unitarity condition
\begin{equation}
{\rm Im}A_4(\nu,0)=\frac{M}{16\pi\nu^2}\left(\sigma_{1/2}(\nu)
-\sigma_{3/2}(\nu)\right)
\label{FixedPole-4}
\end{equation}
we arrive at a generalized GDH sum rule in the form
\begin{equation}
\int^\infty_{\nu_{\rm thr}}\frac{\sigma_{1/2}(\nu)- \sigma_{3/2}(\nu)}
{\nu}d\nu=-\frac{2\pi\alpha_e\kappa^2}{M^2} -\frac{8\pi^2}{M}a_4(0).
\label{FixedPole-5}
\end{equation}
The generalized GDH sum rule (\ref{FixedPole-5}) allows an
under-fulfillment or over-fulfillment of the GDH sum rule depending on
the the sign of $a_4(0)$. Indeed, a first evaluation of the  GDH
sum rule \cite{karliner73} led to deviations from the GDH prediction
which appeared to be isovector, i.e. different in sign for the proton
and the neutron. This evaluation was based on photomeson amplitudes
measured without polarization and estimated corrections for double-pion
photoproduction.
Later on this first evaluation  was confirmed by several
authors \cite{karliner-a} and lead to 
 results ranging
between -289 and -257 $\mu$b, and between -189 and -169 $\mu$b, for the proton
and neutron, respectively, where -- in absolute
numbers -- the proton results were much larger and the neutron results 
much smaller than the GDH integral predictions given in Eq. (\ref{GDHsumrule}).
These findings on the basis of experimental data led to several
investigations of the possible origin of these deviations from
the theoretical value of the GDH integral 
\cite{fox69}. In these investigation a fixed $J=1$ pole entering
into Regge phenomenology as well as
current algebra arguments were taken into consideration.
These speculations provided  part of the motivation to measure 
the GDH sum rule with circularly polarized photons and spin-polarized
nucleons. An other part was provided by the desire to get a better
understanding of the spin-structure of the nucleon.

\subsubsection{Verification of the GDH sum rule}

The GDH integrand on the l.h.s. of
Eq. (\ref{GDHsumrule}) was determined at two electron accelerators.
While the measurement from 0.2 to 0.8 GeV was carried out at MAMI
\cite{ahrens00,ahrens01,ahrens02}, the energy range from 0.68 to 2.9
GeV was covered at the electron stretcher ring ELSA
\cite{dutz03,dutz04}.
In total the range from the resonance region up to the Regge regime
was covered. This range is wide enough to reliably make conclusions on
the validity of the GDH sum rule for the first time. Only the ranges
from 0.14 -- 0.20 GeV and $>$ 2.9 GeV had to be covered using 
model predictions. For the photon energies below 0.20 GeV the unitary
isobar model MAID2002 gives a contribution of (-27.5 $\pm$ 3) $\mu b$
\cite{tiator02}. Above 2.9 GeV the Regge approach of Ref. 
\cite{bianchi99} gives -14 $\mu b$ while the prediction of Ref. 
\cite{simula02} is -13 $\mu b$.
\begin{figure}[h]
\begin{center}
\includegraphics[scale=1.0]{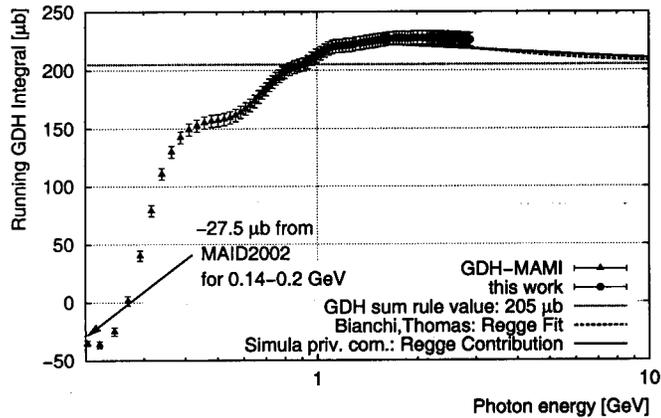}
\end{center}
\caption{Running GDH integral, $I_{\rm GDH}$, for the proton
up to 2.9 GeV. Error
  bars indicate statistical errors only.
$I_{\rm GDH}(\omega)=\int_{\nu_{\rm thr}}^{\omega}(\sigma_{3/2}(\nu)
-\sigma_{1/2}(\nu))d\nu/\nu$ with $\omega$ being the photon energy is shown on
  the abscissa.}
\label{GDHrunning}
\end{figure}
\begin{table}[h]
\caption{Measured values of the GDH integral for the proton 
$I_{\rm GDH}$ and model
predictions for the unmeasured ranges.}
\begin{center}
\begin{tabular}{|l|c|c|}
\hline
\hline
& $E_\gamma$ [GeV]& $I_{\rm GDH}$ [$\mu$ b]\\
\hline
MAID2002 \cite{tiator02}& 0.14-0.20 & -27.5 $\pm$ 3\\
measured (GDH-Collaboration) & 0.20-2.90&  254 $\pm$ 5 $\pm$ 12\\
Bianchi and Thomas \cite{bianchi99} & $>$ 2.9& -14\\
Simula et al.  \cite{simula02} & $>$ 2.9& -13\\
\hline
GDH integral& 0.14 -- $\infty$ & $\approx$ 213 \\
\hline
GDH sum rule & $\nu_{\rm thr} - \infty$ & 205 \\
\hline
\hline
\end{tabular}
\end{center}
\label{GDHtabular}
\end{table}
The results of the MAMI -- ELSA series  of experiments are shown
in Figure  \ref{GDHrunning} and Table \ref{GDHtabular}. The experimental
running GDH-integral clearly overshoots the value  predicted by the
GDH sum rule even when the reduction due to the negative contribution
of the range 0.14--0.20 GeV is taken into account. However, there is a
tendency visible of a decrease of the GDH integral beyond 2 GeV, which
is supported by the predictions from Regge theory \cite{bianchi99,simula02}.
Taking the predicted high-energy 
extrapolation into account the remaining difference 
$213 \mu b - 205 \mu b= 8 \mu
b$ is well below the estimated error $\approx \pm 13 \mu b$ so that the
tentative conclusion is allowed that the GDH sum rule has been 
confirmed for the proton.

\subsection{The Lvov-Nathan sum rule}
The Lvov-Nathan sum rule \cite{lvov99} uses the fixed-angle
$\theta=\pi$ sum rule to make predictions for the backward
spin polarizability $\gamma_\pi$. For the non-Born part of the relevant
amplitude
\begin{equation}
\tilde{A}_2(s,u,t,) \equiv A_2(s,u,t)+
\left( 1-\frac{t}{4M^2}\right)A_5(s,u,t)  
\label{A2tilde}
\end{equation}
we obtain
\begin{eqnarray}
\tilde{A}^{\rm NB}_2(s,u,t,)&=&\frac{1}{\pi}{\cal P} 
\int^\infty_{s_0}\left(\frac{1}{s'-s}
+\frac{1}{s'-u}-\frac{1}{s'}\right)
{\rm Im}_s \tilde{A}_2(s',u',t')ds'\nonumber\\
&+&\frac{1}{\pi}{\cal P}\int^\infty_{t_0}
{\rm Im}_t\tilde{A}_2(s',u',t')\frac{dt'}{t'-t},
\label{nonBorntildeA2} 
\end{eqnarray}
where 
\begin{equation}
s'+u'+t'=2M^2 \quad \mbox{and}\quad s'u'=M^4.
\label{ConStraints}
\end{equation}
When $s=u=M^2$ and $t=0$ , these integrals determine the backward spin 
polarizability of the nucleon
\begin{equation}
\gamma_\pi\equiv \frac{\tilde{A}^{\rm NB}_2(M^2,M^2,0)}{2\pi
  M}=\gamma^s_\pi+\gamma^t_\pi,
\label{gammapi}
\end{equation}
where 
\begin{eqnarray}
&&\gamma^s_\pi=-\frac{1}{2\pi^2 M}\int^\infty_{s_0}\frac{s'+M^2}{s'-M^2}
{\rm Im}_s \tilde{A}_2(s',u',t')\frac{ds'}{s'},\label{gammapis}\\
&&\gamma^t_\pi=- \frac{1}{2\pi^2 M}\int^\infty_{t_0}{\rm Im}_t
\tilde{A}_2(s',u',t')\frac{dt'}{t'}.\label{gammapit}
\end{eqnarray}

\subsubsection{The $s$-channel part of the LN sum rule}

We now replace the Mandelstam variables by the photon energy $\omega$
in the lab frame and the imaginary part of scattering amplitudes by
the appropriate photoabsorption cross sections. 
This is achieved by using the relation derived for ${\rm
  Im}g_\pi(\omega)$ in Eq. (\ref{optical-7}) and by using the relation
between $g_\pi(\omega)$ and ${\tilde A}_2$ derived in Eq.
(\ref{T5}).
For the Mandelstam variable we obtain the replacement
\begin{equation}
s'=m^2+2m\omega \label{s-omega}.
\end{equation}
Then, after some calculation we arrive at \cite{lvov99}
\begin{equation}
\gamma^s_\pi=\int^\infty_{\omega_0}\frac{d\omega}{4\pi^2\omega^3}
\sqrt{1+\frac{2\omega}{M}}\left( 1+\frac{\omega}{M} \right)
\sum_n P_n [\sigma^n_{3/2}(\omega)- \sigma^n_{1/2}(\omega)]
\label{schannel-1}
\end{equation}
with
\begin{equation}
\sum_n P_n [\sigma^n_{3/2}(\omega)- \sigma^n_{1/2}(\omega)]=
\left\{\left( \sigma_{1/2}
(\omega,\Delta P={\rm yes)}- \sigma_{1/2}
(\omega,\Delta P={\rm no})\right)- \left( 1/2\to 3/2\right)\right\}.
\label{schannel-2}
\end{equation}
The dominant contribution to $\gamma^s_\pi$ comes from single-pion 
photoproduction, $\gamma N \to \pi N$, which yields the cross section 
(see Eq. (\ref{optical-7}) and \cite{lvov99})
\begin{equation}
\sum_n P_n [\sigma^n_{3/2}(\omega)- \sigma^n_{1/2}(\omega)]^{\pi N}=
8 \pi \frac{q}{k}\sum^\infty_{l=0} (-1)^l(l+1)\left\{
|A_{l+}|^2-|A_{(l+1)-}|^2-\frac{l(l+2)}{4}(|B_{l+}|^2-|B_{(l+1)-}|^2)\right\}.
\end{equation}
where $q=|{\bf q}|$ and $k=|{\bf k}|$ are the momenta of the pion and
photon in the c.m. system, respectively.
Here a sum over channels with charged and neutral  pions is implied,
where  $A_{l\pm}$ and $B_{l\pm}$ are the standard Walker photoproduction
multipoles \cite{walker69}.

\subsubsection{The $t$-channel part of the LN sum rule}

Each pseudoscalar meson $M$ makes a contribution \cite{lvov99}
\begin{equation}
\tilde{A}^M_2(t)=\frac{g_{MNN}F_{M\gamma\gamma}}{t-m^2_M-i0}\tau_M
\label{poleA2}
\end{equation}
to the amplitude entering into Eq. (\ref{gammapit}). Here $g_{MNN}$ is the 
meson-nucleon coupling constant and $F_{M\gamma\gamma}$ the constant of the 
two-photon decay of the meson. The isospin factor  $\tau_M$ is either 1
or $\tau_3$ for isoscalar and isovector mesons, respectively.
Making use of the symbolic equality
\begin{equation}
\frac{1}{x-i0}={\cal P}\frac{1}{x} +i\pi\delta(x)
\label{shiftedpole}
\end{equation}
we arrive at 
\begin{equation}
\gamma^{(M)}_\pi=-\frac{\tilde{A}^M_2(0)}{2\pi M}
=\frac{g_{MNN}F_{M\gamma\gamma}}{2\pi m^2_M M}\tau_M.
\label{gammapifinal}
\end{equation}
For the further discussion of $\gamma^t_\pi$ we refer to the paper of 
L'vov and Nathan \cite{lvov99}.

\subsubsection{Comparison between theory and experiment}

Table \ref{spin-pol} summarizes the results obtained for the backward
spin polarizabilities of proton and neutron in comparison with the
LN sum rule predictions. The experimental results have been obtained from the
invariant amplitudes adjusted to the experimental Compton differential
cross sections  using the formula
\begin{equation}  
\gamma_\pi({\rm fixed}-t)=- \, \frac{1}{2\pi M}
{\tilde A}^{\rm NB}_2(0,0)=
- \, \frac{1}{2\pi M}\left[A^{\rm NB}_2(0,0)
+  A^{\rm NB}_5(0,0) \right] 
\label{gamma-pi-integral}
\end{equation}
where NB denotes the non-Born part of the scattering amplitude.
During the adjustment procedure the photo-absorption cross sections
as calculated  from the pion photoproduction data and the difference 
$\alpha-\beta$ 
of electromagnetic polarizabilities  
obtained from low-energy Compton scattering experiments
have been kept constant, whereas the 
spin-polarizability $\gamma_\pi({\rm fixed}-t)$ has been used as an
adjustable parameter. In fixed-$t$ dispersion theory the non-Born part
of the  amplitude is a superposition of an integral part and an
asymptotic part
\begin{equation}
{\tilde A}^{\rm NB}_2(\nu,t)=  {\tilde A}^{\rm int}_2(\nu,t)+
{\tilde A}^{\rm as}_2(\nu,t)
\label{gamma-pi-integral-2}
\end{equation}
\begin{table}[h]
\caption{Experimental results obtained for the backward
  spin-polarizabilities of proton and neutron compared with
  predictions from the LN sum rule.
The unit is  $10^{-4} {\rm fm}^4$}
\begin{center}
\begin{tabular}{|l|l|l|l|}
\hline
spin polarizabilities& proton&neutron&\\
\hline
$\gamma_\pi({\rm fixed}-t)$& $-$38.7$\pm$1.8&+58.6$\pm$4.0&experiment 
\cite{camen02,kossert02}\\
$\gamma_\pi{\rm fixed}-\theta$& $-$39.5$\pm$2.4&+52.5$\pm$2.4&sum rule 
\cite{lvov99}\\
\hline
$\gamma^t_\pi\equiv \gamma^{\rm as}_\pi$& $-$46.6 &+43.4&
$\pi^0+\eta+\eta'$ -poles
\cite{lvov99}\\
$\gamma^{\rm int}_\pi$& +7.9 $\pm$ 1.8& +15.2$\pm$ 4.0& 
experiment \cite{camen02,kossert02}\\
$\gamma^s_\pi$& +7.1 $\pm$ 1.8& +9.1$\pm$ 1.8&
sum rule \cite{lvov99}\\
\hline
\end{tabular}
\end{center}
\label{spin-pol}
\end{table}
where the integral part can be calculated from pion photoproduction
data alone, whereas the asymptotic part is not exactly known. However,
it may be assumed  that this part is essentially given by the
pseudoscalar poles which are expected to exhaust the $t$-channel of
fixed-$\theta$ dispersion theory.   This contribution to $\gamma_\pi$ 
is given by
\begin{equation}
\gamma^t_\pi
=\frac{1}{2\pi M}
\left[ \frac{g_{\pi NN}F_{\pi^0 \gamma\gamma}}{m^2_{\pi^0}}\tau_3
+\frac{g_{\eta  NN}F_{\eta \gamma\gamma}}{m^2_{\eta}}
+\frac{g_{\eta'  NN}F_{\eta' \gamma\gamma}}{m^2_{\eta'}} \right].
\label{tchannalgammapi}
\end{equation} 
In Table \ref{spin-pol} the numbers in line 2 and 3 represent the
spin-polarizabilities obtained from the experimental Compton
differential cross sections on the basis of fixed-$t$ dispersion theory
and the LN sum-rule prediction, respectively. For the proton the agreement
of the two different results is well within the errors,
whereas  for the neutron the difference of the two results is of the order   
of the errors. It has to be kept in mind, that for the neutron
the experimental result for
$\gamma_\pi({\rm fixed}-t)$
has a comparatively large error  so that the major part of the 
conclusions has  to rest on the proton data. The numbers in line 4
represent the $t$-channel results as obtained from the pseudoscalar
poles, where it has been assumed that these numbers coincide
with the asymptotic contributions entering into fixed-$t$ dispersion
theory. Subtracting the numbers in line 4 from the corresponding numbers
in lines 2 and 3, we arrive at the numbers in line 5 and 6, respectively.
 
The conclusions to be drawn from Table \ref{spin-pol}
are the following. The LN sum rule has been shown to be fulfilled with good
precision and there is no major difference between $\gamma^t_\pi$ as entering
into fixed-$\theta$ dispersion theory and 
$\gamma^{\rm as}_\pi$ as entering into fixed-$t$ dispersion theory. 
This result
is nontrivial because $\gamma^t_\pi$ has been calculated from the LN
sum rule and  $\gamma^{\rm as}_\pi$ has been obtained by adjustments to
experimental data by using this quantity as the adjustable parameter.
This  finding contrasts with the findings made in the 
following subsection in connection with 
the $t$-channel and asymtotic contributions in case of 
the BEFT  sum rule.

\subsection{The Bernabeu-Ericson-FerroFontan-Tarrach sum rule}

The BEFT sum rule 
\cite{bernabeu74,bernabeu77,guiasu76,guiasu78,budnev79,holstein94}
may be derived from the non-Born part of the
invariant amplitude
\begin{equation}
{\tilde A}_1(s,u,t) \equiv A_1(s,u,t) -\frac{t}{4M^2}A_5(s,u,t)
\label{A-1-tilde}
\end{equation}
by applying the fixed-$\theta$ dispersion relation for
$\theta=180^\circ$ ($a$ = 0).
Then we arrive at
\begin{eqnarray}
{\tilde A}_1(s,u,t) &=& \frac{1}{\pi}{\cal P}
\int^\infty_{s_0}\left(\frac{1}{s'-s}+\frac{1}{s'-u}-\frac{1}{s'}\right)
{\rm Im}_s{\tilde A}_1(s',u',t')ds'\nonumber\\
&+& \frac{1}{\pi}{\cal P}\int^\infty_{t_0}{\rm Im}_t {\tilde A}_1(s',u',t') 
\frac{dt'}{t'-t}\,\, ,
\label{disprelA1}
\end{eqnarray}
where $s_0=(M+m_\pi)^2$, $t_0= 4 m^2_\pi$, $s'+u'+t'=2M^2$ and $s'u'=M^4$.
Using 
\begin{equation}
\alpha-\beta=-\frac{1}{2\pi}{\tilde A}_1(M^2,M^2,0)=(\alpha-\beta)^s+
(\alpha-\beta)^t
\label{alpha-beta-BEFT}
\end{equation}
we arrive at
\begin{eqnarray}
&&(\alpha - \beta)^s = - \frac{1}{2\pi^2} \int^\infty_{s_0}
\frac{s'+M^2}{s'-M^2}{\rm Im}_s {\tilde
  A}_1(s',u',t')\frac{ds'}{s'}\,\, ,\\
&&(\alpha-\beta)^t=
-\frac{1}{2\pi^2}\int^\infty_{t_0}{\rm Im}_t
{\tilde A}_1(s',u',t') \frac{dt'}{t'}.
\label{SR}
\end{eqnarray}

\subsubsection{The $s$-channel part of the BEFT sum rule}

Using the relation for $f_\pi$ derived in Eq. (\ref{T4}) 
and the relation for ${\rm Im}f_\pi$ derived in Eq. (\ref{optical-5})
we arrive at
\begin{equation}
(\alpha-\beta)^{s}=\frac{1}{2\pi^2}
\int^\infty_{m_\pi+\frac{m^2_\pi}{2M}}
\sqrt{1+\frac{2\omega}{M}}
\left[ \sigma(\Delta P= {\rm yes}) -\sigma(\Delta P = {\rm no})
  \right]
\frac{d\omega}{\omega^2}.
\label{alphabetafinal}
\end{equation}
This integral contains the $s$-channel part of the BEFT sum rule
in terms of quantities which are convenient for calculations.
The main contribution to $(\alpha-\beta)^s$ comes from the
photoproduction of $\pi N$ states. For these states, the cross-section
difference can be expressed in terms of the standard CGLN amplitudes
via \cite{levchuk05}
\begin{eqnarray}
\left[ \sigma(\Delta P= {\rm yes}) -\sigma(\Delta P = {\rm no})
  \right]^{\pi N}&=& 4\pi\frac{q}{k}\sum^\infty_{k=0} (-1)^k
(k+1)^2\Big\{ (k+2)\left(|E_{k+}|^2- |M_{(k+1)-}|^2\right)\nonumber\\
&& \hspace{4.3cm}+ k\left(|M_{k+}|^2-|E_{(k+1)-}|^2\right)\Big\}^{\pi N},
\label{CGLN}
\end{eqnarray}
where $q=|{\bf q}|$ and $k=|{\bf k}|$  are the pion and photon 
momenta in the c.m. system, respectively.

\subsubsection{The $t$-channel part of the BEFT sum rule}

The imaginary part of the amplitude ${\tilde A}_1$ in the $t$-channel
can be found using the general unitarity relation
\begin{equation}
{\rm Im}_t T(\gamma\gamma\to N{\bar N})=\frac12 \sum_n (2\pi)^4
\delta^4(P_n-P_i)T(\gamma\gamma\to n)T^*(N{\bar N}\to n),
\label{t-unitarity}
\end{equation}
where the sum on the right-hand side is
taken over all allowed intermediate states $n$ having the  same total
4-momentum as  the initial state. In the following we restrict ourselves
to two-pion intermediate states, i.e. $n=\pi\pi$.

The 
amplitudes $T(\gamma\gamma\to\pi\pi)$ and  $T(\pi\pi\to\bar{N}N)$
are  constructed  making use of available experimental 
information on two different reactions. For the amplitude
$T(\gamma\gamma\to\pi\pi)$ this is  the two-photon fusion reaction
$\gamma\gamma\to\pi\pi$. Since
there are no data on the reaction $\pi\pi\to\bar{N}N$,
the amplitude $T(\pi\pi \to \bar{N}N)$ is constructed in a dispersive approach
using the well known amplitudes of the pion-nucleon scattering reaction
$\pi N \to N \pi$. At $t>m^2_\pi$ unitarity shows that the phases of the
amplitudes 
$T(\gamma\gamma\to\pi\pi)$ and  $T(\pi\pi\to\bar{N}N)$
are the same and equal to the phases of pion-pion scattering, $\delta^J_I$, 
which are known from the data on the reaction $\pi p\to p\pi\pi$.

The phase-dependent factor entering into  the amplitude of a 
narrow resonance is  described  by a 
Breit-Wigner curve, whereas for the present case of a very broad resonance 
as given by the functions $\delta^J_I(t)$, a generalized version of the
Breit-Wigner curve has to be used.
This generalized  phase-dependent  factor is given
by the Omn\`es \cite{omnes58} function
$\Omega^J_I(t)$  defined through
\begin{equation}
\Omega^J_{I}(t)={\rm exp}{\left[
\frac{t}{\pi}\int^\infty_{4m^2_\pi}dt'\frac{\delta^J_{I}(t')} 
{t'(t'-t-i 0)}\right]}\equiv e^{i\delta^J_{I}(t)}
{\rm exp}{\left[\frac{t}{\pi}{\cal P}\int^\infty_{4m^2_\pi}
\frac{\delta^J_{I}(t')dt'}{t'(t'-t)} \right]},
\label{Omnesfunct}
\end{equation}
where use is made of Eq. (\ref{shiftedpole}).
For the discussion, it is helpful to know that 
for a narrow resonance with infinitesimal width $\Gamma$ 
 the Omn\`es function can
be written in the form 
\begin{equation}
\Omega(t)=\frac{m^2}{m^2-t-im\Gamma},
\label{Omnesfunct-2}
\end{equation}
where  $m^2 = t(\delta=\pi/2)$ is the ``bare mass'' of the resonance.
This means that for small $\Gamma$ the Omn\`es function has the structure
 of a pole 
located on the $t$-axes at a position where the phase is equal to
$\delta=\pi/2$. For energies  far below the  resonance energy, i.e.
$t\ll m^2$, the Omn\`es function is only slightly dependent on $t$. Note that
according to Eq. (\ref{Omnesfunct-2})
$\Omega(0)=1$.   In the general case the Omn\`es function retains 
the essential parts of these properties.
Instead of the Omn\`es function in most treatments 
the  function $D^J_I(t)=1/\Omega^J_I(t)$ is in use. 
The procedure of
taking into account the $\pi\pi$ phase relation is based on dispersion
relations  termed the $N/D$
method (see e.g. \cite{hoehler83}).

For the discussion of the properties of the $T(\pi\pi\to \bar{N}N)$
amplitude in the unphysical region  $4m^2_\pi\leq t \leq 4M^4$  
we use  the backward amplitude
$F^{(+)}(t)$ discussed  by Bohannon \cite{bohannon76}. For the pion
scattering process  the quantity $t$ is negative, whereas for positive
$t$ the analytic continuation of $F^{(+)}(t)$ describes the
$N\bar{N}\to \pi\pi$ annihilation process for the case that the
helicities of the nucleon and the antinucleon are the same, 
$\lambda = \bar{\lambda}$. This provides us with a
tool to construct the $N\bar{N}\to \pi\pi$ amplitude from the measured
$\pi N\to \pi N$ amplitude. Since the backward $\pi N\to N\pi$
scattering amplitude implies also backward $N \bar{N} \to \pi\pi$
annihilation, it can be shown that the first two terms of the expansion 
for $F^{(+)}(t)$
at positive $t$ are
\begin{equation}
F^{(+)}(t)= \frac{16 \pi}{M(4M^2-t)}f^{0}_+(t) -\frac{5\pi(t-4m^2_\pi)}
{M} f^2_+(t).
\label{bohannon}
\end{equation}
The amplitudes $f^J_+(t)$ are the partial wave amplitudes
introduced  by Frazer and Fulco 
\cite{frazer60}, where $J$ denotes the angular momentum of the
$\pi\pi$ intermediate state. The $+$ sign denotes that the
two  helicities $\lambda$ and $\bar{\lambda}$ of  $N$ and $\bar{N}$,
respectively, are the same, $\lambda = \bar{\lambda}$.
The construction of these amplitudes follows the  standard $N/D$ procedure
which is described at many places (see e.g. \cite{hoehler83}).

The construction of both amplitudes, $T(\gamma\gamma\to\pi\pi)$ and
$T(\pi\pi \to \bar{N}N)$, in connection with the scalar-isoscalar $t$-channel
of Compton scattering has first been described and worked out in some detail
by K{\"o}berle \cite{koeberle68} and later discussed  by several 
authors, of whom we wish to cite 
\cite{morgan88,drechsel99,drechsel03,levchuk04,levchuk05}. 
For a very broad resonance the appropriate ansatz reads
\begin{equation}
F^J_{I\lambda}(t)=\Omega^J_I(t)   P^J_{I\lambda}(t).
\label{gg-pipi-ansatz}
\end{equation}
where $P^J_{I\lambda}(t)$ is a real amplitude in the 
$\gamma\gamma\to\pi\pi$ physical region  and 
$\Omega^J_I(t)$ the  phase-dependent Omn\`es function discussed above.
In (\ref{gg-pipi-ansatz}) $I$ is the isospin of the
transition, $J$ the angular momentum and 
$\lambda\equiv \Lambda^t_{\gamma\gamma}$ 
the helicity
difference of the two photons. In the present case we 
have $\lambda\equiv \Lambda^t_{\gamma\gamma}\equiv 0$, 
so that we can omit the index
$\lambda$ without loss of generality. 

The amplitudes $F^J_I(t)$ have to be constructed such that they
have the correct low-energy properties,
 reproduce the 
cross section of the photon fusion reaction $\gamma\gamma\to \pi\pi$ and 
incorporate the phases $\delta^J_I(t)$
\cite{colangelo01,kaloshin94,pennington94,hyams73,froggatt77}.
For  $J=0$  the following form of Eq. 
(\ref{gg-pipi-ansatz}) has been obtained \cite{levchuk04,levchuk05}:
\begin{eqnarray}
F^0_{I}(t)&=&\Omega^0_I(t)\Big\{\left[F^{B,0}_{I}(t)+ \Delta
 F^0_{I}(t)\right] 
{\rm Re}\frac{1}{\Omega^0_I(t)} -\frac{t^2}{\pi}\Big[ {\cal P} \int^\infty
_{4 m^2_\pi}[ F^{B,0}_{I}(t')+\Delta
 F^0_{I}(t')] {\rm Im}\frac{1}{\Omega^0_I(t')}
\frac{dt'}{t'^2(t'-t)}\nonumber\\
&+& A^0_I + t B^0_I\Big]\Big\}.
\label{modified-s-wave}
\end{eqnarray}
The expression in Eq. (\ref{modified-s-wave}) is given for  the $s$-wave
amplitude where contributions going beyond the Born approximation
have been taken into account.
The leading term in (\ref{modified-s-wave}) is a superposition of a
Born term, $F^{B,0}_{I}(t)$, and a pion-structure dependent correction,
$\Delta F^{0}_{I}(t)$.
Since the reactions $\gamma\gamma\to \pi^+\pi^-,\pi^-\pi^+,\pi^0\pi^0$
show up with two components having  isospin $I=2$ and one component having
$I=0$ we have to take into account these two isospins with the appropriate 
weights (see e.g. \cite{gourdin60}):
\begin{eqnarray}
&&F_{I=0}=F^C+\frac12 F^N \nonumber\\    
&&F_{I=2}=F^C-F^N
\label{gourdin}
\end{eqnarray}
where $F^C$ is the charged and $F^N$ the neutral component.

For both isospin values, $I=0$ and $I=2$, the Born terms are
\begin{equation}
F^{B,0}_{0}(t)=F^{B,0}_{2}(t)= 2e^2
\frac{1-v^2}{v}\frac12 \ln\frac{1+v}{1-v},
\label{bornterm}
\end{equation}
where $v$ is the pion velocity. At low energies,
the pion-structure dependent corrections can be expressed through the
electromagnetic polarizabilities of the pions in the form
\begin{eqnarray}
\Delta F^0_{0}(t) &=&2\pi m_\pi t [(\alpha_{\pi^{\pm}}- \beta_{\pi^{\pm}})
+\frac12 (\alpha_{\pi^0}- \beta_{\pi^0})],
\nonumber\\
\Delta F^0_{2}(t) &=&2\pi m_\pi t [(\alpha_{\pi^{\pm}}- \beta_{\pi^{\pm}})
- (\alpha_{\pi^0}- \beta_{\pi^0})].
\label{deltaF}
\end{eqnarray}
The last term $A^0_I+t B^0_I$ has been introduced to represent 
those contributions to $F^0_{I}(t)$ which go beyond the Born +
pion-polarizability approximation. They are constructed in a way (i) that
general constraints available for the amplitude $F^0_{I}(t)$ are fulfilled 
and (ii) that experimental data \cite{boyer90,behrend92} on the total
cross sections of the reactions $\gamma\gamma\to \pi^+\pi^-$ and
 $\gamma\gamma\to \pi^0\pi^0$ 
are fitted.
The amplitude $F^2_0(t)$ is constructed analogously.

The amplitudes of interest for the prediction of 
$(\alpha-\beta)^t$
are the $S$-wave amplitude $F^0_{0}(t)$ and the $D$-wave amplitude
$F^2_{0}(t)$ with the $D$-wave amplitude  leading to only a small
correction. It, therefore, is appropriate to restrict the discussion
mainly to the amplitude $F^0_{0}(t)$. The essential
property  of this  amplitude is  provided  by the Omn\`es function
which introduces a zero crossing of the amplitude at about 570 MeV.
This zero crossing may  be considered as a manifestation of 
that part of the $\sigma$ meson which shows up through phase-shift 
$\delta^0_0(t)$ of the correlated $\pi\pi$ pair.

If we restrict ourselves in the calculation
of the $t$-channel absorptive part to intermediate states with two
pions with angular momentum $J\leq 2$, the sum rule (\ref{SR})
takes the convenient form for calculations \cite{bernabeu77}:
\begin{eqnarray}
(\alpha-\beta)^t&=& 
 \frac{1}{16 \pi^2}\int^\infty_{4 m^2_\pi}\frac{dt}{t^2}\frac{16}{4M^2-t}
\left(\frac{t-4m^2_\pi}{t}\right)^{1/2}\Big[f^0_+(t)
  F^{0*}_{0}(t)\nonumber\\
&&-\left(M^2-\frac{t}{4}\right)\left(\frac{t}{4}-m^2_\pi\right)
f^2_+(t) F^{2*}_{0}(t)\Big],\label{BackSR}
\end{eqnarray}
where
$f^{(0,2)}_+(t)$ and $F^{(0,2)}_0(t)$ are the partial-wave helicity
amplitudes of the processes $N\bar{N}\to \pi\pi$ and 
$\pi\pi\to \gamma\gamma$ with angular momentum $J=0$ and $2$,
respectively, and isospin $I=0$.

\subsubsection{Test of the BEFT sum rule}

Though being the first who published the BEFT sum rule in its presently
accepted form, Bernabeu and Tarrach \cite{bernabeu77}
were not aware of the appropriate amplitudes  to calculate this sum rule 
numerically.   Also the first calculation of Guiasu and Radescu
\cite{guiasu76} remained incomplete because the amplitudes were used in the
form of their Born approximation  and, as the major drawback,  the correlation
of pions was not taken into account.  
Table \ref{BEFtable} summarizes those results
\begin{table}[h]
\caption{Numerical evaluation of the BEFT  sum rule, with corrections a) 
and b) supplemented by the present author. The unit is $10^{-4}{\rm fm}^3$.}
\vspace{5mm}
\begin{center}
\begin{tabular}{|l|l|l|l|}
\hline
$(\alpha_p-\beta_p)^s$&$(\alpha_p-\beta_p)^t$&$(\alpha_p-\beta_p)^{\rm BEFT}$
&authors\\
\hline
$-$4.92&+9.28$^{a)}$&+4.36&Guiasu, Radescu,1978 \cite{guiasu78}\\
$-$4  & +10.4$^b)$ & +6.4 & Budnev, Karnakov, 1979 \cite{budnev79}\\
$-$5.42&+8.6&$+(3.2^{+2.4}_{-3.6})$&{Holstein,Nathan,1994}\cite{holstein94}\\
$-$5.56&+16.46&+(10.7$\pm$0.2)$^{c)}$&{Drechsel,Pasquini,
Vanderhaeghen,2003}\cite{drechsel03}\\
\hline
$-$(5.0 $\pm 1.0)$ &+ (14.0 $\pm$ 2.0)& + (9.0$\pm$2.2)&{Levchuk, et al. 
}\cite{levchuk05}\\
\hline
\end{tabular}
\end{center}
a) corrected for the $D$-wave 
contribution $(-1.7)$ included. In an
earlier work Guiasu and Radescu \cite{guiasu76} used the Born
approximation without $\pi\pi$ correlation for both 
amplitudes $N\bar{N}\to \pi\pi$ and
$\pi\pi\to\gamma\gamma$ 
and arrived at $(\alpha-\beta)^t= +17.51$.\\
b) correction for the polarizability of the pion (+3.0) included.\\
c) best value from a range of results given by the authors \cite{drechsel03}.
\label{BEFtable}
\end{table}
\begin{figure}[h]
\includegraphics[scale=0.4]{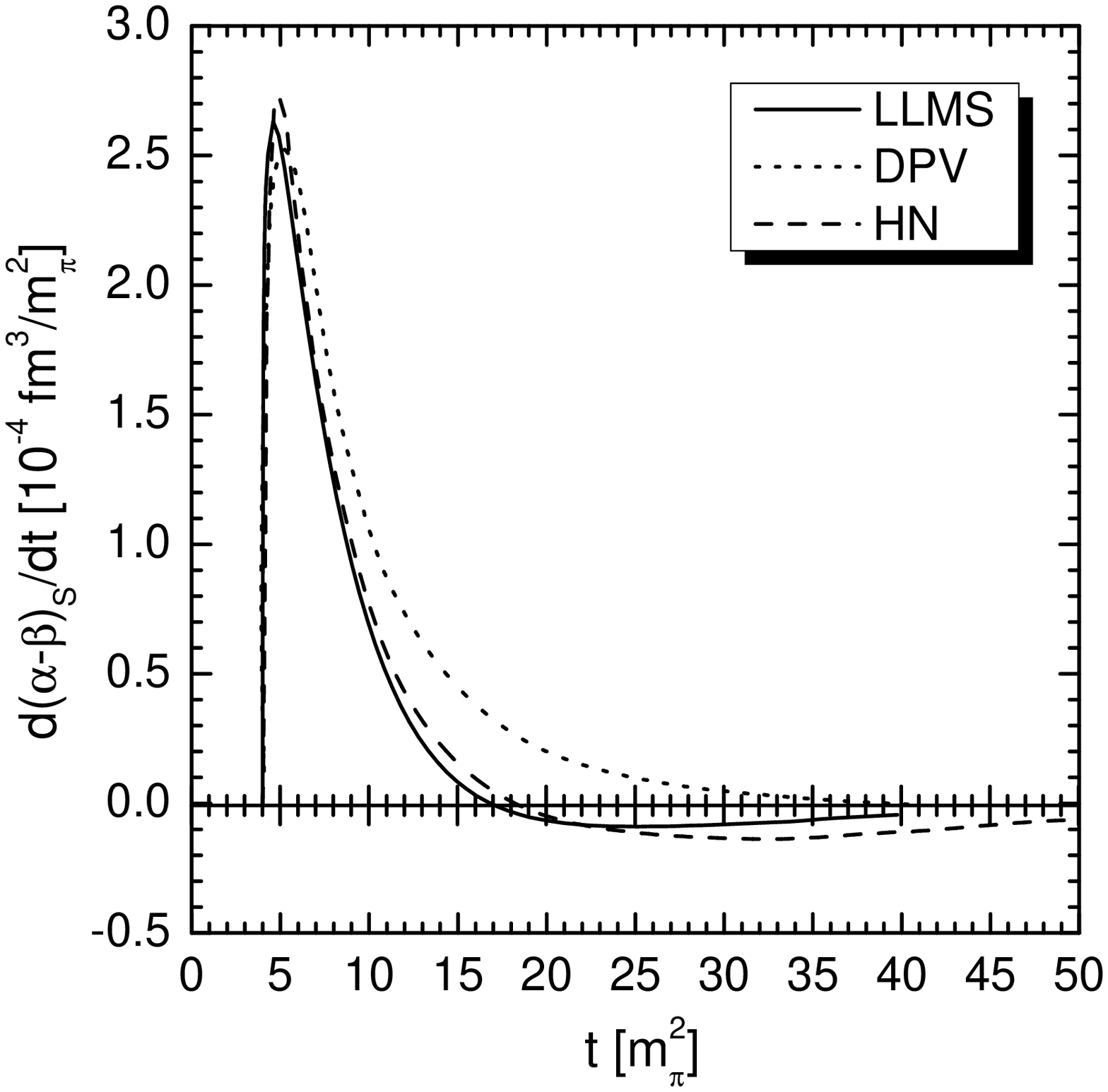}
\includegraphics[scale=0.4]{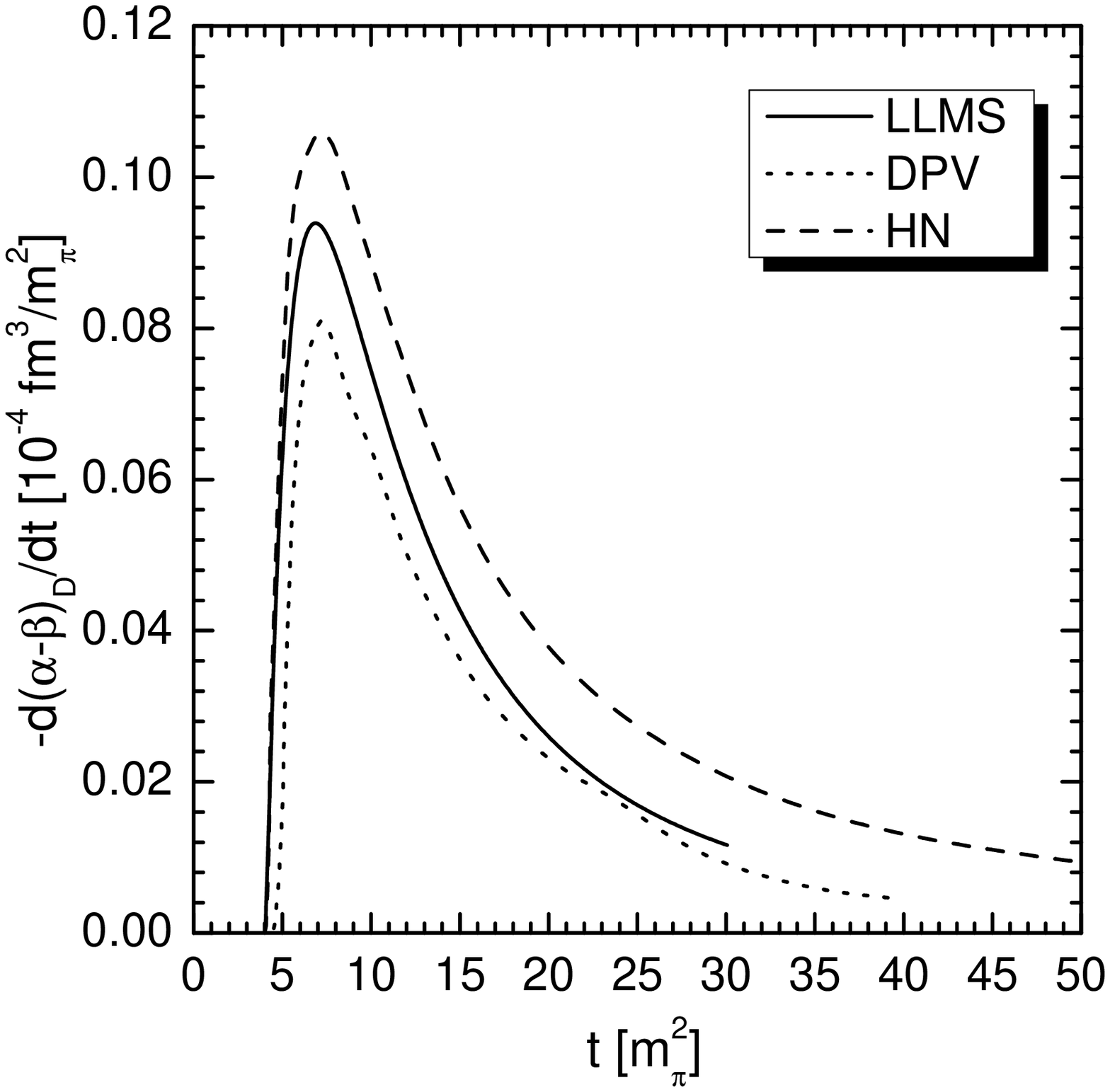}
\caption{Left panel: The integrand for the $t$-channel $S$-wave
  contribution to $\alpha-\beta$ as given by a recent calculation
 \cite{levchuk04} (solid: LLMS), Ref. \cite{holstein94}
(dashed: HN), and Refs. \cite{drechsel99,drechsel03} (dotted: DPV). Right
  panel:
Same for the $t$-channel $D$-wave contribution. Units are 
$10^{-4} {\rm fm}^3/m^2_\pi$. The dotted curve in the left panel 
may be considered as an upper
limit the dashed  curve as a lower limit of a band of possible results
\cite{levchuk05}.}
\label{LevchukBEF}
\end{figure}
of tests of the BEFT sum rule where at least the $\pi\pi$ correlation is taken
into account. In the early works of Guiasu and Radescu \cite{guiasu78}
and Budnev and Karnakov \cite{budnev79} some missing pieces in the results
were identified which are supplemented in Table \ref{BEFtable} on the basis
of the results given by Holstein and Nathan \cite{holstein94}.
In case of the Holstein and
Nathan result \cite{holstein94} we interpret the 
estimated upper and lower bounds as errors.   In case of the Drechsel
et al. result \cite{drechsel03} we quote the $s$-channel and $t$-channel
contributions calculated at $\theta=180^\circ$. The total result
$(\alpha_p-\beta_p)^{\rm BEFT}$ is the best value for this quantity 
extracted by the authors from results obtained in the angular region
$140^\circ \leq \theta_{\rm lab} \leq 180^\circ$. The result obtained
in   \cite{levchuk04} (see Figure \ref{LevchukBEF}) 
 confirmes the result of Holstein and Nathan
\cite{holstein94}
when using using a cut-off in  the integration
in (\ref{BackSR}) at $t=30m^2_\pi$. 

There is consistency  between the five 
results as far as the  $s$-channel contribution is concerned but 
an apparent discrepancy in case of the $t$-channel 
between the results of Holstein and Nathan
\cite{holstein94} on the one hand 
and of Drechsel et al. \cite{drechsel03} on the other.
In order to find an explanation for this discrepancy we extended
the  calculation of \cite{levchuk04} where a cut-off in  the integration
in (\ref{BackSR}) at $t=30m^2_\pi$ was used by a more complete calculation.
This investigation showed that the available data to be used as input
are not very precise and lead to large differences in the predictions
for $(\alpha-\beta)^{t}$. The curves obtained for $d(\alpha-\beta)_S/dt$
were located in a broad band between the dotted curve and the dashed curve 
shown in the left panal of Figure \ref{LevchukBEF}. Nevertheless, is appears
possible to give a tentative result for the $t$ channel part
of the BEFT sum rule prediction
in the form $(\alpha-\beta)^{t}=14.0\pm 2.0$. This result is in agreement
with the result of Drechsel et al. \cite{drechsel03} but has a considerably
larger error.  We will use the results given in line 6 of Table  
\ref{BEFtable} in the further discussion.

\subsubsection{Summary on predicted information on $\alpha-\beta$ 
compared with the experimental result}

In the foregoing the BEFT sum rule has been discussed making predictions
for the $s$-channel part, $(\alpha-\beta)^s$, and the $t$-channel part,
$(\alpha-\beta)^t$, of the difference of electromagnetic polarizabilities.
In fixed-$t$ dispersion theory the $s$-channel part is replaced by the
integral (along the $\nu$ axis) part,  $(\alpha-\beta)^{\rm int}$,  and the 
$t$-channel part by the asymptotic (contour integral) part, 
$(\alpha-\beta)^{\rm as}$,  where the two versions  are closely related 
to each other but are not completely identical. 
For illustration it is of interest to derive expressions for
$(\alpha-\beta)^{\rm int}$ in a similar form as those obtained for
$(\alpha-\beta)^s$.

In fixed-$t$ dispersion theory we may write
\begin{equation}
(\alpha-\beta)^{\rm int}=-\frac{1}{2\pi}A^{\rm int}_1(0,0)
=-\frac{1}{\pi^2} \int^\infty_{\nu_{thr(0)}}
{\rm Im}A_1(\nu',0)\frac{d\nu'}{\nu'}.
\label{alpha-beta-fixed-2}
\end{equation}
In (\ref{alpha-beta-fixed-2}) the finite upper limit $\nu_{\rm max}(0)$ 
of the integral is 
replaced by $\infty$.
From the general formula given in (\ref{invarianthelicity})
we may derive the relevant special case for $t=0$
\begin{equation}
{\rm Im}A_1(s,0)=\frac{1}{(s-M^2)^2}\left[-\frac{s}{M}{\rm Im}\,\tau_4
-\frac{\sqrt{s}}{M}{\rm Im}\,\tau_5\right].
\label{alpha-beta-fixed-3}
\end{equation}
with
\begin{equation}
\nu = \frac{s-M^2}{2M}.
\label{s-nu-relation}
\end{equation}
Restricting the further discussion to the $1\pi$ channel we may write
\begin{eqnarray}
&&{\rm Im}[\tau_4]^{1\pi}_{t=0}=8\pi q \sqrt{s}\sum_{k\geqslant 1}
\frac{k(k+1)^2(k+2)}{2}\left(A_{k+}B^*_{k+}-A_{(k+1)-}B^*_{(k+1)-}\right),
\label{alpha-beta-fixed-4}\\
&&{\rm Im}[\tau_5]^{1\pi}_{t=0}=8\pi q \sqrt{s}\sum_{k\geqslant 0}
2(k+1)^2 \left(|A_{k+}|^2 - |A_{(k+1)-}|^2 \right)
\label{alpha-beta-fixed-5}
\end{eqnarray}
with
\begin{eqnarray}
&&A_{k+}B^*_{k+}=\frac12\big[(k+2)|E_{k+}|^2+kM_{k+}E^*_{k+}
-(k+2)E_{k+}M^*_{k+}-k|M_{k+}|^2   \big], \nonumber\\
&&A_{(k+1)-}B^*_{(k+1)-}=\frac12
    \big[-k|E_{(k+1)-}|^2+(k+2)M_{(k+1)-}E^*_{(k+1)-}\nonumber\\
&&\hspace{3.2cm}-kE_{(k+1)-}M^*_{(k+1)-} +(k+2)|M_{(k+1)-}|^2\big],\nonumber\\
&&|A_{k+}|^2=\frac14 \big[ (k+2)^2|E_{k+}|^2+(k+2)kE_{k+}M^*_{k+}
+k(k+2)M_{k+}E^*_{k+}+k^2|M_{k+}|^2\big],\nonumber\\
&&|A_{(k+1)-}|^2=\frac14 \big[k^2|E_{(k+1)-}|^2
-(k+2)kM_{(k+1)-}E^*_{(k+1)-}\nonumber\\&&\hspace{2.3cm}
-k(k+2)E_{(k+1)-}M^*_{(k+1)-} +(k+2)^2|M_{(k+1)-}|^2\big]. 
\end{eqnarray}
This shows
that the integrand entering into Eq. (\ref{alpha-beta-fixed-2}) 
has no such a simple decomposition into photo-absorption cross-sections
of definite electromagnetic multipolarity as found for the 
corresponding integrand (\ref{alphabetafinal})
valid in  case of fixed-$\theta=\pi$ dispersion theory.
For further illustration we  simplify the expressions by taking only the
multipoles $E_{0+}(E1)$ and $M_{1+}(M1)$ into account which are the 
most prominent
ones at low energies. Then some algebra shows that in this approximation
\begin{equation}
(\alpha-\beta)^{\rm int}=\frac{1}{2\pi^2}
\int^\infty_{\omega_0}\sqrt{1+\frac{2\omega}{M}}\left[
\sigma_{E1}(\omega)-\left( \frac32 \sqrt{1+\frac{2\omega}{M}}-\frac12\right)
\sigma_{M1}(\omega)\right]\frac{d\omega}{\omega^2}
\label{alpha-beta-int}
\end{equation}
where $\omega$ is the photon energy in the lab system,
to be compared with
\begin{equation}
(\alpha-\beta)^s=\frac{1}{2\pi^2}
\int^{\infty}_{\omega_0}
    {\sqrt{1+\frac{2\omega}{M}}\left[(\sigma_{E1}(\omega)
-\sigma_{M1}(\omega)\right]}\frac{d\omega}{{\omega}^2}.
\end{equation}
For infinitely heavy particles both expressions converge against the well 
known relation
\begin{equation}
(\alpha-\beta)_{M\to \infty}= \frac{1}{2\pi^2}\int^\infty_{\omega_0}
\left(\sigma_{E1}(\omega)-\sigma_{M1}(\omega)\right) \frac{d\omega}{\omega^2} .
\label{infiniteMass}
\end{equation}
For numerical predictions, 
calculations on the basis of the complete expressions
are required. For fixed-$t$ dispersion theory this has latest been carried
out by Wissmann \cite{wissmann04}
leading to the result 
\begin{equation}
(\alpha_p-\beta_p)^{\rm int}= +7.1 (1\pi) +2.0 (2\pi) -12.2 (\Delta)=-3.1.
\label{alpha-beta-numerical}
\end{equation}
As a summary Table \ref{shortdistance} shows  differences between calculated
values for 
$(\alpha-\beta)_p$ and the  experimental value for this quantity obtained 
under conditions explained in the caption of the table.
\begin{table}[h]
\caption{Predicted information  $(\alpha-\beta)^{\rm calc}_p$ on the 
polarizability difference compared with the
experimental result $(\alpha-\beta)^{\rm exp}_p=10.5\pm 1.1$ (global average
of \cite{olmos01}) or $(\alpha-\beta)^{\rm exp}_p=10.1\pm 0.9$ (adopted
average
including all existing data \cite{baranov00,olmos01}). In
fixed-$t$ dispersion theory the prediction $(\alpha-\beta)^{\rm calc}_p$
corresponds to the integral part $(\alpha-\beta)^{\rm int}_p$,
in fixed-$\theta$ dispersion theory $(\alpha-\beta)^{\rm calc}_p$ is
either chosen to be the $s$-channel contribution only (line 3) or the
predicted $s$-channel contribution supplemented by the predicted 
$\gamma\gamma\to \pi\pi\to N\bar{N}$ $t$-channel contribution according to the
BEFT sum rule (line 4). The unit is $10^{-4}{\rm fm}^3$.}
\begin{center}
\begin{tabular}{|l|l|l|}
\hline
disp. theory & $(\alpha-\beta)^{\rm calc}_p$& $(\alpha-\beta)^{\rm exp}_p
-(\alpha-\beta)^{\rm calc}_p$\\
\hline
fixed-$t$&$(\alpha-\beta)^{\rm int}_p=-3.1$ \cite{wissmann04}
&$(\alpha-\beta)^{\rm
  as}_p=13.2 \pm 1.3$\\
fixed-$\theta$&$(\alpha-\beta)^s_p=-(5.0\pm 1.0)$
&$(\alpha-\beta)^{t\mbox{-}{\rm exp}}_p=15.1\pm 1.3$\\
fixed-$\theta$&$(\alpha-\beta)^{s+t}_p=+(9.0\pm 2.2)$
&$(\alpha-\beta)^{t\mbox{-}{\rm miss}}_p=1.1\pm 2.4$\\
\hline
\end{tabular}
\end{center}
\label{shortdistance}
\end{table}
We see that in fixed-$t$ dispersion theory we have to explain 
$(\alpha-\beta)^{\rm as}_p= 13.2\pm 1.3$ 
through a
contribution which has no  interpretation in terms of the integral
part in fixed-$t$ dispersion theory. This value, therefore, may be interpreted
as an empirical result for the asymptotic contribution to ($\alpha-\beta$).
In case of fixed-$\theta$ dispersion theory the quantity 
$(\alpha-\beta)^{t\mbox{-}{\rm miss}}_p$ is compatible with zero. 
This  means that the BEFT sum-rule is likely to be confirmed.

An interesting alternative for the prediction of $(\alpha-\beta)$
which deserves further consideration has been proposed and evaluated 
by Akhmedov and Fil'kov \cite{akhmedov81}. In this approach $(\alpha-\beta)$
is expressed through a dispersion relation at fixed $u=M^2$ in the point $t=0$.
Due to the different dispersion theory a different partition 
of $(\alpha-\beta)$ into $s$ and $t$ channel parts is obtained.  Nevertheless,
the conclusion is drawn that a  $\sigma$ meson in the intermediate state
is responsible for the largest part of $(\alpha-\beta)$.

\subsubsection{The effective $\sigma$ pole and an  outlook on
  an interpretation of $(\alpha-\beta)$}

According to \cite{scadron04}  we have reasons to assume that the $\sigma$
meson has a composite structure of the type
\begin{equation}
|\sigma\rangle= \cos\theta_\sigma  
|\pi\pi\rangle  + \sin\theta_\sigma |q{\bar q}\rangle,
\label{sigmacomponents}
\end{equation}
implying that  the $\sigma$ meson has two possibilities to couple
to two photons, i.e. via a $\pi$ loop and via a quark loop. 
There is ongoing work \cite{levchuk05}  where this ansatz is used to make   
predictions on $\alpha-\beta$ and the relevant invariant amplitude.
It is too early to refer to this ongoing work at the present 
stage of development. 
Instead we tentatively make use of an evaluation of the reaction
$\gamma\gamma\to\pi^0\pi^0$ \cite{filkov99} which has led to a determination
of the position of the $\sigma$ pole as well as to a determination of the
photon decay width $\Gamma_{\sigma\to\gamma\gamma}$ (Table
\ref{filkovtable}). 
We start from Eq. (21) of \cite{lvov97}
\begin{equation}
\frac{g_{\sigma NN}F_{\sigma\gamma\gamma}}{2\pi m^2_\sigma}
=(\alpha-\beta)^{\sigma\mbox{-}{\rm pole}} 
\label{AlphaBetaPole}
\end{equation}
and first insert the nominal value of the $\sigma$ mass
$m_\sigma = 600\,\, {\rm MeV} \hat{=} 3.04\,\,{\rm fm}^{-1}$ 
which has been found to be consistent with experimental differential cross
sections in the second resonance region \cite{galler01,wolf01}
(see also Section 3.8). 
The relation  to the decay width is given by
\begin{equation}
g_{\sigma NN} F_{\sigma\gamma\gamma}=+16 \pi 
\sqrt{\frac{g^2_{\sigma NN}}{4\pi}\frac{\Gamma_{\sigma\to 2\gamma}}
{m^3_\sigma}}
\label{AlphaBetaPole-1}
\end{equation}
where the approximate equality of the two coupling constants in
(\ref{AlphaBetaPole-1})
\begin{equation}
\frac{g^2_{\sigma NN}}{4\pi}\simeq \frac{g^2_{\pi NN}}{4\pi}=13.75
\label{AlphaBetaPole-2}
\end{equation}
may be justified through the linear $\sigma$ model. Using
\begin{equation}
\Gamma_{\sigma\to \gamma\gamma} = 0.68\pm 0.19 \,\,{\rm keV}
\label{decaywidth}
\end{equation} 
as evaluated by Fil'kov and Kashevarov \cite{filkov99}, we arrive at 
\begin{equation}
(\alpha-\beta)^{\sigma\mbox{-}{\rm pole}}=10.7 \pm 1.7.
\label{sigma-pole-prediction}
\end{equation}
The same calculation may be carried out using the experimental value 
$m_\sigma=(547 \pm 45)$ MeV as obtained by 
Fil'kov and Kashevarov \cite{filkov99}. Then we arrive at 
\begin{equation}
(\alpha-\beta)^{\sigma\mbox{-}{\rm pole}}= 14.8^{+2.1}_{-2.5}
(\Delta\Gamma_{\sigma\to 2\gamma}){}^{+5.2}_{-3.6}(\Delta m_\sigma).
\label{sigma-pole-prediction2}
\end{equation}
It is satisfactory to see that the
numbers obtained for $(\alpha-\beta)^{\sigma\mbox{-}{\rm pole}}$ 
are compatible with $(\alpha-\beta)^{\rm as}=13.2\pm 1.3$
given in the third row of Table \ref{shortdistance}.

 In a quark-level linear $\sigma$ model (L$\sigma$M) \cite{scadron04}
the decay widths of the $\sigma\to\gamma\gamma$ decay has been obtained in the
form 
\begin{equation}
\Gamma_{\sigma\to\gamma\gamma}=\frac{m^3_\sigma}{64\pi}\left[
\frac53 \frac{\alpha_e}{\pi f_\pi}+ \frac13 \frac{\alpha_e}{\pi f_\pi}
\right]^2 \approx 3.5 \,\,\, \mbox{keV}
\label{quarklevel}
\end{equation}
for $m_\sigma=650$ MeV. Here, the first term is due to the nonstrange
quark triangle, while the second term stems from charged-kaon and -pion
triangle graphs. This result was found to be compatible with the data estimate
given in \cite{boglione99}.
We cite this result as a possible guidance for future research in connection
with attempts of including  the physics of the scalar 
$t$ channel of $(\alpha-\beta)$
into a model of the nucleon.

\section{Summary  and Discussion}

It has been shown that Compton scattering by the nucleon at energies 
below 1 GeV provides insight into the structure of the nucleon which
barely can be seen by other methods. It has been found advantageous to 
analyze the results of Compton scattering experiments in terms of
four fundamental sum rules:
\begin{enumerate}
\item
The Baldin or Baldin-Lapidus (BL) sum rule for the sum of
electromagnetic polarizabilities $(\alpha+\beta)$, related to
spin-independent forward scattering.
\item
The Gerasimov-Drell-Hearn (GDH) sum rule for the square of the
anomalous magnetic moment $\kappa^2$, related to  spin-dependent 
forward scattering.
\item
The Bernabeu-Ericson-FerroFontan-Tarrach (BEFT) sum rule for
the difference of the electromagnetic polarizabilities
$(\alpha-\beta)$, related to spin-independent backward scattering.
\item
The L'vov-Nathan (LN) sum rule for the backward spin-polarizability
$\gamma_\pi$, related to spin-dependent backward scattering. 
\end{enumerate}

The BL sum rule has been found to be fulfilled within rather small
experimental errors. The largest part of the BL-integral is related to
the conventional constituent-quark-meson structure of the nucleon.
A smaller fraction of the BL integral is related to 
tensor-meson and pomeron exchanges. This part can be taken
into account by making a Regge ansatz to describe the high-energy
total photo-absorption cross section $\sigma_{\rm tot}(\omega)$.

The result for the GDH sum rule is very similar to the one for the BL
sum rule.  The largest part of the GDH-integral is related to the 
conventional constituent-quark-meson structure of the nucleon. For the 
mesonic part of the nucleon structure showing up in the energy range from 
0.14 -- 0.20 GeV, the cross section $\sigma_{1/2}$
is larger than the cross section $\sigma_{3/2}$ whereas for  the resonant
part   due to the excitation of the constituent-quark structure the 
opposite is true.
In the Regge range the cross section $\sigma_{1/2}$ is again larger
than the cross section $\sigma_{3/2}$ thus making contributions to the 
GDH integral negative. Taking all these parts together we find the
GDH sum rule likely to be fulfilled.

For backward scattering the conventional constituent-quark-meson
structure is of minor importance. Instead, intermediate states of the
scattering process are observed where the production and annihilation
of mesons is essential. For the BEFT sum rule the relevant  meson is
the scalar-isoscalar $\sigma$ meson, whereas for the LN sum rule
the pseudoscalar $\pi^0$ is the most relevant one with minor 
contributions from the $\eta$ and $\eta'$ mesons. 
It has been shown that the LN sum rule is fulfilled whereas
the BEFT sum rule is likely to be fulfilled.
Furthermore, there are arguments that the $\sigma$ meson
has a composite structure
with a $|\pi\pi\rangle$ and a $|q\bar{q}\rangle$ component.
In order to arrive at a consistent description of the observations 
made so far for the $t$-channel part of the BEFT sum rule, this sum
rule  may  be investigated
in terms of a quark-level linear $\sigma$ model where the 
coupling of the $\sigma$
meson to two photons  takes place via a pion loop as well as via a quark loop.

If we start from the reasonable supposition that the electromagnetic
polarizabilities and spin polarizabilities are quantities which are
related to the internal structure of the nucleon, we  come to
the conclusion that the $\pi^0$ and $\sigma$ $t$-channel exchanges  have
to be considered as part of the nucleon structure. Since these
intermediate states cannot be understood in terms of an excitation of
the constituent-quark-meson structure of the nucleon they have to be
considered as part of the constituent quarks or of the surrounding 
QCD vacuum. This view certainly provides a fascinating aspect for
further investigations. \\

\noindent
Acknowledgement:\\
The author is indebted to M.I. Levchuk, A.I. L'vov and A.I. Milstein for
valuable discussions, for carefully reading the manuscript, and for providing
results of their work prior to publication. He is indebted to F. Smend  
and F. Wissmann for carefully reading the manuscript and many valuable
comments. He thanks T.R. Hemmert and an anonymous  referee
for clarifying information on chiral perturbation
theory.

\newpage


\begin{thebibliography}{99}

\bibitem{drechsel03}
D. Drechsel, B. Pasquini, M. Vanderhaeghen, {\it Phys. Rept.} 378 (2003) 99

\bibitem{wissmann04}
F. Wissmann, {\it Springer Tracts in Modern Physics}  Volume 200 (2004)1


\bibitem{lvov93}
A.I. L'vov, {\it Int. J. Mod. Phys.} A 8 (1993) 5267


\bibitem{petrunkin81}
V.A. Petrun'kin, {\it Fiz. Elem. Chastits At. Yadra} 12 (1981) 692
[{\it Sov. J. Part. Nucl.} 12 (1981) 278]



\bibitem{sachs50}
R.G. Sachs, L.L. Foldy, {\it Phys. Rev.} 80 (1950) 824

\bibitem{aleksandrov56}

Yu. A. Aleksandrov, P.I. Bondarenko, {\it Zh. Eksp. Teor. Fiz.}
31 (1956) 726  [{\it Sov. Phys. - JETP} 4 (1957) 612]



\bibitem{low54}
F.E. Low, {\it Phys. Rev.} 96 (1954) 1428

\bibitem{gell-mann54}
M. Gell-Mann, M.L. Goldberger, 
{\it Phys. Rev.} 96 (1954) 1433


\bibitem{klein55}
A.Klein, {\it Phys. Rev.} 99 (1955) 998


\bibitem{baldin60}
A.M. Baldin, {\it Nucl. Phys.} 18 (1960) 310



\bibitem{petrunkin61}
V.A. Petrun'kin, {\it Zh. Eksp. Teor. Fiz.} 40 (1961) 1148
[{\it Sov. Phys.-JETP} 13 (1961) 808]

\bibitem{petrunkin64}
V.A. Petrun'kin, {\it Nucl. Phys. 55} (1964) 197

\bibitem{petrunkin68}
V.A. Petrun'kin, {\it Tr. P.N. Lebedev Phys. Inst.} (in Russian) 41 (1968) 165

\bibitem{shekhter68}
V.M. Shekhter, {\it Yad. Fiz.} 7 (1968) 1272 
[{\it Sov. J. Nucl. Phys.} 7 (1968) 756]


\bibitem{goldansky60}
V.I. Gol'danskii, O.A. Karpukhin, A.V. Kutsenko, V.V. Pavlovskaya,
{\it Zh. Eksp Teor. Fiz} 38 (1960) 1695 
[{\it Sov. Phys.-JETP} 11 (1960) 1223]; Nucl. Phys. 18 (1960) 473

\bibitem{baranov74} 
P.S. Baranov, L. Fil'kov, L.N. Shtarkov, {\it ZhETF Pis. Red.} 20 (1974)
762 [{\it JETP Lett.} 20 (1974) 353];
P.S. Baranov, et al., {\it Yad. Fiz.} 21 (1975)  689
[{\it Sov. J. Nucl. Phys.} 21 (1975) 355];
P.S. Baranov et al., {\it Phys. Lett.} B 52 (1974) 22; {\it Sov. J. Phys.} 21
(1975) 355


\bibitem{baranov00}
P.S. Baranov, A.I. L'vov, V.A. Petrun'kin, L.N. Shtarkov,
{\it 9th International Seminar on Electromagnetic Interactions of
  Nuclei at Low and Medium Energies, Moscow, Russia, 20-22 Sep 2000}, 
{\it e-Print Archive:} nucl-ex/0011015, {\it Phys. Part. Nucl.} 32 (2001) 376


\bibitem{aleksandrov92}
Yu. A. Aleksandrov, {\it Fundamental Properties of the Neutron}
(Clarendon Press, Oxford 1992)


\bibitem{alexandrov86}
Y.A. Alexandrov, et al., {\it Sov. J. Nucl. Phys.} 44 (1986) 900

\bibitem{koester86}
L. Koester, et al., {\it Physika} B 137 (1986) 282

\bibitem{schmiedmayer88}
J. Schmiedmayer, et al., {\it Phys. Rev. Lett.} 61 (1988) 1065

\bibitem{koester88}
L. Koester, et al., {\it Z. Phys.} A 329 (1988) 229


\bibitem{hagiwara02}
K. Hagiwara, et al. (Particle Data Group), {\it Phys. Rev.} D 66 (2002)
010001
(URL:http://pdg.lbl.gov)

\bibitem{eidelman04}
S. Eidelman, et al. (Particle Data Group), {\it Phys. Lett.} 
B 592 (2004) 1; URL:
http://pdg.lbl.gov


\bibitem{landau90}

L.D.  Landau, E.M. Lifschitz, {\it Lehrbuch der Theoretischen Physik IV,
Quantenelektrodynamik} (Akademieverlag, Berlin 1990); 
V.B. Berestetskii, E.M. Lifshitz, L.P. Pitaevskii, 
{\it Quantum Electrodynamics}  (Pergamon, New York 1982)


\bibitem{hecking81}
P.C. Hecking, G.F. Bertsch, {\it Phys. Lett.} B 99 (1981) 237; 
A. Sch\"afer,
B. M\"uller, D. Vasak, W. Greiner, {\it Phys. Lett.} B 143 (1984) 323;


\bibitem{dattoli77}
G. Dattoli, G. Matone, D. Prosperi, {\it Lett. Nuovo Cim.} 19 (1977) 601;
D. Drechsel, A. Russo, {\it Phys. Lett.} B 137 (1982) 295;
M. De Sanctis, D. Prosperi, {\it Nuovo Cim.} A 103 (1990) 1301;
S. Capstick,
B. D. Keister, {\it Phys. Rev.} D 46 (1992)  84 ;
H. Liebl, G.R. Goldstein, {\it Phys. Lett.} B 343 (1995) 363;
M. Traini, R. Leonardi,
{\it Phys. Lett.} B 334 (1994) 7 


\bibitem{weiner85}
R. Weiner, W. Weise, {\it Phys. Lett.}  B 159  (1985)  85;
N.N. Scocola, W. Weise, {\it Nucl. Phys.} A 517 (1990) 495



\bibitem{scocola89}
N.N. Scocola, W. Weise, {\it Phys. Lett.} B 232 (1989) 287;
W. Broniowski, M.K. Banerjee, T.D. Cohen, {\it Phys. Lett.} B 283 (1992) 22


\bibitem{nyman84}
E.M. Nyman, {\it Phys. Lett.} B 142 (1984) 388; 
M. Chemtob, {\it Nucl. Phys.} A
473 (1987)  613;
S. Scherer, P.J. Mulders, {\it Nucl. Phys.} A 549 (1992)   521



\bibitem{bernard88}
V. Bernard, B. Hiller, W. Weise, {\it Phys. Lett.} B 205 (1988) 16;
V. Bernard, D. Vautherin, {\it Phys. Rev.} D 40 (1989) 1615;
B.R. Holstein, {\it Comment Nucl. Part. Phys.} 19 (1990) 221

\bibitem{ericson73}
T.E.O. Ericson, J. H\"ufner, {\it Nucl. Phys.} B 57 (1973) 604

\bibitem{friar75}
J.L. Friar, {\it Ann Phys. (N.Y.)} 95 (1975) 170 

\bibitem{lee01}
R.N. Lee, A.I. Milstein, M. Schumacher, 
{\it Phys. Rev. Lett.} 87 (2001) 051601;
R.N. Lee, A.I. Milstein, M. Schumacher,
{\it Phys. Rev.} A 64 (2001) 032507;
R.N. Lee, A.I. Milstein, M. Schumacher,
{\it Phys. Lett.} B 541 (2002) 87

\bibitem{schumacher94}
M. Schumacher, et al., 
{\it Nucl. Phys.} A 576 (1994) 603

\bibitem{huett00}
M.-Th. H\"utt, A.I. L'vov, A.I. Milstein, M. Schumacher,
{\it Physics Reports} 323 (2000) 457


\bibitem{pilkuhn79}
H.M. Pilkuhn, {\it Relativistic Particle Physics} (Springer Verlag, New York,
1979)

\bibitem{bernard92}
V. Bernard, N. Kaiser, U.-G. Meissner,
{\it Phys. Rev. Lett.} 67 (1991) 1515; 
{\it Nucl. Phys.} B  373 (1992) 346

\bibitem{lvov93a}
A.I. L'vov, {\it Phys. Lett.} B 304 (1993)  29 

\bibitem{thomas01}
A.W. Thomas, W. Weise, 
{\it The Structure of the Nucleon}
(WILEY-VCH Verlag Berlin GmbH 2001) 

\bibitem{bernard95}
V. Bernard, N. Kaiser, U.-G. Meissner, 
{\it Int. Journ. Mod. Phys.} E 4 (1995) 193


\bibitem{bernard92a}
V. Bernard, N. Kaiser, J. Kambor, U.G. Meissner, 
{\it Nucl. Phys.} B 388 (1992) 315 


\bibitem{bernard93}
V. Bernard, N. Kaiser, A. Schmidt, U.-G. Meissner,
{\it Phys. Lett.} B. 319 (1993) 269;
V. Bernard, N. Kaiser, U.-G. Meissner, A. Schmidt,
{\it Z. Phys.} A 348 (1994) 317


\bibitem{hemmert98}
T.R. Hemmert, B.R. Holstein, J. Kambor, G. Kn\"ochlein, 
{\it Phys. Rev.} D 57 (1998) 5746;
T.R. Hemmert, B.R. Holstein, J. Kambor,
{\it Phys. Rev.} D 55 (1997) 5598

\bibitem{pascalutsa04}
V. Pascalutsa,
{\it e-Print Archive:} nucl-th/0412008



\bibitem{hildebrandt03}
R.P. Hildebrandt, H.W. Griesshammer, T. R. Hemmert, B. Pasquini,
{\it Eur. Phys.} J. A 20 (2004) 293;
{\it e-Print Archive:} nucl-th/0307070

\bibitem{pascalutsa03}
V. Pascalutsa, D.R. Phillips, {\it Phys. Rev.} 67 (2003) 055202; 
{\it e-Print Archive:} nucl-th/0212024; J.A. McGovern, {\it
Phys. Rev.} C 63 (2001) 064608;
{\it e-Print Archive:} nucl-th/0101057;
S.R. Beane et al., {\it e-Print Archive:} nucl-th/0403088

\bibitem{low58}
F.E. Low, Proc. 1958 {\it Ann. Intern. Conf. on High Energy Physics at
CERN}, p.98

\bibitem{hearn62}
A.C. Hearn, E. Leader, {\it Phys. Rev.} 126 (1962) 789


\bibitem{koeberle68}
R. K\"oberle, {\it Phys. Rev.} 166 (1968) 1588

\bibitem{scadron04}
M.D. Scadron, et al., {\it Phys. Rev. D 69 (2004) 014010}; Erratum-ibid.
D 69 (2004) 059901; {\it e-Print Archive:} hep-ph/0309109

\bibitem{levchuk05}
M.I. Levchuk, A.I. L'vov, A.I. Milstein, M. Schumacher (to be submitted)


\bibitem{holstein00}
B. Holstein, et al.,  {\it Phys. Rev.} 61 (2000) 034316


\bibitem{babusci98}
D. Babusci, G. Giordano, A.I. L'vov, G. Matone,
A.M. Nathan,  {\it Phys. Rev.} C 58 (1998) 1013


\bibitem{walker69}
R.L. Walker, {\it Phys. Rev.} 182 (1969) 1729


\bibitem{gilman72}
F.J. Gilman, {\it Phys. Rep.} 4 (1972) 95


 \bibitem{roman}
P. Roman, {\it Advanced Quantum Theory}  (Addison-Wesley Publishing
Company, Inc. Reading, Massachusetts 1965)

\bibitem{hoehler83}
G. H\"ohler, in: 
{\it Landolt-B\"ornstein, New Series}, Group I, Vol. 9, Subvolume b: Pion
Nucleon Scattering, Part 2: Methods and Results of Phenomenological
Analysis.



\bibitem{lapidus62}
L.I. Lapidus, {\it Zh. Eksp. Teor. Fiz.} 43 (1962) 1358
[{\it Sov. Phys. JETP}
16 (1963)  964] 


\bibitem{gerasimov66}
S.B. Gerasimov, {\it Sov. J. Nucl. Phys.} 2 (1966) 430 

\bibitem{drell66}
S.D. Drell and A.C. Hearn, {\it Phys. Rev. Lett.} 16 (1966) 908 

\bibitem{hosada66}
M. Hosada and K. Yamamoto, {\it Prog. Theor. Phys.} 36 (1966) 425 



\bibitem{jacob59}
M. Jacob, G.C. Wick, {\it Ann. Phys. (NY)} 7 (1959) 404

\bibitem{hara72}
Y. Hara, {\it Prog. Theor. Phys. Suppl.} 51 (1972) 96




\bibitem{bernabeu74}
J. Bernabeu, T.E.O. Ericson, C. Ferro Fontan, {\it Phys. Lett.} 49B (1974)
381
  
\bibitem{bernabeu77}
J. Bernabeu and B. Tarrach, {\it Phys. Lett.} 69B (1977) 484



\bibitem{lvov99}
A.I. L'vov, A.M. Nathan, {\it Phys. Rev.} C 59 (1999) 1064


\bibitem{lvov97}
A.I. L'vov, V.A. Petrun'kin, M. Schumacher, {\it Phys. Rev.} C 55 (1997) 359


\bibitem{landau48}
L.D. Landau, {\it Dokl. Akad. Nauk.}, USSR 60 (1948) 207

\bibitem{yang50}
C.N. Yang, {\it Phys. Rev.} 77 (1950) 242



\bibitem{schwinger57}
J. Schwinger, {\it Ann. Phys.} 2 (1957) 407

\bibitem{gellmann60}
M. Gell-Mann and M. Levy, {\it Nuovo Cim.} 16 (1960) 705

\bibitem{taketani67}
M. Taketani, et al., {\it Prog. Theor. Phys. Suppl.} No. 39 (1967) 1;
K. Erkelenz, {\it Phys. Rep.} 5 (1974) 191;
R. Machleidt, K. Holinde, Ch. Elster, {\it Phys. Rep.} 149
(1987) 1

\bibitem{nambu61}
Y. Nambu, G. Jona-Lasinio, {\it Phys. Rev.} 122 (1961) 345; 
R. Delbourgo, M.D. Scadron, {\it Phys. Rev. Lett.} 48 (1982) 379;
T. Hatsuda, T. Kunihiro, {\it Phys. Lett.} B 145 (1984) 7;
{\it Prog. Theor. Phys.} 74 (1985) 765;
{\it Phys. Rep.} 247 (1994) 221;
V. Elias, M.D. Scadron, {\it Phys. Rev. Lett.} 53 (1984) 1129; 
S. Klimt, M. Lutz, U. Vogl, W. Weise, {\it Nucl. Phys.} A 516 (1990) 429 

\bibitem{morgan87}
K.L. Au, D. Morgan, M.R. Pennington, {\it Phys. Rev.} D 35 (1987) 1633;
D. Morgan, M.R. Pennington, {\it Phys. Rev.} D 48 (1993) 1185

 \bibitem{ishida96}
M. Ishida, {\it Prog. Theor. Phys.} 96 (1996) 853; 
{\it e-Print Archive:} hep-ph/9905261



\bibitem{achasov94}
N.N. Achasov, G.N. Schestakov, {\it Phys. Rev.} D 49 (1994) 5779;
R. Kaminski, L. Lesniak, J.-P. Maillet, {\it Phys. Rev.} D 50 (1994) 3145;
N.A. Tornqvist, M. Roos, {\it Phys. Rev. Lett.} 76 (1996) 1575;
M. Svec, {\it Phys. Rev.} D 53 (1996) 2343;
M. Harada, F. Sannio, J. Schechter, {\it Phys. Rev.} D 54 (1996) 1991;
S. Ishida, M. Ishida, H. Takahashi, T. Ishida, K. Takamatsu, T. Tsuru,
{\it Prog. Theor. Phys.} 95 (1996) 745; 98 (1997) 1005; I.G. Alekseev, 
et al., {\it Phys. At. Nucl.} 61 (1998) 174



\bibitem{colangelo01}
G. Colangelo, J. Gasser, H. Leutwyler, {\it Nucl. Phys.} B 603 (2001) 125 

\bibitem{ishida03}
M.  Ishida, {\it Prog. Theor. Phys. Suppl.} 149 (2003) 190; 
{\it e-Print Archive:} hep-ph/0212383 


\bibitem{drechsel99}
D. Drechsel, M. Gorchtein, B. Pasquini, M. Vanderhaeghen,
{\it Phys. Rev.} C 61 (1999) 015204


\bibitem{filkov99}
L.V. Fil'kov, V. L. Kashevarov, {\it Eur. Phys.} J A 5 (1999) 285

\bibitem{marsiske90}
H. Marsiske, et al., {\it Phys. Rev.}
D 41 (1990) 3324

\bibitem{ahrens04}
J. Ahrens, et al., {\it e-Print Archive:}
nucl.-ex/0407011



\bibitem{rostovtsev01}
A. Rostovtsev, 
{\it Surveys High Energy Phys.} 16 (2001) 209; 
{\it e-Print Archive:} hep-ph/0108019;
C.W. Akerlof, et al., {\it Phys. Rev.} D 14 (1976) 2864; 
V. N. Bolotov, et al., {\it } Nucl. Phys. B 73 (1974) 365

\bibitem{donnachie92} 
A. Donnachie, P.V. Landshoff, 
{\it Phys. Lett.} B 296 (1992) 227


\bibitem{hesse70}
W.P. Hesse, et al., 
{\it Phys. Rev. Lett.} 25 (1970) 613


\bibitem{bianchi99}
N. Bianchi and E. Thomas, {\it Phys. Lett.} B 450 (1999)  439  



\bibitem{prange58}
R.E. Prange, {\it Phys. Rev.} 110 (1958) 240

\bibitem{bardeen68}
W.A. Bardeen and W.K. Tung, {\it Phys. Rev.} 173 (1968) 1423

\bibitem{lvov81}
A.I. L'vov, {\it Sov. J. Nucl. Phys.} 34 (1981) 597


\bibitem{lvov04}
A.I. L'vov, private communication (2004)



\bibitem{milstein03}
A.I. Milstein, private communication (2003)


\bibitem{chew57}
G.F. Chew, M.L. Goldberger, F.E. Low, Y Nambu,
{\it Phys. Rev.} 106 (1957) 1345



\bibitem{karliner73}
I. Karliner, {\it } Phys. Rev. D 7 (1973) 2717




\bibitem{arndt96}
R.A. Arndt, I.I. Strakovsky, R.L. Workman, {\it Phys. Rev.} C 53 (1996) 430


\bibitem{arndt95}
Computer code SAID, solution SM95 (1995)
 

\bibitem{peise96}
J. Peise, et al., {\it Phys. Lett.}  B 384 (1996) 37;  
A. H\"unger, et al., Nucl. Instr. Meth. A 372 (1996) 135;
A. H\"unger, et al.,
{\it Nucl. Phys.} A 620 (1997)  385 


\bibitem{MAID}
http://www.kph.uni-mainz.de/MAID/; S.S. Kamalov, S.N. Yang, D. Drechsel,
O. Hanstein, L. Tiator, {\it Phys. Rev.} C 64 (2001) 032201; 
http://gwdac.phys.gwu.edu; R.A. Arndt, W.J. Briscoe, I.I. Strakovsky,
R.L. Workman, {\it Phys. Rev.} C 66 (2002) 055213



\bibitem{federspiel91}
F.J. Federspiel, et al., {\it Phys. Rev. Lett.} 67 (1991) 1551


\bibitem{lvov79}
A.I. L'vov, V.A. Petrun'kin, S.A. Startsev, {\it Sov. J. Nucl. Phys.} 29 (1979)
651 



\bibitem{zieger92}
A. Zieger, et al., {\it Phys. Lett.} B 278 (1992) 34

\bibitem{hallin93}
E.L. Hallin, et al., {\it Phys. Rev.} C 48 (1993) 1497

\bibitem{macgibbon95}
B.E. MacGibbon, et al., {\it Phys. Rev.} C 52 (1995) 2097

\bibitem{olmos01}
V. Olmos de Le\'on, et al., {\it Eur. Phys. J.} A 10 (2001) 207

\bibitem{tonnison98}
J. Tonnison, et al.,
{\it Phys. Rev. Lett.}
80  (1998)  4382 


\bibitem{wissmann99}
F. Wissmann, et al., {\it Nucl. Phys.} A 660 (1999) 232
 


\bibitem{galler01}
G. Galler, et al., {\it Phys. Lett.} B 501 (2001) 245

\bibitem{wolf01}
S. Wolf, et al., {\it Eur. Phys. J.} A 12 (2001) 231



\bibitem{drechsel99a}
D. Drehsel, O. Hanstein, S.S. Kamalov, L. Tiator,
{\it Nucl. Phys.} A 645 (1999)  145 

\bibitem{camen02}
M. Camen, et al., {\it Phys. Rev.} C 65 (2002) 032202



\bibitem{blanpied01}
G. Blanpied, et al., {\it Phys. Rev.} C 64 (2001)  025203 

\bibitem{schmiedmayer91}
J. Schmiedmayer, et al., {\it Phys. Rev. Lett.} 66 (1991) 1015

\bibitem{koester95}
L. Koester, et al., {\it Phys. Rev.} C 51 (1995) 3363



\bibitem{enic97}
T.L. Enik, et al., {\it Sov. J. Nucl. Phys.} 60 (1997) 567 

\bibitem{wissmann98}
F. Wissmann, M.I. Levchuk, M. Schumacher, {\it Eur. Phys. J.} A 1 (1998) 193


\bibitem{levchuk86}
M.I. Levchuk, A.I. L'vov, V.A. Petrun'kin, {\it Lebedev Physical
Institute Preprint FIAN} No. 86;
{\it Few-Body Syst.} 16 (1994) 101


\bibitem{rose90}
K.W. Rose et al., {\it Phys. Lett.} B 234 (1990) 460; 
{\it Nucl. Phys.} A 514 (1990) 621



\bibitem{kossert02}
K. Kossert et al., {\it Phys. Rev. Lett.} 88 (2002) 162301; 
{\it Eur. Phys. J.}
A 16 (2003)  259, {\it e-Print Archive:} nucl-ex/0210020



\bibitem{levchuk00}
M.I. Levchuk, A.I. L'vov, {\it Nucl. Phys.} A 674 (2000) 449;
A 684 (2001) 490


\bibitem{kolb00}
N.R. Kolb, et al., {\it Phys. Rev. Lett.} 85 (2000)1388

\bibitem{lundin03}
M. Lundin, et al., {\it Phys. Rev. Lett.} 90 (2003) 192501-1

\bibitem{hildebrandt04}
R.P. Hildebrandt, H.W. Griesshammer, T.R. Hemmert, D.R. Phillips,
{\it e-Print Archive:}
nucl-th/0405077; Accepted for publication in {\it Eur. Phys. J.} A

\bibitem{beane04}
S.R. Beane, M. Malheiro, J.A. McGovern, D.R. Phillips, U. van Kolck,
{\it e-Print Archive:} nucl-th/0403088

\bibitem{molinari96}
C. Molinari, et al., Phys. Lett. B 371 (1996) 181



\bibitem{baranov66}
P.S. Baranov, et al., {\it JETP} 23 (1966)  242 ; 
{\it Sov. J. Nucl. Phys.}
3 (1966)  3791 

\bibitem{gray67}
E.R. Gray, A.O. Hanson, {\it Phys. Rev.} 160 (1967) 121 

\bibitem{genzel76}
H. Genzel, M. Jung, R. Wedemeyer, H.J. Weyer, {\it Z. Phys.} A 279 
(1976) 399



\bibitem{ishii80}
T. Ishii, et al., {\it Nucl. Phys.} B 165 (1980) 189 


\bibitem{damashek70}
M. Damashek, F.J. Gilman, {\it Phys. Rev.} D 1 (1970) 1319 


\bibitem{karliner-a}
R.L. Workman, R.A. Arndt,
{\it Phys. Rev.} D 45 (1992) 1789; A.M. Sandorfi, C.S. Whisnant,
M. Khandaker, {\it Phys. Rev.} D 50 (1994) R6681; R.A, Arndt,
I.I. Strakovsky, R. Workman, {\it Phys. Rev.} C 53 (1996) 430;
D. Drechsel, G. Krein, {\it Phys. Rev.} D 58 (1998) 116009

\bibitem{fox69}
G.C. Fox, D.Z. Freedman, {\it Phys. Rev.} 182 (1969) 1628; 
S. B. Gerasimov, {\it Yad. Fiz.} 5 (1967) 1263 
[{\it Sov. J. Nucl. Phys.} 5 (1969) 902];
H.D. Abarbanel, M.L. Goldberger, {\it Phys. Rev.} 165 (1968) 1594; J.B.
 Bronzan, I.S. Gerstein, B.W. Lee, F.E. Low,
{\it Phys. Rev. Lett.} 18 (1967) 32; {\it Phys. Rev.} 157 (1967) 1448;
K. Kawarabayashi, M. Suzuki, {\it Phys. Rev.} 150 (1966) 1362; 152 (1966)
1383; K. Kawarabayashi, W. Wada, {\it Phys. Rev.} 152 (1966) 1286;
L.N. Chang, Y. Liang, {\it Phys. Lett.} B 268 (1991) 64; 
{\it Phys. Rev.} D 45
(1992) 2121; L. N. Chang, Y. Liang, R.L. Workman, {\it Phys. Lett.} B 329
(1994) 514

\bibitem{ahrens00}
J. Ahrens, et al., {\it Phys. Rev. Lett.} 84   (2000)   5950 

\bibitem{ahrens01}
J. Ahrens, et al., {\it Phys. Rev. Lett.} 87 (2001)  022003 
\bibitem{ahrens02}
J. Ahrens, et al., {\it Phys. Rev. Lett.} 88 (2002)   232002 

\bibitem{dutz03}
H. Dutz, et al., {\it Phys. Rev. Lett.} 91 (2003)  192001 

\bibitem{dutz04}
H. Dutz, et al., {\it Phys. Rev. Lett.} (to be published)

\bibitem{tiator02}
L. Tiator, Proc. GDH2002, p. 27, M. Anghinolfi, M. Battaglieri, 
R. De Vita Eds. (World Scientific, New Jersey  2002)



\bibitem{simula02}
S. Simula, et al., {\it Phys. Rev.} D 65 (2002)  034017 


\bibitem{guiasu76}
I. Guiasu and E.E. Radescu, {\it Phys. Rev.} D14 (1976) 1335;
{\it Phys. Lett.} 62B (1976) 193

\bibitem{guiasu78}
I. Guiasu and E.E. Radescu,
{\it Phys. Rev.} D 18 (1978) 1728

\bibitem{budnev79}
V.M. Budnev, V.A. Karnakov, {\it Yad. Fiz.} 30 (1979) 440
[{\it Sov. J. Nucl. Phys.} 30 (1979) 228]


\bibitem{holstein94}
B.R. Holstein, A.M. Nathan, {\it Phys. Rev.} D 49 (1994) 6101


\bibitem{omnes58}
R. Omn\`es, {\it Nuovo Cimento} 8 (1958) 316


\bibitem{bohannon76}
G.E. Bohannon, {\it Phys. Rev.} D 14 (1976) 126


\bibitem{frazer60}
W.R. Frazer and Fulco, {\it Phys. Rev.} 117 (1960) 1603


\bibitem{morgan88}
D. Morgan, M.R. Pennington, {\it Z. Phys.} C 37 (1988) 431



\bibitem{levchuk04}
M.I. Levchuk, private communication (2004)



\bibitem{kaloshin94}
A.E. Kaloshin, V.V. Serebryakov, {\it Z. Phys.} C 64 (1994) 691

\bibitem{pennington94}
M.R. Pennington, {\it Nucl. Phys.} A 623 (1994) 189c

\bibitem{hyams73}
B. Hyams, et al., {\it Nucl. Phys.} B 64 (1973) 134

\bibitem{froggatt77}
C.D. Froggatt, J.L. Peterson, {\it Nucl. Phys.} B 129 (1977) 89

\bibitem{gourdin60}
M. Gourdin, A. Martin, Nuovo. Cim. 17 (1960) 224


\bibitem{boyer90}
J. Boyer, et al., {\it Phys. Rev.} D 42 (1990) 1350

\bibitem{behrend92}
H.J. Behrend, et al., {\it Z. Phys.} C 56 (1992) 381


\bibitem{akhmedov81}
D.M. Akhmedov, L.V. Fil'kov, 
{\it Yad. Fiz.} 33 (1981) 1083
[{\it Sov. J. Nucl. Phys.} 33 (1981) 573 ]

\bibitem{boglione99}
M. Boglione and M.R. Pennington, {\it Eur. Phys. J.} C 9 (1999) 11;
{\it e-Print Archive:} hep-ph/9812258

\end{thebibliography}
\end{document}